\documentclass[11pt]{article}
\usepackage{epsfig,cite}
\usepackage{amsmath}
\usepackage{amssymb}
\usepackage{amsfonts}
\usepackage{latexsym}
\usepackage{wasysym}
\usepackage{url}
\usepackage{hyperref}
\usepackage{bm}
\usepackage{graphics}
\usepackage{multirow}
\usepackage{rotating}
\usepackage{booktabs}

\def\bea{\begin{eqnarray}}
\def\eea{\end{eqnarray}}
\def\beq{\begin{equation}}
\def\eeq{\end{equation}}
\catcode`@=11 
\def\slash#1{\mathord{\mathpalette\c@ncel#1}}
 \def\c@ncel#1#2{\ooalign{$\hfil#1\mkern1mu/\hfil$\crcr$#1#2$}}
\def\lsim{\mathrel{\mathpalette\@versim<}}
\def\gsim{\mathrel{\mathpalette\@versim>}}
 \def\@versim#1#2{\lower0.2ex\vbox{\baselineskip\z@skip\lineskip\z@skip
       \lineskiplimit\z@\ialign{$\m@th#1\hfil##$\crcr#2\crcr\sim\crcr}}}
\catcode`@=12 

\def\({\left(}
\def\){\right)}
\def\[{\left[}
\def\]{\right]}

\relax

\def    \hepph  #1 {{\tt hep-ph/#1}}
\def    \hepex  #1 {{\tt hep-ex/#1}}

\newcommand{\la}{\left\langle}
\newcommand{\ra}{\right\rangle}
\newcommand{\lc}{\left[}
\newcommand{\rc}{\right]}
\newcommand{\lp}{\left(}
\newcommand{\rp}{\right)}
\newcommand{\be}{\begin{equation}}
\newcommand{\ee}{\end{equation}}

\begin{document}
\begin{figure}[h]
\epsfig{width=0.32\textwidth,figure=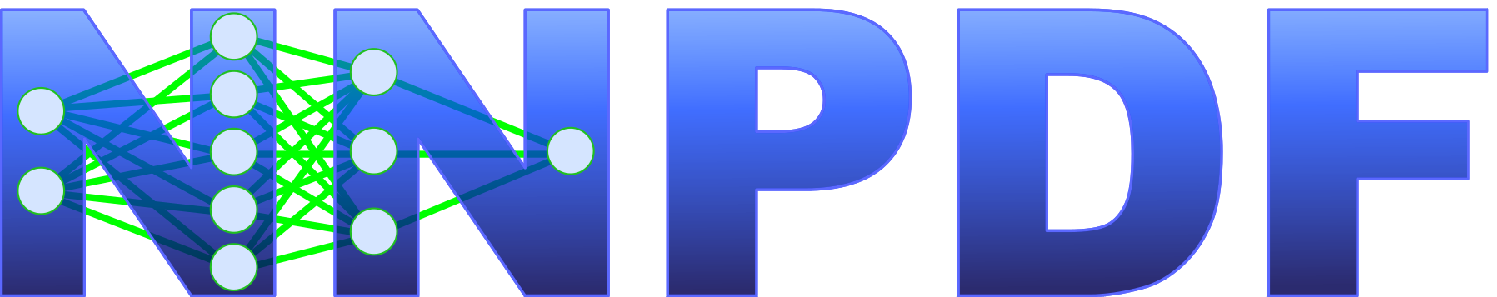}
\end{figure}
\vspace{-2.0cm}

\begin{flushright}
CERN-PH-TH-2014-106 \\
IFUM-1028-FT\\
Edinburgh-14/11  \\
OUTP-14-06P \\
\end{flushright}
\begin{center}
{\Large \bf A first unbiased global determination\\ of polarized PDFs and their uncertainties}
\vspace{0.8cm}

{\bf The NNPDF Collaboration}: Emanuele~R.~Nocera,$^1$, Richard~D.~Ball,$^{2}$ Stefano~Forte,$^1$ \\
  Giovanni~Ridolfi$^3$ and
Juan~Rojo.$^{4,5}$

\vspace{1.cm}

{\it 
~$^1$ Dipartimento di Fisica, Universit\`a di Milano and
INFN, Sezione di Milano,\\ Via Celoria 16, I-20133 Milano, Italy\\
~$^2$ Higgs Centre, University of Edinburgh,\\
JCMB, KB, Mayfield Rd, Edinburgh EH9 3JZ, Scotland\\
~$^3$ Dipartimento di Fisica, Universit\`a di Genova and
INFN, Sezione di Genova,\\ Via Dodecaneso 33, I-16146 Genova, Italy\\
~$^4$ PH Department, TH Unit, CERN, CH-1211 Geneva 23, Switzerland \\
~$^5$ Rudolf Peierls Centre for Theoretical Physics, 1 Keble Road,\\ University of Oxford, OX1 3NP Oxford, United Kingdom \\
}

\end{center}

\vspace{0.8cm}

\begin{center}
{\bf \large Abstract:}
\end{center}

We present a first global determination of spin-dependent
parton distribution functions (PDFs) and their uncertainties 
using the NNPDF methodology: {\tt NNPDFpol1.1}.
Longitudinally polarized
deep-inelastic scattering data, already used for the previous {\tt
  NNPDFpol1.0} PDF set,  are supplemented with the most
recent polarized hadron collider data for inclusive jet and
$W$ boson production from the STAR and PHENIX experiments at RHIC, 
and with open-charm production data from the COMPASS experiment,
thereby allowing for a separate determination of the  
polarized quark and anti-quark PDFs, and an improved determination of the 
medium- and large-$x$ polarized gluon PDF.
We study the phenomenological implications of the {\tt NNPDFpol1.1}
set, and we provide predictions for the longitudinal double-spin asymmetry
for semi-inclusive  pion production at RHIC.

\clearpage

\tableofcontents

\clearpage

\section{A global polarized PDF determination}
\label{sec:intro}

In a recent paper, we presented the {\tt NNPDFpol1.0} 
parton set~\cite{Ball:2013lla}, a first unbiased determination of 
polarized parton distribution functions (PDFs) of the proton 
and their associated uncertainties based on the NNPDF 
methodology~\cite{DelDebbio:2004qj,DelDebbio:2007ee,Ball:2008by,Ball:2010de}. 
This methodology differs from that used in other recent 
next-to-leading order (NLO)
analyses~\cite{deFlorian:2009vb,Leader:2010rb,Hirai:2008aj,Blumlein:2010rn,Jimenez-Delgado:2013boa},
in that it relies on a Monte Carlo sampling and representation of PDFs, 
and it uses a parametrization of PDFs based on neural networks with a very large
number of free parameters.

The \texttt{NNPDFpol1.0} parton set
was determined from all available 
inclusive deep-inelastic scattering (DIS) data
with longitudinally polarized beams.
One important limitation of only using (neutral-current) inclusive DIS
is that only the quark PDF
combinations $\Delta u^+=\Delta u + \Delta\bar{u}$, 
$\Delta d^+=\Delta d + \Delta\bar{d}$, 
$\Delta s^+=\Delta s + \Delta\bar{s}$, 
and the gluon  $\Delta g$ are accessible.
Furthermore, in DIS the gluon  is mostly
determined by scaling violations, and thus subject to sizable
uncertainties  due to the
restricted lever-arm in $Q^2$ of polarized DIS measurements.

In recent years, the set of experimental data which may be used for
the determination of longitudinally polarized PDFs has been extended
impressively. 
They include now semi-inclusive DIS (SIDIS) in fixed-target 
experiments~\cite{Ackerstaff:1999ey,Adeva:1997qz,Airapetian:2004zf,Alekseev:2007vi,Alekseev:2009ab},
one- or two-hadron and open-charm production in lepton-nucleon 
scattering~\cite{Airapetian:1999ib,Airapetian:2010ac,Adeva:2004dh,Adolph:2012vj,Adolph:2012ca}, and
semi-inclusive particle 
production~\cite{Adler:2004ps,Adler:2006bd,Adare:2007dg,Adare:2008qb,Adare:2008aa,Adamczyk:2013yvv,Adare:2014hsq},
high-$p_T$ jet production~\cite{Adare:2010cc,Adamczyk:2012qj,Adamczyk:2014ozi} and 
parity-violating $W^\pm$ boson production~\cite{Aggarwal:2010vc,Adare:2010xa,Adamczyk:2014xyw}
in polarized proton-proton collisions.
Many of these  data probe individual quark flavors separately, or  
combinations of them.
For instance, semi-inclusive DIS and $W^\pm$ production data allow one
to determine the light quark-antiquark separation
of polarized PDFs.
In addition, inclusive jet and pion production in polarized proton-proton
collisions, as well as hadron or open-charm electroproduction 
provide a handle on the  polarized gluon.
Available measurements that provide information on
polarized PDFs,  the corresponding leading partonic subprocesses,
the PDFs that are being probed, and the approximate  range of $x$ and
$Q^2$ covered by available data, are summarized in Tab.~\ref{tab:processes}.

It is clear from  Tab.~\ref{tab:processes} that currently available
processes 
do not extend significantly the  kinematic coverage of
polarized DIS data, even though they do provide
 independent information, and, in some cases,
first information on some PDF combinations.
It follows that only moderate improvements are expected from these
data on the first moments of polarized PDFs, which
are limited by the extrapolation to the unconstrained small-$x$
region. Only future accelerators, such as a high-energy polarized 
Electron-Ion Collider (EIC)~\cite{Deshpande:2005wd,Accardi:2012hwp,Boer:2011fh} 
or a neutrino factory~\cite{Mangano:2001mj}, could extend significantly
the coverage of the small-$x$ regime and
improve knowledge of the first 
moments of polarized PDFs~\cite{Forte:2001ph,Aschenauer:2012ve,Ball:2013tyh}.

The goal of this paper is to include in the NNPDF determination of 
polarized PDFs the new experimental information provided by 
polarized hadron collider data. This will lead to
our first global 
polarized PDF set: \texttt{NNPDFpol1.1}.
We will not include in our global fit processes which 
require knowledge of fragmentation
functions for light quarks, such as for instance
SIDIS or pion production. Fragmentation functions are on the
same footing as PDFs: they can only be determined from a fit to
experimental
data~\cite{Kretzer:2000yf,Kniehl:2000fe,Bourhis:2000gs,Hirai:2007cx,deFlorian:2007aj,Albino:2008fy,Albino:2008aa},
and as such they are subject to the same potential sources of
bias. Because our methodology aims at reducing bias,
and a determination  of fragmentation functions based on our
methodology is not yet available, we prefer not to use data which
require their use for the time being. 
We do, however, include the 
 open-charm leptoproduction data, because the fragmentation function
 for heavy quarks is almost computable in perturbation theory and
 only introduces a very moderate uncertainty. 

\begin{figure}[!t]
\begin{center}
\Large{\texttt{NNPDFpol1.1} data set}\\
\epsfig{width=0.75\textwidth,figure=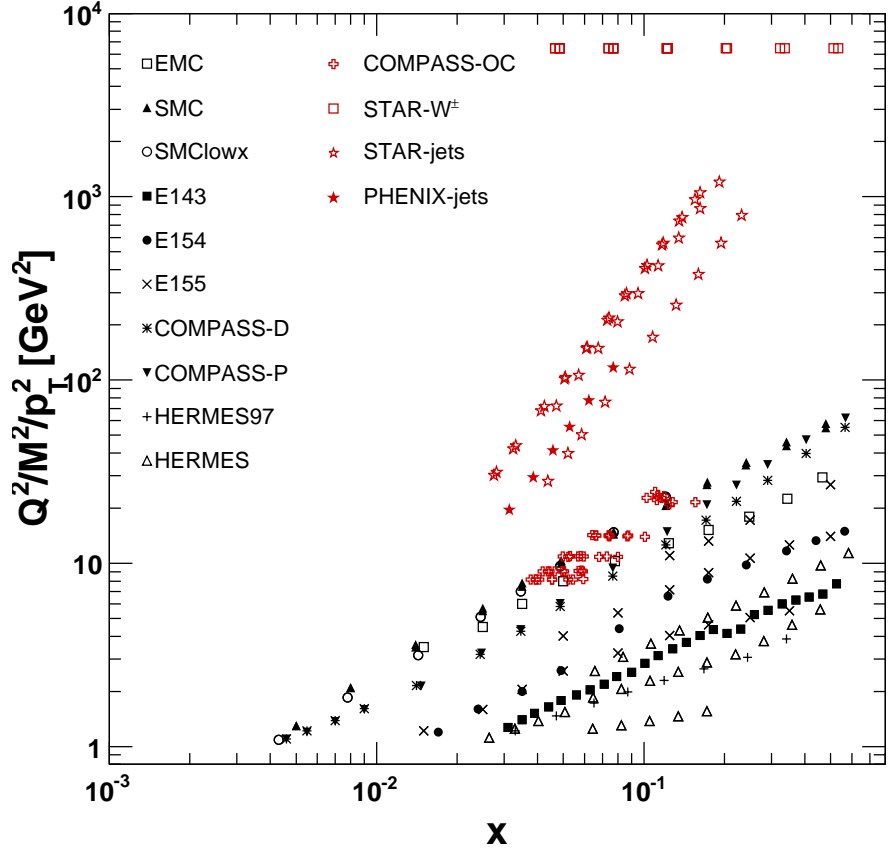}
\caption{\small Kinematic coverage in the $(x,Q^2)$ plane of
the new experimental data included in \texttt{NNPDFpol1.1} (red points, listed
in Tab.~\ref{tab:processes})
together with that of the inclusive DIS data already in \texttt{NNPDFpol1.0}
(black points).
The new experiments are listed in the second column of the legend.
For hadronic data, LO kinematics are assumed.
}
\label{fig:NNPDFpol11-kin}
\end{center}
\end{figure}

Our PDF determination thus includes, in addition to DIS data, 
open-charm production data 
from the COMPASS experiment at CERN and the most recent high-$p_T$
inclusive jet and $W^\pm$ production data from the STAR and PHENIX experiments 
at RHIC. The kinematic coverage of these data is shown
in Fig.~\ref{fig:NNPDFpol11-kin}, together with that of 
the fixed-target inclusive DIS data
already included in~\cite{Ball:2013lla}.

Other available polarized PDF sets 
include some of the non-DIS data of
Tab.~\ref{tab:processes}: 
in particular, the fits from the DSSV family 
(\texttt{DSSV08}~\cite{deFlorian:2009vb}
and \texttt{DSSV+}/\texttt{DSSV++}~\cite{deFlorian:2011ia})
include SIDIS data, inclusive jet and
identified hadron production measurements 
from polarized proton-proton collisions at RHIC, while the 
\texttt{LSS10} fit~\cite{Leader:2010rb} includes SIDIS data.

The new data sets will be added to those already included in the {\tt
  NNPDFpol1.0} polarized PDF determination~\cite{Ball:2013lla}
using the Bayesian reweighting method described in 
Refs.~\cite{Ball:2010gb,Ball:2011gg}.
This methodology consists of updating the representation of the
probability distribution in the space of PDFs provided by an 
available PDF set by means of Bayes' theorem in such a way that the
information contained in the new data sets is included. The method has
the dual advantage of not involving any further approximation once the
starting Monte Carlo set is given  
- no parametrization or minimization is necessary - and
also of being computationally rather light, in that the predictions to
be compared to the new data can be made only once, rather than at
each iteration of a minimization algorithm. On the other hand, the
method becomes impractical if the new data bring in a large amount of
new information. Indeed reweighting has the effect of \textit{zooming in} on the
part of the space of PDFs which is compatible with the new data, by
giving small weights to replicas which have little compatibility with
them. As a consequence, after reweighting the number of
replicas in the Monte Carlo set is effectively smaller than the
starting one, and thus only if the starting number of replicas was
sufficiently large will the final representation of the probability density
remain accurate. If too much new information is brought in by the new
data, the method becomes impractical because a very large number of
starting replicas would be required in order to obtain accurate
results after reweighting

\begin{table}[!t]
\centering
\small
\begin{tabular}{c|c|c|c|c|c}
\toprule
\textsc{reaction} & \textsc{partonic subprocess} & \textsc{pdf probed}
& $x$ & $Q^2$ [GeV$^2$] & \textsc{ref.}\\
\midrule
\multirow{2}*{$\ell^\pm\{p,d,n\}\to\ell^\pm X$} &
\multirow{2}*{$\gamma^*q\to q$} &
$\Delta q +\Delta\bar{q}$ &
\multirow{2}*{$0.003\lesssim x \lesssim 0.8$} &
\multirow{2}*{$1\lesssim Q^2 \lesssim 70$} &
\multirow{2}*{\cite{Ashman:1989ig,Adeva:1998vv,Adeva:1999pa,Alexakhin:2006oza,Alekseev:2010hc,Abe:1998wq,Abe:1997cx,Anthony:2000fn,Ackerstaff:1997ws,Airapetian:2007mh}}\\
& & $\Delta g$ & &
\\
\multirow{2}*{$\overrightarrow{p}\overrightarrow{p}\to {\rm jet(s)} X$} &
$gg\to gq$ &
\multirow{2}*{$\Delta g$} &
\multirow{2}*{$0.05\lesssim x \lesssim 0.2$} &
\multirow{2}*{$30\lesssim p_T^2 \lesssim 800$}&
\multirow{2}*{\cite{Adare:2010cc,Adamczyk:2012qj,Adamczyk:2014ozi}}\\
& $qg\to qg$ & & &
\\
\multirow{2}*{$\overrightarrow{p}p\to W^\pm X$} &
$u_L\bar{d}_R\to W^+$ &
$\Delta u$ $\Delta\bar{u}$ &
\multirow{2}*{$0.05\lesssim x \lesssim 0.4$} &
\multirow{2}*{$\sim M_W^2$} &
\multirow{2}*{\cite{Aggarwal:2010vc,Adare:2010xa,Adamczyk:2014xyw}}\\
& $d_L\bar{u}_R\to W^-$ & $\Delta d$ $\Delta\bar{d}$ & & 
\\
\midrule
$\ell^\pm\{p,d\}\to \ell^\pm D X$ &
$\gamma^*g\to c\bar{c}$ &
$\Delta g$ &
$0.06\lesssim x \lesssim 0.2$ &
$0.04 \lesssim p_T^2 \lesssim 4$ &
\cite{Adolph:2012ca}
\\
\multirow{3}*{$\ell^\pm\{p,d\}\to \ell^\pm h X$} &
\multirow{3}*{$\gamma^*q\to q$} &
$\Delta u$ $\Delta\bar{u}$  &
\multirow{3}*{$0.005\lesssim x \lesssim 0.5$} &
\multirow{3}*{$1\lesssim Q^2\lesssim 60$} &
\multirow{3}*{\cite{Ackerstaff:1999ey,Adeva:1997qz,Airapetian:2004zf,Alekseev:2007vi,Alekseev:2009ab}}
\\
 & & $\Delta d$ $\Delta\bar{d}$ & & \\
 & & $\Delta g$  & & & \\
\multirow{2}*{$\overrightarrow{p}\overrightarrow{p}\to \pi X$} &
$gg\to gg$ &
\multirow{2}*{$\Delta g$} &
\multirow{2}*{$0.05\lesssim x \lesssim 0.4$} &
\multirow{2}*{$1\lesssim p_T^2 \lesssim 200$} &
\multirow{2}*{\cite{Adler:2004ps,Adler:2006bd,Adare:2007dg,Adare:2008qb,Adare:2008aa,Adamczyk:2013yvv,Adare:2014hsq}}\\
& $qg\to qg$ & & & \\
\bottomrule
\end{tabular}
\caption{\small Summary of available processes that allow for  the determination
of polarized PDFs. 
For each process, we show the leading partonic subprocesses, polarized
PDFs, and  the approximate  ranges of $x$ and $Q^2$ that are
accessible using the data from the references given.
Processes listed in the upper part of the table do not depend on
fragmentation functions, while those in the lower part of the table do.
}
\label{tab:processes}
\end{table}
The reweighting method is especially useful in our case because on
the one hand, fast interfaces for the computation of hadronic
observables, such as the \texttt{FastNLO}
framework~\cite{Kluge:2006xs}, the general-purpose interface
\texttt{APPLgrid}~\cite{Carli:2010rw}, and the \texttt{FastKernel}
method~\cite{Ball:2010de}, used for the unpolarized PDF determinations
in Ref.~\cite{Ball:2013lla} are not yet available in the polarized case.\footnote{
Note however that recent progress on 
interfacing {\tt MadGraph5\_aMC@NLO}~\cite{Alwall:2014hca} to \texttt{APPLgrid}
will provide the required fast computations in the near future.
}
On the other hand, we wish to add to our data set only
a few dozen hadronic data points, out of
a total of approximately 300 data points. Nevertheless, as we shall
see, the construction of the prior probability distribution will
require some care, due to the fact that the new data affect some
combinations of PDFs which were completely undetermined before reweighting.

As a consequence, we will start in Sect.~\ref{sec:prior} with a 
discussion on how to set up the reweighting method, and in particular
how to construct a suitable prior. 
We will then analyze separately the
impact of each of the 
new  processes which are included in our determination on  polarized PDFs:
first,  in Sect.~\ref{sec:gluon}, open-charm and jet 
production data, which affect the  gluon PDF $\Delta g$, and then
in Sect.~\ref{sec:flavour},
$W^\pm$ production which allows for a determination of the  light
antiquark PDF $\Delta\bar{u}$ and $\Delta\bar{d}$.
Finally, in Sect.~\ref{sec:pheno}, we will discuss a simultaneous reweighting 
with all new data sets and present our final PDF set,
\texttt{NNPDFpol1.1}.  Specifically, we will discuss the
phenomenological implications of \texttt{NNPDFpol1.1} 
compared to \texttt{NNPDFpol1.0}: 
we will reassess the spin content of the proton by computing PDF first moments,
and give an illustrative example of the predictive power of our new PDF set
for the case of longitudinal double-spin asymmetry for single-inclusive
particle production in proton-proton collisions recently measured at RHIC.

\section{Reweighting: construction of the prior}
\label{sec:prior}

We wish to include the polarized hadronic data into our polarized PDF
determination 
by means of Bayesian reweighting.
As a starting point, we construct a set of $N_{\mathrm{rep}}=1000$
\texttt{NNPDFpol1.0} PDF
replicas according to the procedure of Ref.~\cite{Ball:2013lla}: a
large prior set is needed because reweighting always entails
some loss of efficiency, so the final reweighted replica set
corresponds to a smaller effective number of unweighted
replicas~\cite{Ball:2010gb,Ball:2011gg}. 

However, this is not sufficient, because
the NNPDF polarized parton set
{\tt NNPDFpol1.0}~\cite{Ball:2013lla} does not
allow for a separation between quark and antiquark parton distributions,
since it was determined from a fit to inclusive DIS data
only. Because the new data that we wish to include are
sensitive to such separation, we need to supplement the prior based on {\tt
  NNPDFpol1.0}  with some assumption on the $\Delta q-\Delta
\bar q$ PDF combinations which are left  unspecified in it.

Clearly, choosing a completely unbiased flat prior for the PDF
combinations which are undetermined in {\tt NNPDFpol1.0} would be extremely
inefficient, given that PDFs span a space of functions. We will thus
choose a prior based on an existing PDF set in which these PDF
combinations are determined, and then check that our reweighted results 
are independent of this choice by varying the prior. In practice, 
we construct the prior by supplementing the PDFs which are determined 
in the  {\tt NNPDFpol1.0} set, namely
 $\Delta u^+$, $\Delta d^+$, $\Delta s^+$ and $\Delta g$, with
$\Delta\bar{u}$ and $\Delta\bar{d}$  from  
the {\tt DSSV08}~\cite{deFlorian:2009vb} set, but with the uncertainties
inflated by a given factor. Of course, if the uncertainty was infinite this
would be equivalent to an (unbiased) flat prior. Hence, we will verify
that our results are independent of the choice of prior by inflating 
the uncertainty by an increasingly large factor, until the results stabilise.

\begin{figure}[!t]
\begin{center}
\epsfig{width=0.40\textwidth,figure=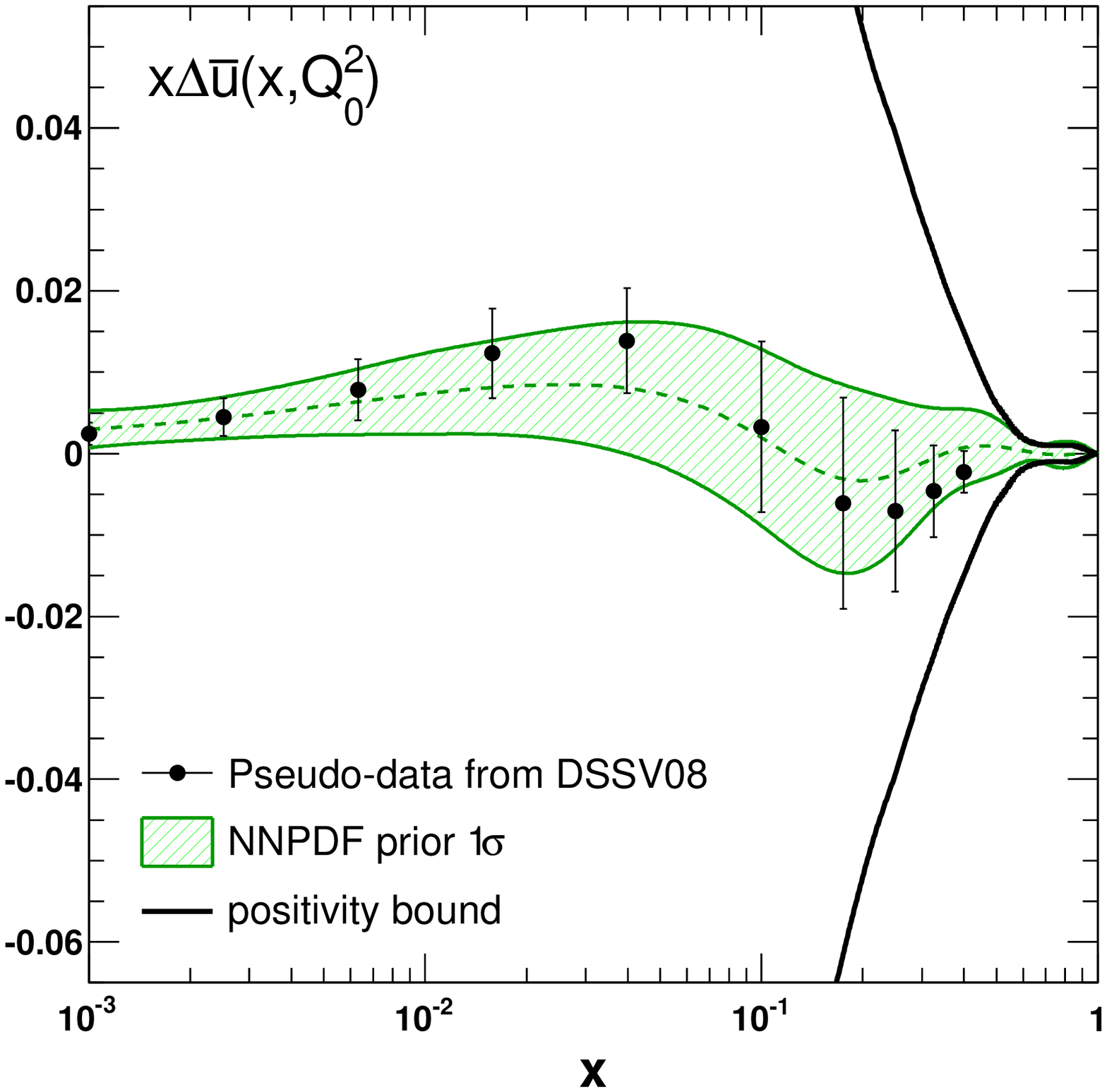}
\epsfig{width=0.40\textwidth,figure=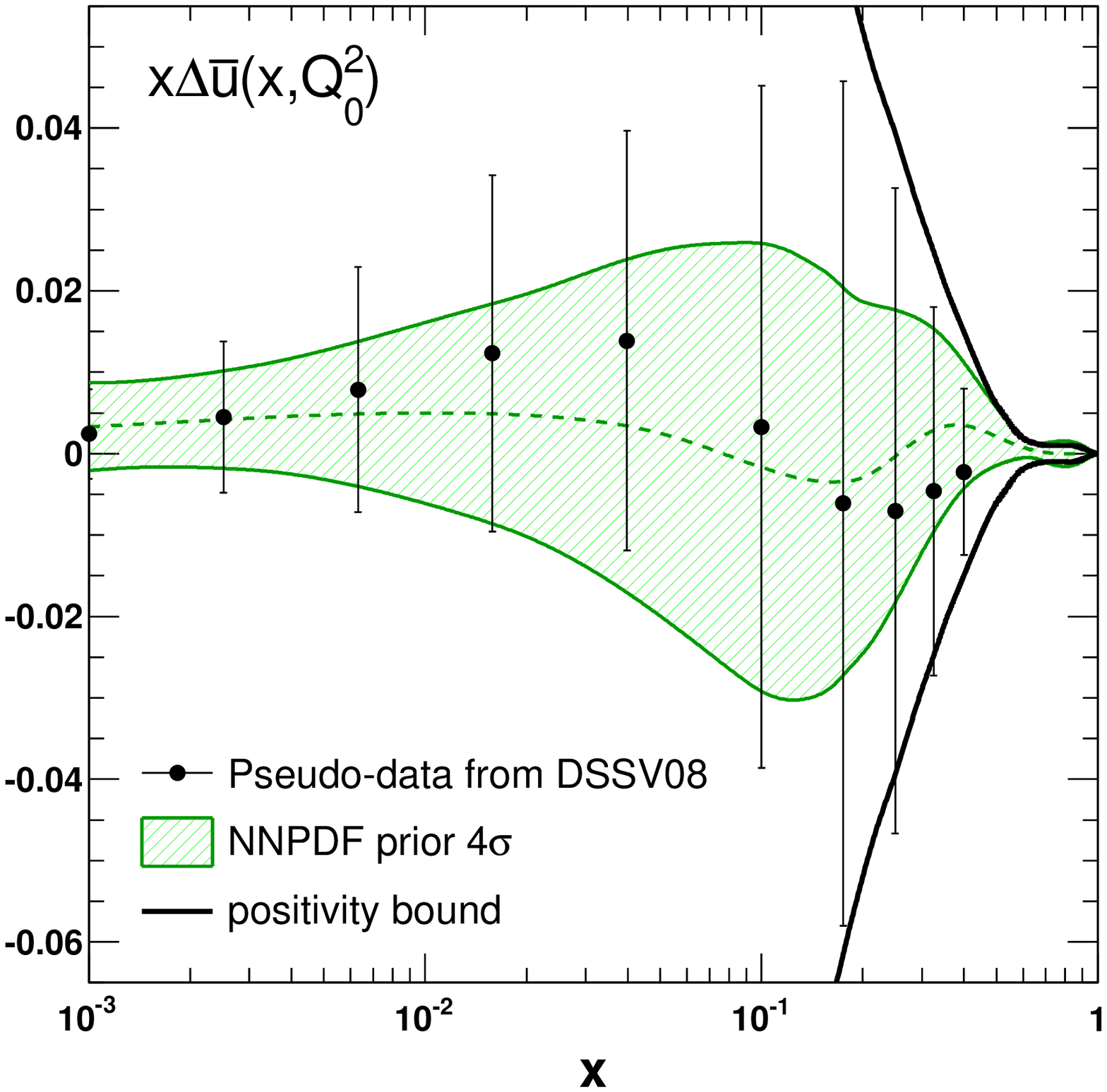}\\
\epsfig{width=0.40\textwidth,figure=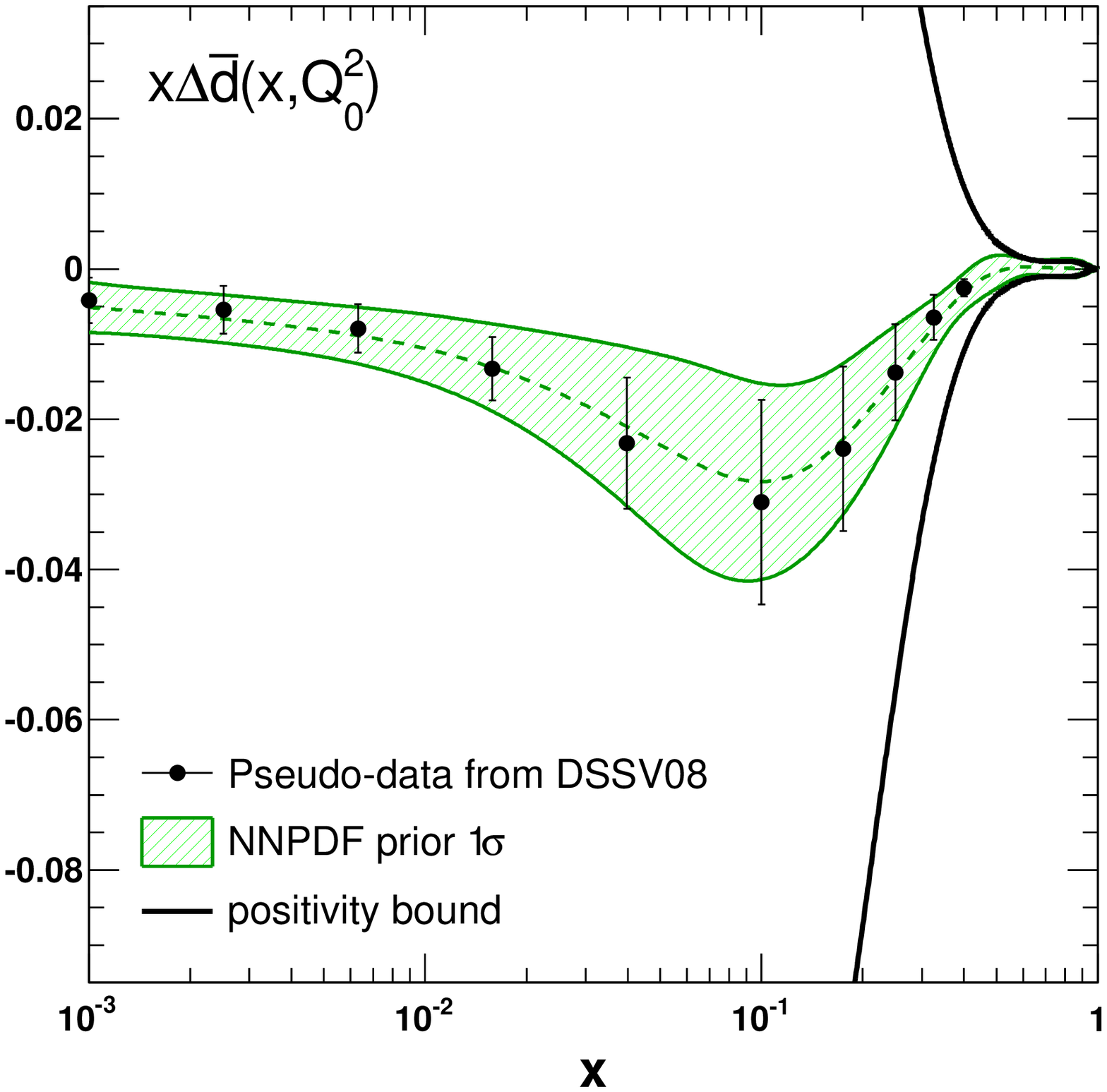}
\epsfig{width=0.40\textwidth,figure=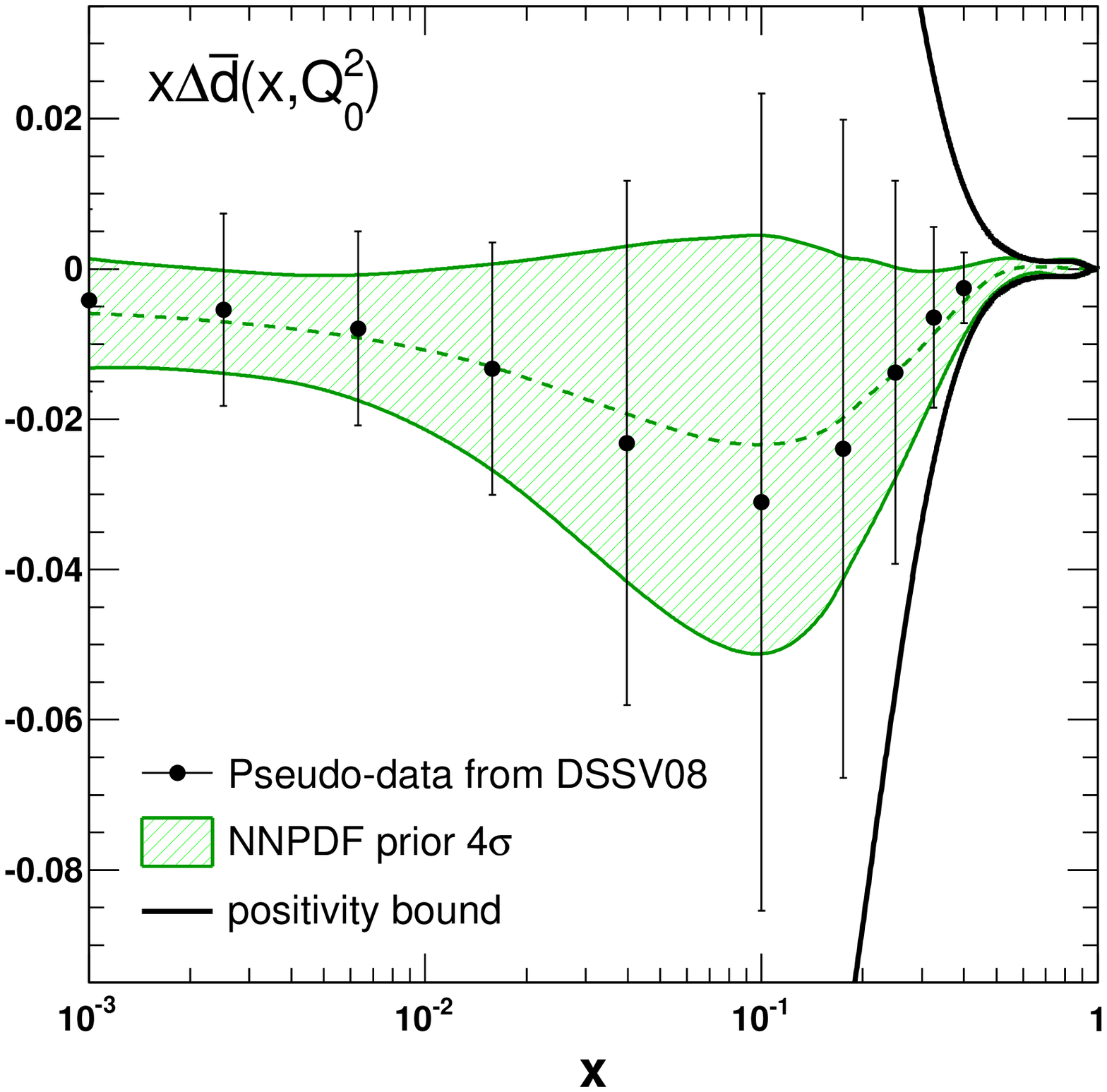}
\caption{\small The polarized sea quark densities,
 $x\Delta\bar{u}(x,Q_0^2)$ (upper plots)
and the $x\Delta\bar{d}(x,Q_0^2)$ (lower plots) at the initial energy scale
$Q_0^2=1$ GeV$^2$ from the neural network fit (green full band) to
the {\tt DSSV08} pseudo-data (points with uncertainties). 
Results are shown for the $1\sigma$ (left plots) and $4\sigma$ 
(right plots) prior ensembles (see text). 
The PDF positivity bounds from the corresponding
unpolarized \texttt{NNPDF2.3} counterpart are also shown.
}
\label{fig:prior}
\end{center}
\end{figure}

In order to do this,  we sample the {\tt DSSV08} $\Delta\bar{u}$ and 
$\Delta\bar{d}$  distributions at a fixed reference scale 
$Q_0^2=1$ GeV$^2$. 
We then select ten points, half logarithmically and half linearly spaced 
in the interval of momentum fraction $10^{-3}\lesssim x \lesssim 0.4$, 
which roughly corresponds to the range covered by SIDIS experimental 
data relevant for separating quark-antiquark contributions.
These points with the corresponding 
PDF uncertainties are treated
as sets of experimental pseudo-observables.
Henceforth, they will be labeled as DSSV$_{U}$ and DSSV$_{D}$ 
respectively.
In these pseudo-data prior fits, the  experimental data 
are always taken as the central value from
the {\tt DSSV08} best fit, while the experimental uncertainties
are the corresponding nominal $\Delta\chi^2=1$ Hessian uncertainties,
multiplied by a factor one, two, three and four respectively in order
to obtain the sets with inflated uncertainty as explained above. 

We then generate $N_{\rm rep}=1000$ replicas of
the original pseudo-data, following the procedure described 
in Sect.~2 of Ref.~\cite{Ball:2013lla}, and 
we fit each replica with a set of neural networks. To this purpose, we
supplement 
the input PDF basis given in Sect.~3 of Ref.~\cite{Ball:2013lla},
namely $\Delta \Sigma$, $\Delta T_3$, $\Delta T_8$ and $\Delta g$,  
with two new linearly independent light quark combinations:
the total valence, $\Delta V$, and the valence isotriplet, $\Delta V_3$,
\begin{eqnarray}
\Delta V(x,Q_0^2) &=& \Delta u^-(x,Q_0^2)+\Delta d^-(x,Q_0^2)
\,\mbox{,} 
\label{eq:PDFbasis1}
\\
\Delta V_3(x,Q_0^2) &=& \Delta u^-(x,Q_0^2)-\Delta d^-(x,Q_0^2)
\,\mbox{,} 
\label{eq:PDFbasis2}
\end{eqnarray}
where $\Delta q^-=\Delta q-\Delta\bar{q}\mbox{, } q=u,d$.
Equation~(\ref{eq:PDFbasis1}) holds under the assumption that
$\Delta s = \Delta\bar{s}$, i.e. $\Delta V_8= \Delta V$.
This assumption is not based on a theoretical motivation, but simply
on the observation that present data are  insufficient to determine
$\Delta s^-$: hence this PDF combination should be simply viewed as
undetermined in our fit.\footnote{In the corresponding
{\tt LHAPDF} grids, which require input of both $\Delta s$ and
$\Delta \bar{s}$, we will assume $\Delta s^-=0$.   }

Each of the PDF combinations in 
Eqs.~(\ref{eq:PDFbasis1})-(\ref{eq:PDFbasis2}) 
is  parametrized as usual by means of a neural network supplemented 
with a preprocessing function, 
\begin{eqnarray}
\Delta V(x,Q_0^2) 
&=& 
(1-x)^{m_{\Delta V}} x^{n_{\Delta V}} \textrm{NN}_{\Delta V}(x)
\,\mbox{,}
\label{eq:NNparam1} 
\\
\Delta V_3(x,Q_0^2) 
&=& 
(1-x)^{m_{\Delta V_3}} x^{n_{\Delta V_3}} \textrm{NN}_{\Delta V_3}(x)
\,\mbox{,}
\label{eq:NNparam2}
\end{eqnarray}
where $\textrm{NN}_{\Delta \textrm{pdf}}$, $\textrm{pdf}=V, V_3$ is the output of 
the neural network, and the preprocessing exponents $m,n$ are linearly 
randomized for each
Monte Carlo replica within the ranges given in Tab.~\ref{tab:prepexp}.
We have checked that our choice of preprocessing exponents does not bias the
fit, according to the procedure discussed in Sect.~4.1 of 
Ref.~\cite{Ball:2013lla}.
The neural network architecture is the same as in {\tt NNPDFpol1.0}, namely 
2-5-3-1.

The {\tt DSSV08} PDFs provide us with pseudo-data for  $\Delta \bar{u}$
and $\Delta \bar{d}$ which in terms of the PDF basis are given by
\begin{eqnarray}
\Delta \bar{u}(x,Q_0^2) 
&=& \frac{1}{12}\lc 2\Delta \Sigma  + 3\Delta T_3
+\Delta T_8 -3\Delta V -3\Delta V_3  \rc
(x,Q_0^2)
\,\mbox{,} 
\label{eq:relbasis1}
\\
\Delta \bar{d}(x,Q_0^2) 
&=& \frac{1}{12}\lc 2\Delta \Sigma  - 3\Delta T_3
+\Delta T_8 -3\Delta V +3\Delta V_3  \rc
(x,Q_0^2)
\,\mbox{.}
\label{eq:relbasis2}
\end{eqnarray}
Each {\tt DSSV08} pseudo-data replica is then combined at random with
an {\tt NNPDFpol1.0} PDF replica, and the two missing basis combinations
Eqs.~(\ref{eq:PDFbasis1})-(\ref{eq:PDFbasis2}) 
 are determined by fitting with the standard NNPDF methodology, 
 including the theoretical
 constraints which are relevant in the polarized case, as discussed in
Ref.~\cite{Ball:2013lla}.
In particular, the positivity constraints 
Eqs.~(61)-(62) of Ref.~\cite{Ball:2013lla} have been enforced
by letting $f=u, \bar{u}, d, \bar{d}$ separately. 
Note that no additional sum rules affect $\Delta V$ and $\Delta
V_3$. As a consequence, the new PDF combinations $\Delta V$ and $\Delta
V_3$ are completely uncorrelated to the PDF combinations of the {\tt
  NNPDFpol1.0} set, as they are based on completely independent
information and there is no further theoretically-induced cross-talk.
The quality of the pseudo-data fits is quantitatively assessed 
by the $\chi^2$ values per data point quoted in Tab.~\ref{tab:chi2ud}
which are close to one for both DSSV$_{U}$ and DSSV$_{D}$ 
data sets, and their combination.

\begin{table}[!t]
 \centering
 \small
 \begin{tabular}{ccc}
  \toprule
   PDF & $m$ & $n$ \\
  \midrule
   $\Delta V(x,Q_0^2)$   & [1.5,3.0] & [0.05,0.60]\\
   $\Delta V_3(x,Q_0^2)$ & [1.5,3.0] & [0.01,0.60]\\
  \bottomrule
 \end{tabular}
 \caption{\small Ranges for the small- and large-$x$ preprocessing exponents 
in Eqs.~(\ref{eq:NNparam1})-(\ref{eq:NNparam2}).}
 \label{tab:prepexp}
\end{table}

We thus  end up with four separate prior PDF ensembles,
labeled as $1\sigma$, $2\sigma$, $3\sigma$
and $4\sigma$, corresponding to the different factors 
by which the {\tt DSSV08} nominal
PDF uncertainty has been enlarged.
In Sects.~\ref{sec:gluon}-\ref{sec:flavour} 
we will explicitly show that this is sufficient to obtain reweighted 
results which are 
independent of the choice of prior, the $3\sigma$ and $4\sigma$ sets both 
being effectively unbiased priors.

In Fig.~\ref{fig:prior}, we show the 
$x\Delta\bar{u}(x,Q_0^2)$ and $x\Delta\bar{d}(x,Q_0^2)$ PDFs
at the initial energy scale $Q_0^2=1$ GeV$^2$
from the $1\sigma$ and $4\sigma$ sets. 
The other priors, $2\sigma$ and $3\sigma$, 
consistently provide intermediate results.
In these plots, the positivity bound discussed in Ref.~\cite{Ball:2013lla}
and pseudo-data points sampled from \texttt{DSSV08} are also shown.
\begin{table}[!t]
\centering
\small
\begin{tabular}{ll|c|cccc}
\toprule
& & & \multicolumn{4}{c}{$\chi^2_{\mathrm{tot}}/N_{\mathrm{dat}}$}\\
\midrule
Experiment & Set & $N_{\mathrm{dat}}$ & $1\sigma$ & $2\sigma$ & $3\sigma$ & $4\sigma$ \\
\midrule
DSSV &            & $20$ & $1.04$ & $1.10$ & $1.09$ & $0.97$ \\
& DSSV$_{U}$ & $10$ & $1.13$ & $1.09$ & $1.08$ & $0.97$ \\
& DSSV$_{D}$ & $10$ & $0.96$ & $1.10$ & $1.09$ & $0.96$ \\
\bottomrule
\end{tabular}
\caption{\small The value of the $\chi^2_{\mathrm{tot}}$ per data point for both 
separate and combined $\Delta\bar{u}$ and $\Delta\bar{d}$ data sets
after the neural network fit to $\Delta \bar{u}$ and $\Delta \bar{d}$ 
pseudo-data sampled from \texttt{DSSV08}~\cite{deFlorian:2009vb}.}
\label{tab:chi2ud}
\end{table}

These priors will be used in the next section for the inclusion of the
new data sets, listed in the upper part of Tab.~\ref{tab:processes}.

\section{The  polarized gluon: open charm and jet production}
\label{sec:gluon}

In this section, we include by reweighting the information coming from
the data of Tab.~\ref{tab:processes}
which mostly affect the gluon distributions:
open-charm production data from the COMPASS experiment at CERN, 
and high-$p_T$ jet production measurements in 
proton-proton collisions from the STAR and PHENIX experiments at RHIC.
We discuss the inclusion of each of these two data sets in turn.

\subsection{Open-charm muoproduction at COMPASS}
\label{subsec:COMPASS}

Open-charm production in 
polarized DIS~\cite{Frixione:1996ym} directly probes the gluon
distribution because at leading-order (LO) this process proceeds  through
photon-gluon fusion (PGF), $\gamma^*g\to c\bar{c}$, 
followed by the fragmentation of the charm quarks
into charmed mesons,
typically $D^0$ mesons. 
The corresponding measurable photon-nucleon asymmetry  
$A_{LL}^{\gamma N \to D^0 X}$ at LO is thus given by
\begin{equation}
A_{LL}^{\gamma N \to D^0 X}\equiv\frac{d\Delta\sigma_{\gamma N}}{d\sigma_{\gamma N}}
      =\frac{d\Delta\hat{\sigma}_{\gamma g}\otimes\Delta g\otimes D_c^{D^0}}
            {d\hat{\sigma}_{\gamma g}\otimes g\otimes D_c^{D^0}}
\,\mbox{,}
\label{eq:ALLgammaN}
\end{equation}
where $\Delta\hat{\sigma}_{\gamma g}$ ($\hat{\sigma}_{\gamma g}$) is 
the spin-dependent (spin-averaged) partonic cross-section, 
$\Delta g$ ($g$) is the polarized (unpolarized) gluon PDF, 
and $D_{c}^{D^0}$ is the fragmentation
function for a charm quark into 
a $D^0$ meson, assumed to be spin independent. 

The COMPASS collaboration has measured the photon-nucleon asymmetry
$A_{LL}^{\gamma N \to D^0 X}$, Eq.~(\ref{eq:ALLgammaN}), obtained 
by the scattering of polarized muons of energy $E_{\mu}=160$ GeV 
off a fixed target of longitudinally polarized 
protons or deuterons~\cite{Adolph:2012ca}.
Three different data sets are available, depending on the
$D^0$ decay mode used to reconstruct the charmed hadron 
in the final state: 
$D^0\to K^-\pi^+$, $D^0\to K^-\pi^+\pi^0$ or $D^0\to K^-\pi^+\pi^+\pi^-$.
In the following, these will be referred to as
COMPASS~$K1\pi$, COMPASS~$K2\pi$ and COMPASS~$K3\pi$
respectively.
Assuming LO kinematics, the polarized gluon PDF is being probed
at intermediate momentum fraction values, $0.06 \lesssim x \lesssim 0.22$,
and at energy scale $Q^2 = 4(m_c^2+p_T^2) \sim 13$ GeV$^2$, 
where $m_c$ is the charm quark mass 
and $p_{T}$ is the transverse momentum of the produced charmed hadron,
as shown in Fig.~\ref{fig:NNPDFpol11-kin} and Tab.~\ref{tab:processes}.

In order to include COMPASS open-charm muoproduction data in our
polarized PDF determination~\cite{Ball:2013lla} via reweighting,
we need to compute the predictions for the virtual photon-nucleon
asymmetry $A_{LL}^{\gamma N \to D^0 X}$ given in Eq.~(\ref{eq:ALLgammaN}).
We perform this computation, separately for the numerator and the
denominator of Eq.~(\ref{eq:ALLgammaN}), using the LO expressions
in Ref.~\cite{Simolo:2006iw}, to which we refer the reader 
for more details.
The NLO prediction is also available~\cite{Riedl:2012qc};  however, as
we shall see shortly, these data have a negligible impact, so much so
that results would be essentially unchanged by the use of NLO theory.

Notice that here we do not need the prior Monte Carlo
samples constructed in
Sect.~\ref{sec:prior}, since the asymmetry
Eq.~(\ref{eq:ALLgammaN}) only depends on the polarized gluon.
We use the Peterson parametrization of the
fragmentation function $D_c^{D^0}$~\cite{Peterson:1982ak}
with $\epsilon=0.06$; 
we checked that results
are unaffected by reasonable variations of the
fragmentation function~\cite{Colangelo:1992kh}, and indeed it was  
pointed out in Ref.~\cite{Riedl:2012qc} that the dependence on the
fragmentation function is weaker than scale uncertainties of the NLO
computation. 

\begin{figure}[!t]
\begin{center}
\epsfig{width=0.32\textwidth,figure=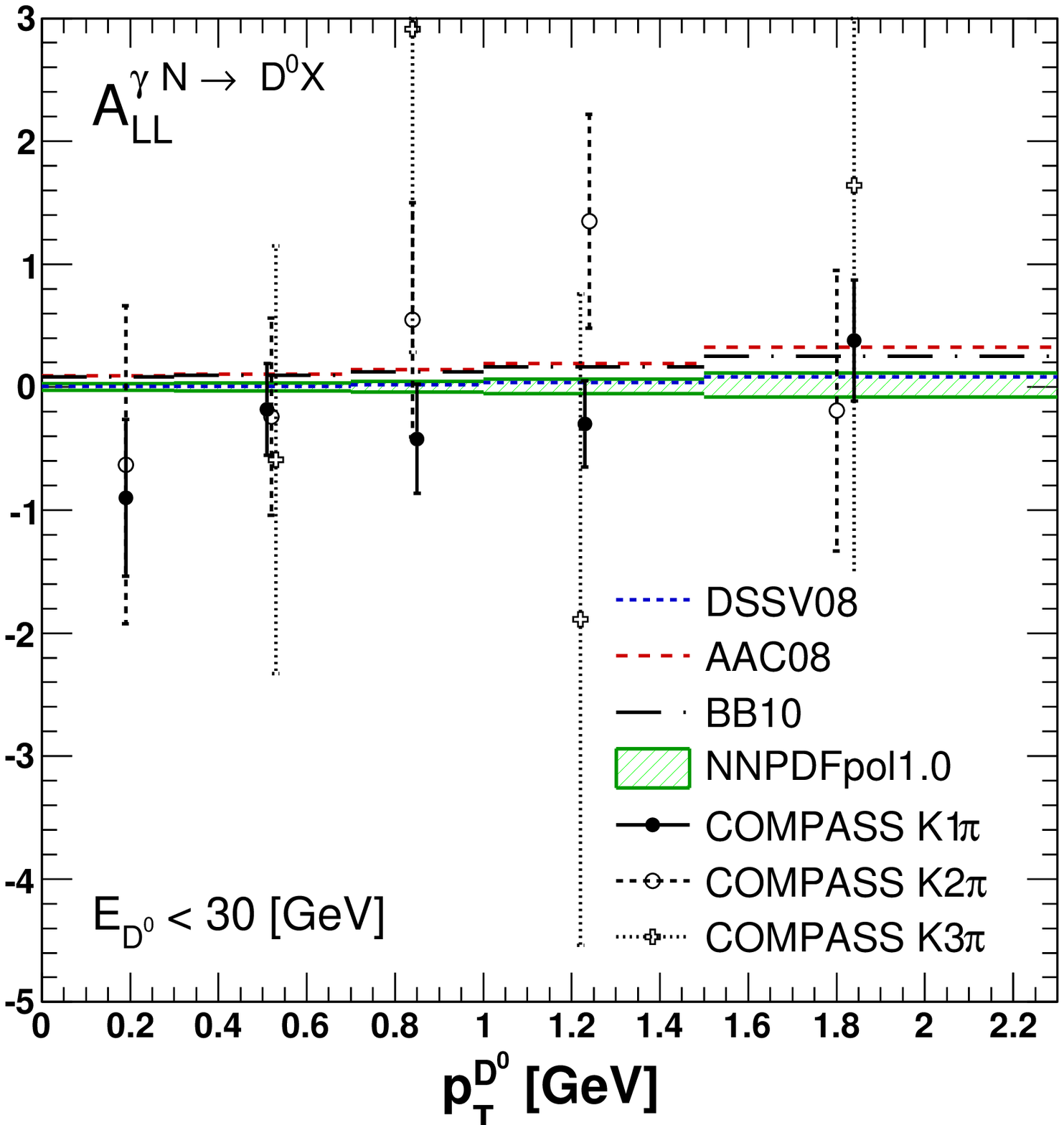}
\epsfig{width=0.32\textwidth,figure=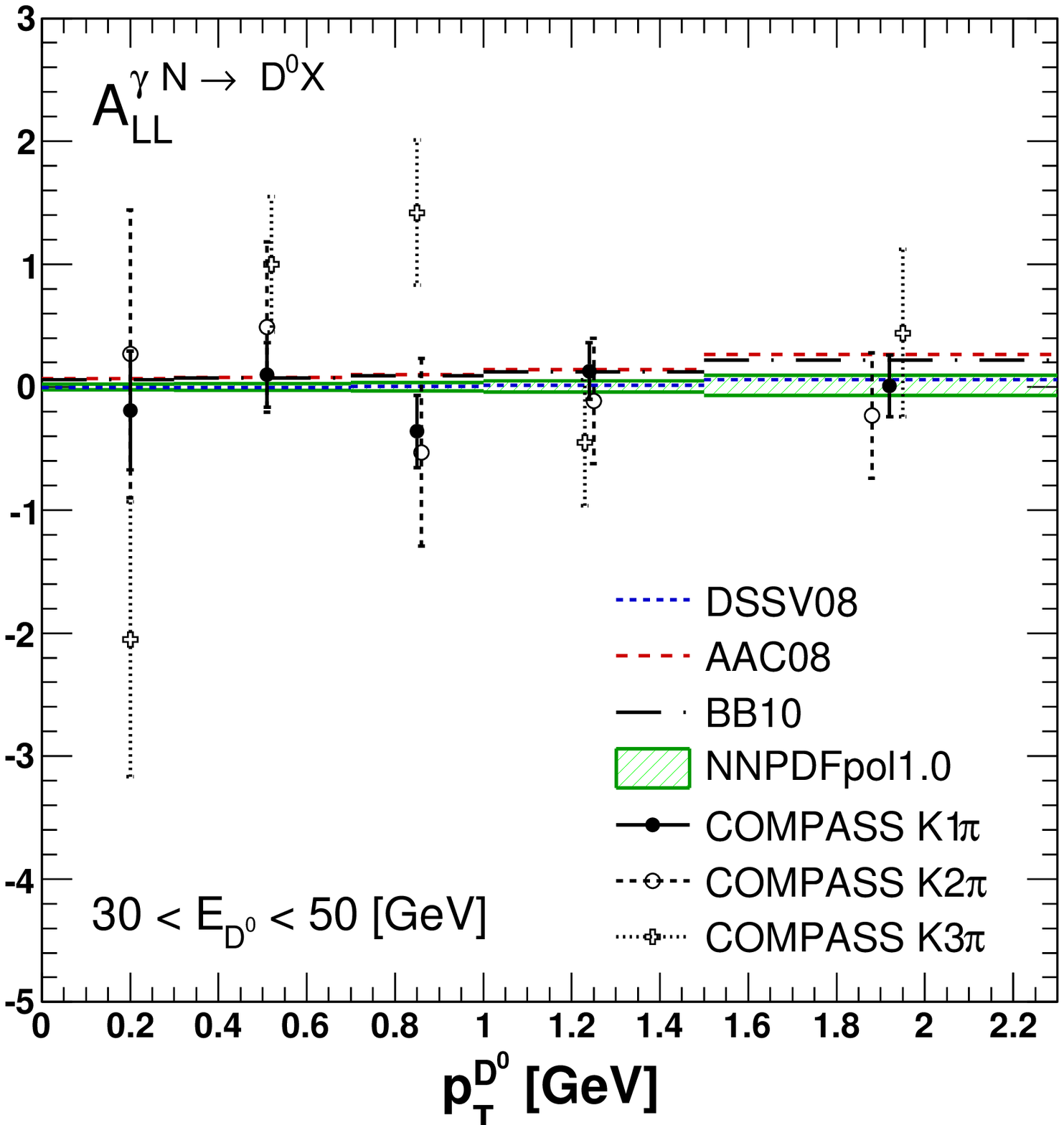}
\epsfig{width=0.32\textwidth,figure=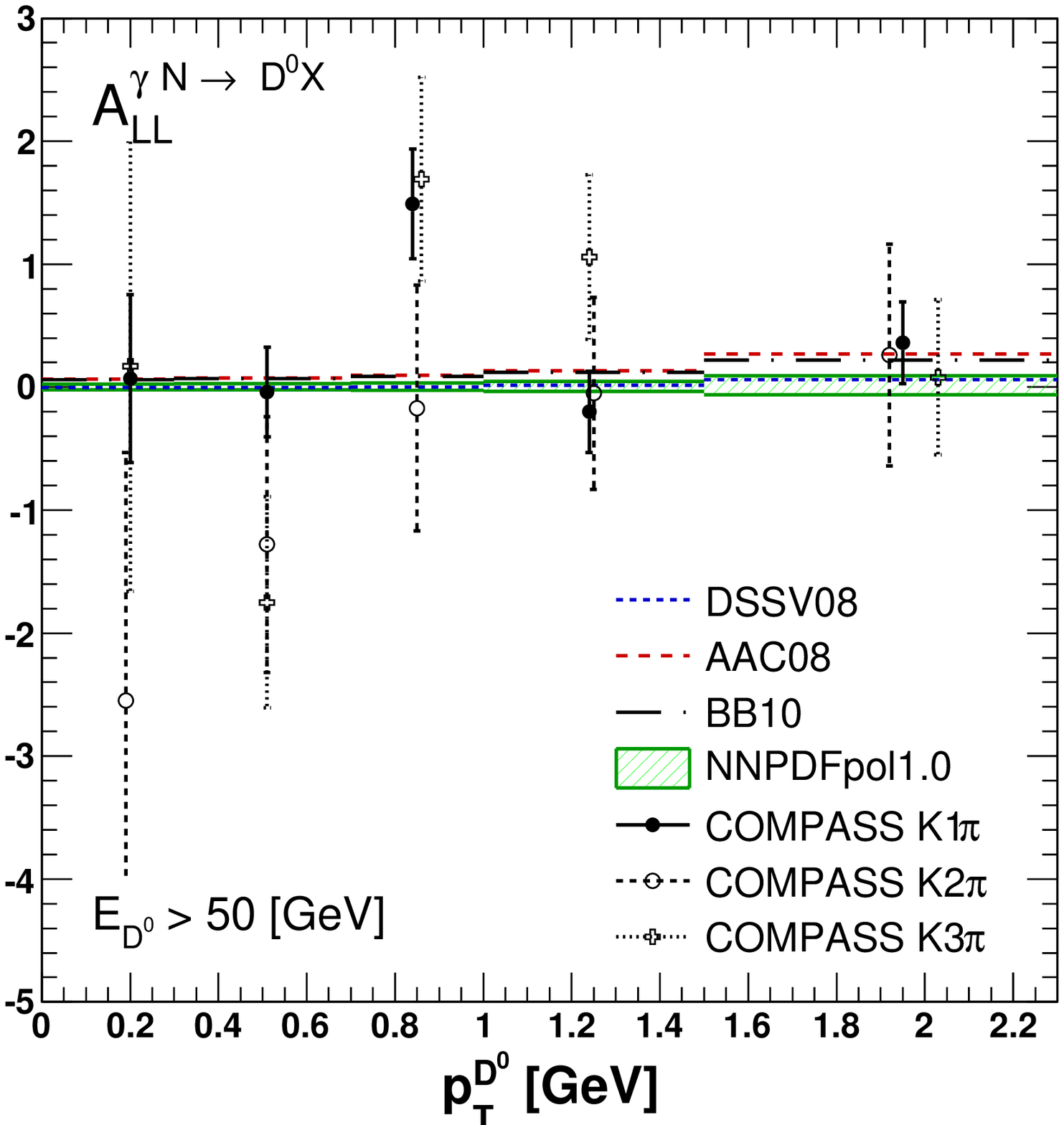}\\
\caption{\small Double-spin asymmetry for $D^0$ meson 
photoproduction $A_{LL}^{\gamma N \to D^0 X}$, Eq.~(\ref{eq:ALLgammaN}), as
measured by COMPASS~\cite{Adolph:2012ca} from the 
three different decay channels,
compared to the corresponding LO theoretical prediction obtained
using  {\tt NNPDFpol1.0} 
{\tt DSSV08}, {\tt ACC08} and {\tt BB10} polarized parton sets,
supplemented in each case by a suitable unpolarized set (see text).
Results are presented for three bins of the $D^0$ meson energy, 
$E_{D^0}$, and in five bins of its transverse momentum, $p_{T}^{D_0}$.}
\label{fig:COMPASS_beforerw}
\end{center}
\end{figure}

To get a feeling for the potential impact of these data,
in Fig.~\ref{fig:COMPASS_beforerw} we compare the LO  predictions
obtained using various PDF sets  to
the COMPASS data, 
separated into individual decay channels and  three  bins for the energy 
$E_{D^0}$ of the charmed hadron. Specifically, we show results
obtained using
{\tt DSSV08}, {\tt AAC08}~\cite{Hirai:2008aj}, {\tt
  BB10}~\cite{Blumlein:2010rn}   and {\tt NNPDFpol1.0} 
polarized sets
supplemented with the following unpolarized sets:
\texttt{CTEQ6}~\cite{Pumplin:2002vw} for \texttt{DSSV08}; 
\texttt{MRST2004}~\cite{Martin:2006qz} for {\tt AAC08} and  {\tt
  BB10}; and \texttt{NNPDF2.3}~\cite{Ball:2012cx} (for {\tt NNPDFpol1.0}).
In all cases,
the  PDF uncertainties shown are obtained neglecting the uncertainties
due to the unpolarized sets. 
 It is clear from
Fig.~\ref{fig:COMPASS_beforerw} that the COMPASS data have very large
uncertainties on the scale of the gluon PDF uncertainty, despite the
fact that the PDF uncertainty is very large, especially for 
the {\tt NNPDFpol1.0} set~\cite{Ball:2013lla}.

The  COMPASS data have been included into the {\tt NNPDFpol1.0} set by
Bayesian reweighting. The values of $\chi^2$ per data point before and after
reweighting are shown in Tab.~\ref{tab:chi2before}, along with those
obtained using other PDF sets in which the COMPASS data are not
included; in each case the unpolarized set is the same as that used in 
Fig.~\ref{fig:COMPASS_beforerw}. We also list the number of
data points, and the effective number of replicas after reweighting
a prior of $N_{\mathrm{rep}}=1000$ replicas.
Note that, because information on the correlation of systematics is unavailable,
statistical and systematic uncertainties are added in quadrature
when computing the $\chi^2$.

It is clear from Tab.~\ref{tab:chi2before} that the inclusion of the
COMPASS data has a negligible impact on the fit, as shown by the fact
that the $\chi^2$ values before and after reweighting are either  the
same or extremely close, and the effective number of replicas after
reweighting is always very close to the number of replicas in the
prior set. Also, despite very significant differences in the shape of
the central gluon distribution for the various PDF sets considered
here, in each case the $\chi^2$ values are essentially the same for
all sets. This means that the $\chi^2$ is mostly determined by the
mutual consistency or inconsistency of the data themselves, rather than by the
actual shape of the gluon.
\begin{table}[!t]
 \centering
 \small
  \begin{tabular}{ll|c|c|ccccc}
   \toprule
   \multirow{2}*{Experiment} & \multirow{2}*{Set} & \multirow{2}*{$N_{\mathrm{dat}}$}& \multirow{2}*{$N_{\mathrm{eff}}$} &
   \multicolumn{4}{c}{$\chi^2/N_{\mathrm{dat}}$} \\
  &  &  &  & \texttt{NNPDFpol1.0} & reweighted  & \texttt{DSSV08} & \texttt{AAC08} & \texttt{BB10}\\
   \midrule
     COMPASS &   &  45 &  980&1.23 & 1.23 &1.23 & 1.27 & 1.25 \\
   & COMPASS~$K1\pi$ &  15 & 990 & 1.27 &1.27 & 1.27 & 1.43 & 1.38 \\
   & COMPASS~$K2\pi$ &  15 & 990 &0.51 &0.51  & 0.51 & 0.56 & 0.55 \\
   & COMPASS~$K3\pi$ &  15 & 970 & 1.90 &1.89 & 1.90 & 1.81 & 1.82 \\
   \bottomrule
  \end{tabular}
\caption{\small Quality of the fit to combined and individual 
COMPASS open-charm data sets
obtained using the polarized PDF sets (and unpolarized counterparts)
of Fig.~\ref{fig:COMPASS_beforerw}, as well as a set (denoted as
reweighted) obtained starting with
\texttt{NNPDFpol1.0} and including the COMPASS data by reweighting. 
In each case, we show the number of
data points, the effective number of replicas after reweighting for
the reweighted set, and the $\chi^2$
per data point obtained using each set.}
\label{tab:chi2before}
\end{table}

\begin{figure}[!t]
\begin{center}
\epsfig{width=0.32\textwidth,figure=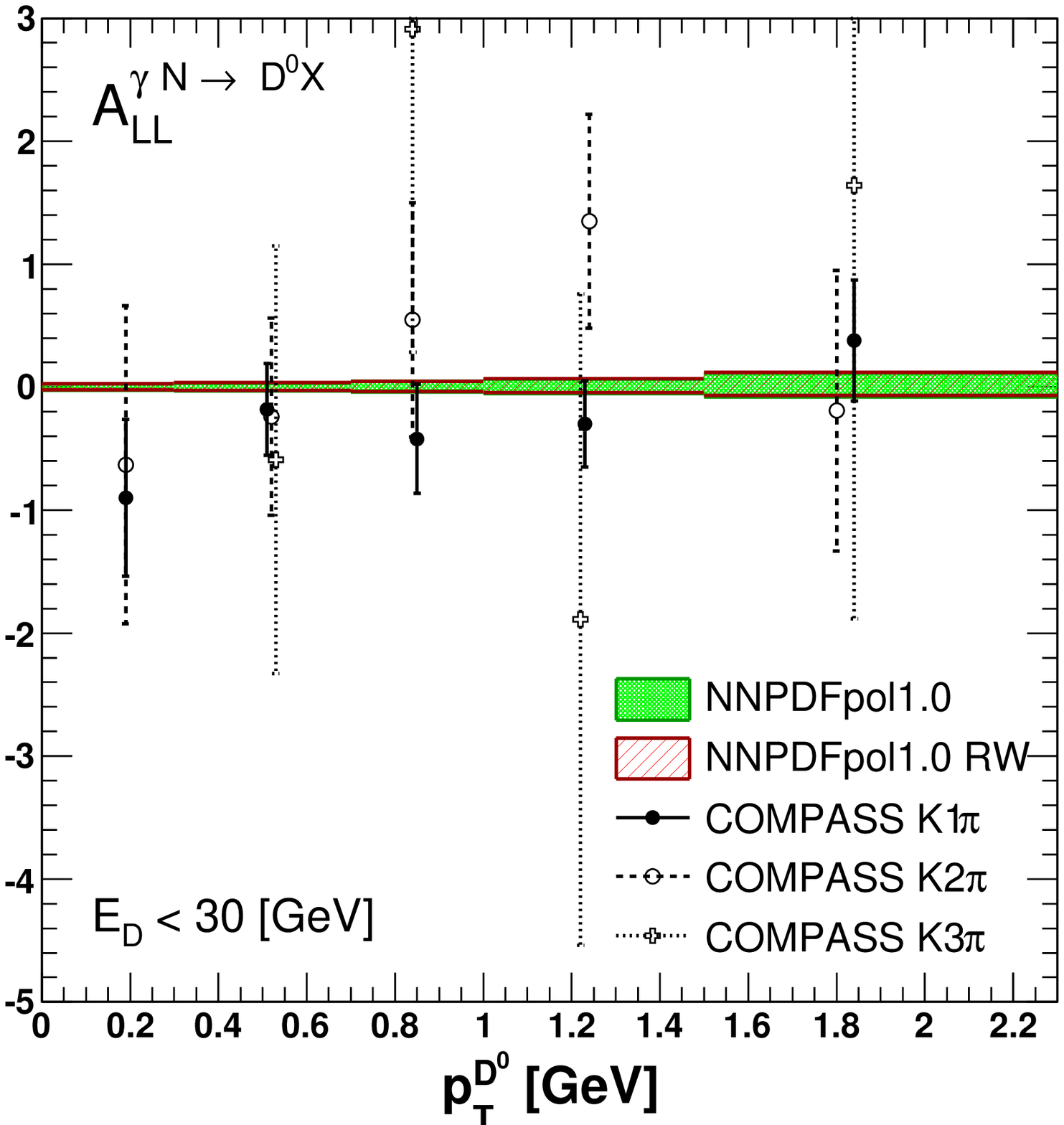}
\epsfig{width=0.32\textwidth,figure=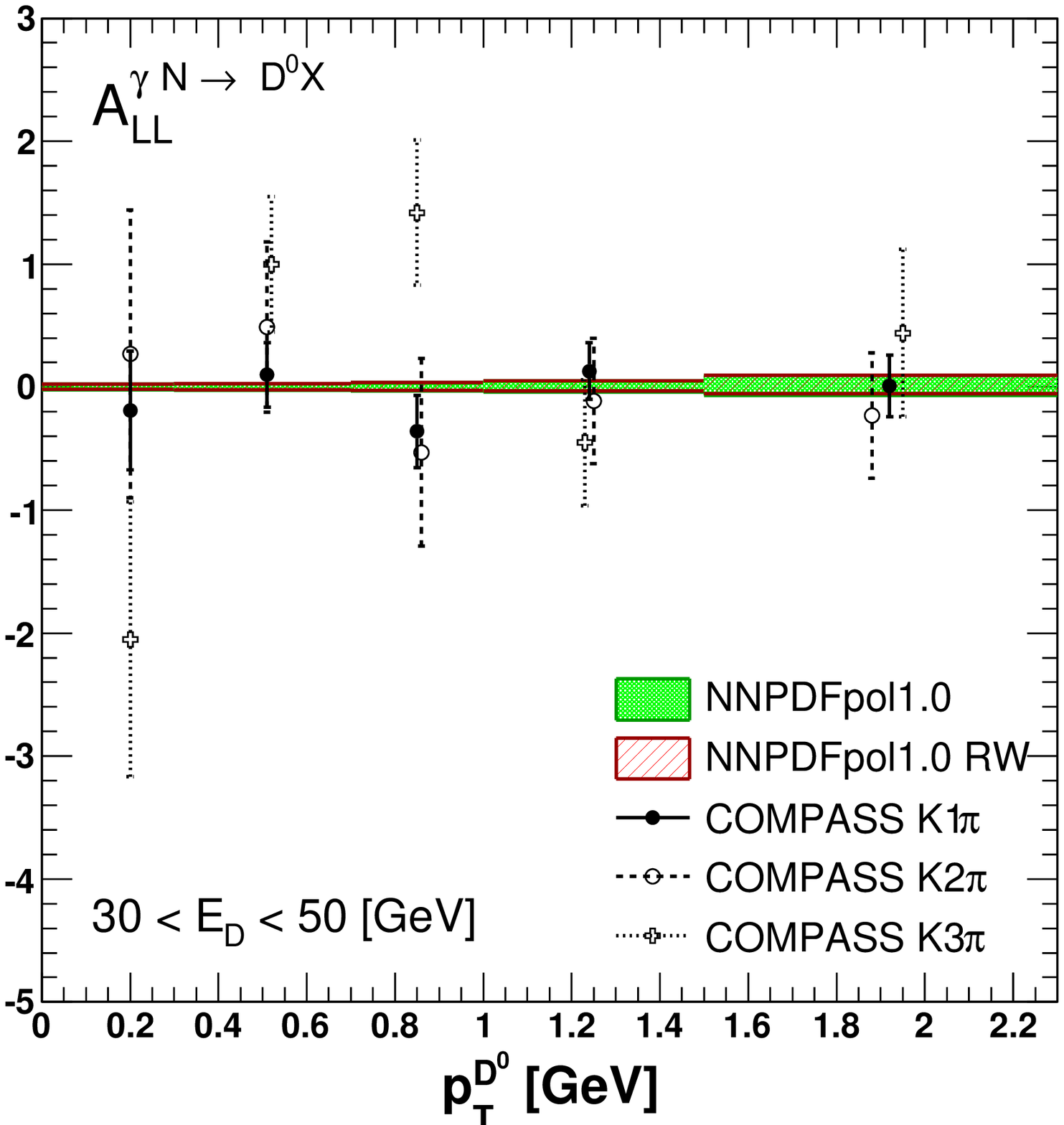}
\epsfig{width=0.32\textwidth,figure=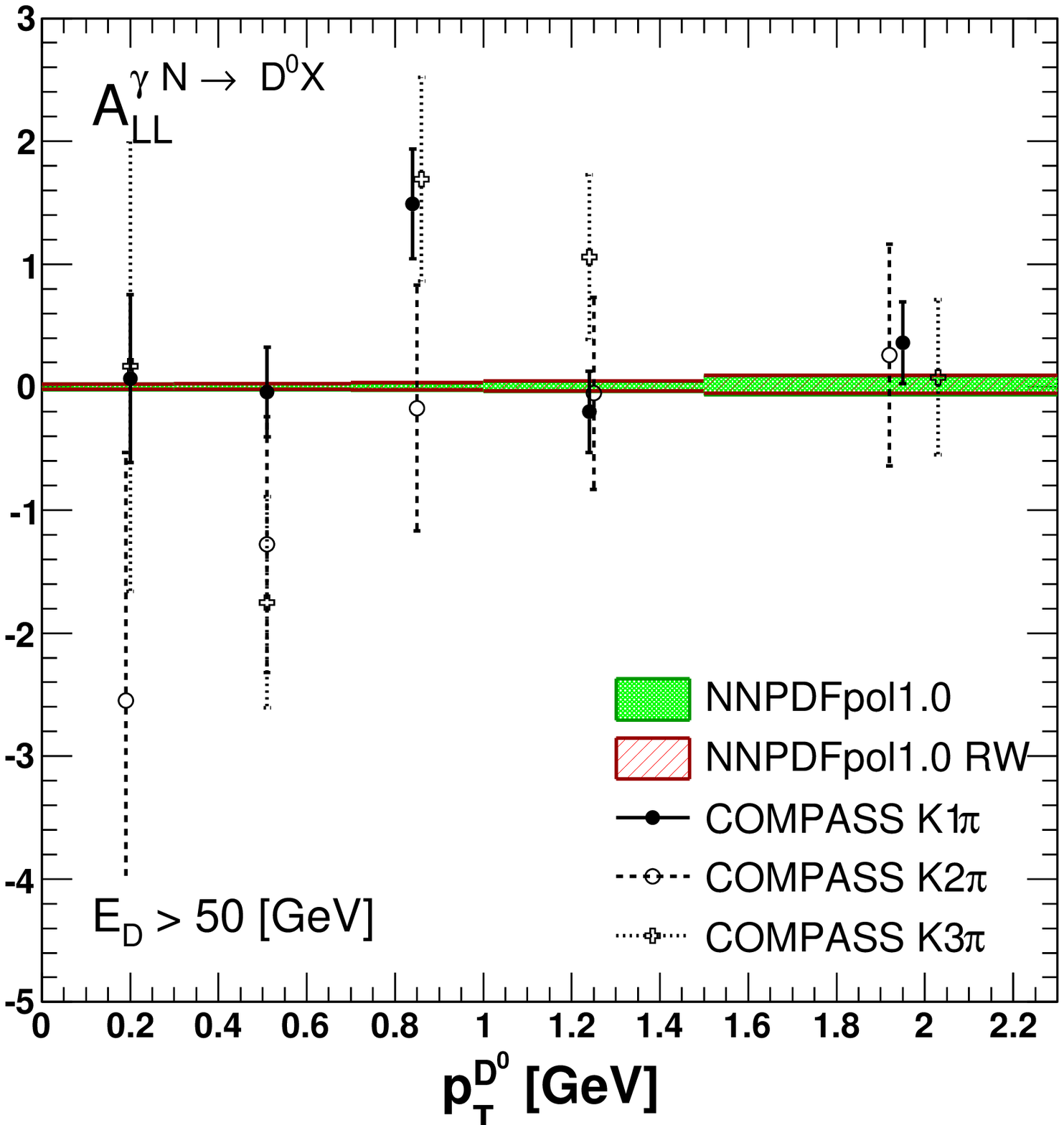}\\
\caption{\small The double-spin asymmetry $A_{LL}^{\gamma N \to D^0 X}$
determined  before and after
reweighting compared to the COMPASS data.
}
\label{fig:COMPASS_afterrw}
\end{center}
\end{figure}

\begin{figure}[!t]
\begin{center}
\epsfig{width=0.4\textwidth,figure=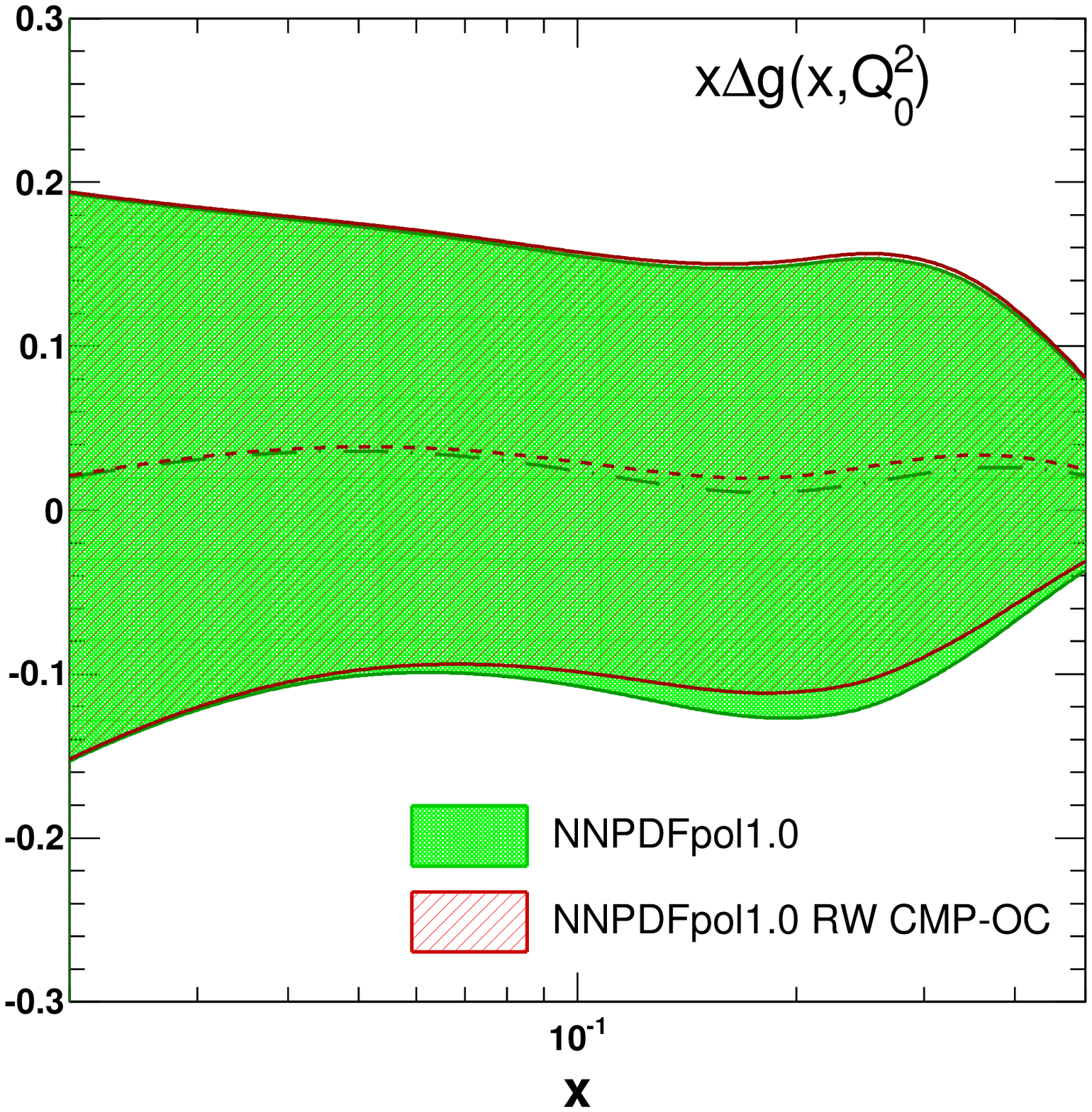}
\epsfig{width=0.4\textwidth,figure=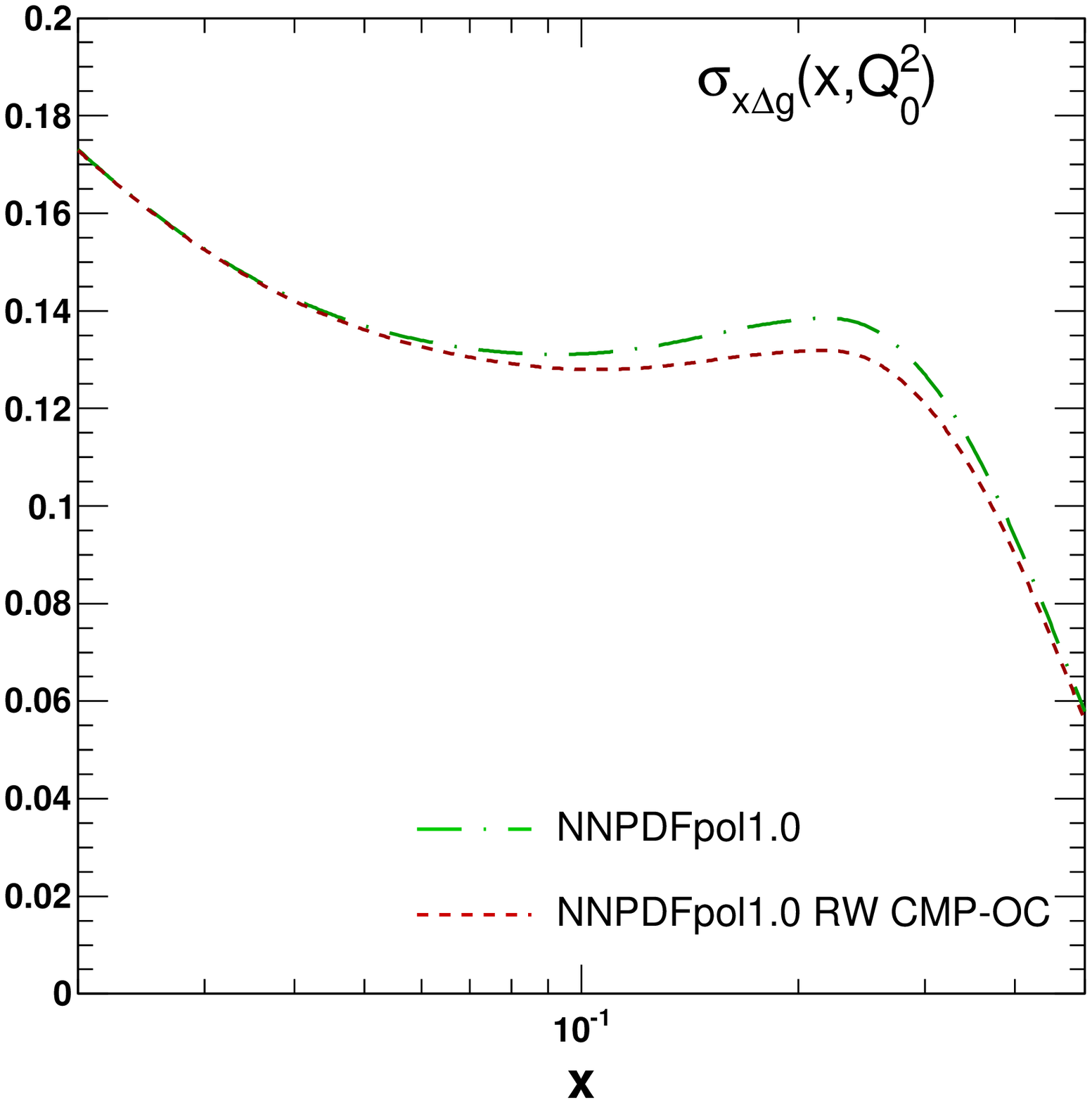}
\\
\caption{\small The polarized gluon 
 $x\Delta g(x,Q_0^2)$ at $Q_0^2=1$ GeV$^2$
 before and after reweighting with the COMPASS data
(left); the absolute uncertainty on the
gluon is also shown (right).}
\label{fig:COMPASS-gluon}
\end{center}
\end{figure}

The observable $A_{LL}^{\gamma N \to D^0 X}$,
Eq.~(\ref{eq:ALLgammaN}) and the
polarized gluon PDF  $x\Delta g(x,Q_0^2)$ at $Q_0^2=1$ GeV$^2$
 before and after reweighting, are shown in
 Fig.~\ref{fig:COMPASS_afterrw} and Fig.~\ref{fig:COMPASS-gluon}
 respectively: as expected,
they are  essentially unaffected by the inclusion of the COMPASS
data. Note in particular that the uncertainty on the gluon, also shown
in Fig.~\ref{fig:COMPASS-gluon}, is essentially unchanged. 
We conclude that the COMPASS data have little or no effect on
polarized PDFs, and specifically the polarized 
gluon PDF.



\subsection{High-$p_{T}$ jet production at STAR and PHENIX}
\label{subsec:STARjet}

We now turn to inclusive jet production in 
longitudinally polarized proton-proton collisions for which RHIC
data are available (see e.g. \cite{Bourrely:1993dd,Bunce:2000uv}). 
This data is expected
to have a significant impact on the gluon PDF because of the 
dominance of $gg$ and $qg$ initiated subprocesses in 
the accessible kinematic range (see
e.g. \cite{Bourrely:1990pz,Chiappetta:1992cp}). 
Semi-inclusive hadron production in polarized 
collisions~\cite{Adare:2008aa,Adare:2008qb,Adare:2012nq,Adler:2004ps,Adler:2006bd,Adare:2007dg,Adamczyk:2013yvv}
is also sensitive
to the gluon PDF, but it requires knowledge of fragmentation
functions, which should be consistently determined along with parton
distributions. Predictions for some semi-inclusive processes 
will be provided
in Sect.~\ref{sec:pionpheno} below with the goal of assessing their
potential relevance, but we do not include them in our PDF
determination in order not to have to rely on poorly known 
fragmentation functions.

We consider specifically data for  the longitudinal double-spin asymmetry,
\begin{equation}
A_{LL}^{\rm 1-jet}=\frac{\sigma^{++}-\sigma^{+-}}{\sigma^{++}+\sigma^{+-}}
\mbox{ ,}
\label{eq:ALL-jet}
\end{equation}
defined as the ratio of the difference to the sum of inclusive 
jet cross-sections with equal ($\sigma^{++}$) or opposite ($\sigma^{+-}$)
proton beam polarizations.
For dijet production, 
 the leading-order parton kinematics is 
\begin{equation}
x_{1}=\frac{p_T}{\sqrt{s}}\lp e^{\eta_3} + e^{\eta_4} \rp \, , \qquad
x_{2}=\frac{p_T}{\sqrt{s}}\lp e^{-\eta_3} + e^{-\eta_4} \rp
\mbox{ ,}
\label{eq:part-kin-jet}
\end{equation}
where $p_T$ is the transverse jet momentum, $\eta_{3,4}$ are the rapidities of 
the two jets and $\sqrt{s}$ is the center-of-mass energy.
In single-inclusive jet production, the underlying Born kinematics 
is not fixed uniquely because the second jet is being integrated over (in
Fig.~\ref{fig:NNPDFpol11-kin} we have conventionally assumed
equal longitudinal momentum of the incoming partons, so $\eta_3=-\eta_4\equiv \eta$
and
$x_{1,2}=\frac{2p_T}{\sqrt{s}} e^{\pm \eta}$).

The NLO QCD computation for inclusive high-$p_T$ jet 
production in polarized hadron collisions was first presented in 
Ref~\cite{deFlorian:1998qp}, based on the subtraction method of 
Refs.~\cite{Frixione:1995ms,Frixione:1997np}, along with a code for
parton-level event generation. More recently,
 spin-dependent and spin-averaged cross-sections
for single-inclusive high-$p_T$ jet production have been determined
in Ref.~\cite{Jager:2004jh} using the so-called \textit{narrow-cone} 
approximation, which holds in the limit of not too large jet radius $R$.
In this approximation, analytical results for the corresponding NLO partonic 
cross-sections can be derived, leading to a faster and more efficient
computer code, as all singularities arising in the intermediate steps 
of the calculation  explicitly cancel.
The narrow-cone approximation was shown to be close to the result of Ref.~\cite{deFlorian:1998qp}
in the particular case of RHIC kinematics~\cite{Jager:2004jh}. We will
therefore use the code of Ref.~\cite{Jager:2004jh}, rather than
that of Ref.~\cite{deFlorian:1998qp}. The approach of Ref.~\cite{Jager:2004jh} 
has been recently extended in Ref.~\cite{Mukherjee:2012uz} to 
$k_t$-type jet algorithms, used for the latest STAR data.

The two general purpose experiments at RHIC, STAR and PHENIX, have presented
measurements of the longitudinal double-spin asymmetry
for inclusive jet production, Eq.~(\ref{eq:ALL-jet}). 
Results from STAR are available for the 2005 and 2006 runs and,
since very recently, also for the 2009 run.
On the other hand, a single data set is available from PHENIX, 
corresponding to data taken in 2005, while further jet measurements from this
experiment are not foreseen due to detector limitations
in angular coverage.
In the following, these data sets will be referred to as STAR 1j-05, STAR 1j-06,
STAR 1j-09A, STAR 1j-09B and PHENIX 1j respectively.

The features of these data sets are 
summarized in Tab.~\ref{tab:jet-data} and the corresponding Born-level 
kinematic coverage is shown in Fig.~\ref{fig:NNPDFpol11-kin}.
The experimental covariance matrix is available only for the 
STAR 1j-09A and STAR 1j-09B data sets: we included it in our analysis
through a routine provided by the STAR collaboration~\cite{web:Gagliardi},
which takes into account also additional fully correlated systematics
arising from relative luminosity and jet energy scale uncertainties.
For the other data sets, systematic and statistical uncertainties
are added in quadrature. 
For the STAR 1j-05 and STAR 1j-06 data sets, we have to account for the fact that
the data are taken in bins of $p_T$, whereas the corresponding theoretical 
predictions are computed for the center of each bin.
We estimate the 
corresponding uncertainty as the maximal variation of the observable 
within each bin and take that value as a further uncorrelated systematic
uncertainty. Furthermore, although these data are provided with asymmetric systematic 
uncertainties, we  symmetrize them, according to
Eqs.~(7)-(8) of Ref.~\cite{DelDebbio:2004qj}.
\begin{table}[!t]
\small 
\centering
\begin{tabular}{lccccccc}
\toprule
Data set & $N_{\mathrm{dat}}$ & jet-algorithm & $R$  & 
$[\eta_{\mathrm{min}},\eta_{\mathrm{max}}]$ &
$\sqrt{s}$ [GeV] & $\mathcal{L}$ [pb$^{-1}$] & Ref. \\
\midrule
STAR 1j-05 & $10$ & midpoint-cone & $0.4$ & $[+0.20,+0.80]$ &
$200$ & $2.1$ & \cite{Adamczyk:2012qj} \\
STAR 1j-06 & $9$ & midpoint-cone & $0.7$ & $[-0.70,+0.90]$ &
$200$ & $5.5$ & \cite{Adamczyk:2012qj} \\
STAR 1j-09A   & $11$ & anti-$k_t$ & $0.6$ & $[-0.50,+0.50]$ &
$200$ & $25$ & \cite{Adamczyk:2014ozi} \\
\multirow{2}*{STAR 1j-09B }  & \multirow{2}*{$11$} & 
\multirow{2}*{anti-$k_t$} & \multirow{2}*{$0.6$} & $[-1.00,-0.50]$ &
\multirow{2}*{$200$} & \multirow{2}*{$25$} & \multirow{2}*{\cite{Adamczyk:2014ozi}} \\
& & & & $[+0.50,+1.00]$ & & & \\
\midrule
PHENIX 1j & $6$ & seeded-cone & $0.3$ & $[-0.35,+0.35]$ &
$200$ & $2.1$ & \cite{Adare:2010cc} \\
\bottomrule
\end{tabular}
\caption{\small Features of the RHIC inclusive jet data included in 
the present analysis:  the number of data points, $N_{\mathrm{dat}}$, 
the algorithm used for jet reconstruction and the  value of the jet radius, 
$R$, the range over which the rapidity $\eta$ is integrated, 
the center-of-mass energy of the collisions,
$\sqrt{s}$, and the integrated luminosity for each run, $\mathcal{L}$.}
\label{tab:jet-data}
\end{table}

For each data set in Tab.~\ref{tab:jet-data}, we have computed the 
longitudinal double-spin asymmetry Eq.~(\ref{eq:ALL-jet}), at NLO,
using the narrow-cone approximation code of
Ref.~\cite{Jager:2004jh,Mukherjee:2012uz}, suitably modified 
in order to use the NNPDF polarized parton sets via
the {\tt LHAPDF} interface~\cite{Whalley:2005nh,web:LHAPDF}. In each case 
we use the  jet algorithm and cone radius  which
are appropriate for the given data set, as listed in Tab.~\ref{tab:jet-data}.
Polarized and unpolarized PDFs are taken respectively from the prior 
ensembles constructed in Sect.~\ref{sec:prior} and from the 
\texttt{NNPDF2.3} NLO reference parton set.
As for  open-charm muoproduction, the 
numerator in Eq.~(\ref{eq:ALL-jet}) is computed for each replica in the
prior ensembles ($N_{\mathrm{rep}}=1000$), while the denominator is evaluated 
only once for the central unpolarized replica.
This is justified because  uncertainties of the polarized 
PDF completely dominate over those of the unpolarized ones.

\begin{table}[!t]
\small 
\centering
\begin{tabular}{ll|c|cccc|cccc}
\toprule
Experiment & Set & $N_{\mathrm{dat}}$ 
& \multicolumn{4}{c|}{$\chi^2/N_{\mathrm{dat}}$}
& \multicolumn{4}{c}{$\chi^2_{\mathrm{rw}}/N_{\mathrm{dat}}$}\\
\midrule
& & & $1\sigma$ & $2\sigma$ & $3\sigma$ & $4\sigma$
    & $1\sigma$ & $2\sigma$ & $3\sigma$ & $4\sigma$\\
STAR &            & $41$ & $1.50$ & $1.49$ & $1.50$ & $1.50$
                         & $1.05$ & $1.04$ & $1.04$ & $1.04$\\
     & STAR 1j-05 & $10$ & $1.04$ & $1.05$ & $1.04$ & $1.04$
                         & $1.01$ & $1.02$ & $1.02$ & $1.02$\\
     & STAR 1j-06 &  $9$ & $0.75$ & $0.76$ & $0.76$ & $0.76$
                         & $0.59$ & $0.58$ & $0.59$ & $0.59$\\
     & STAR 1j-09A& $11$ & $1.40$ & $1.39$ & $1.39$ & $1.40$
                         & $0.98$ & $0.99$ & $0.98$ & $0.98$\\
     & STAR 1j-09B& $11$ & $3.04$ & $3.05$ & $3.03$ & $3.05$
                         & $1.18$ & $1.17$ & $1.17$ & $1.18$\\                         
\midrule
PHENIX & & & & & & & & & & \\
     &  PHENIX 1j &  $6$ & $0.24$ & $0.24$ & $0.24$ & $0.24$
                         & $0.24$ & $0.24$ & $0.24$ & $0.24$ \\
\midrule
     &            & $47$ & $1.35$ & $1.35$ & $1.35$ & $1.36$
                         & $1.00$ & $1.01$ & $1.01$ & $1.00$ \\
\bottomrule
\end{tabular}
\caption{\small The value of the $\chi^2$ per data point $\chi^2/N_{\mathrm{dat}}$
($\chi^2_{\mathrm{rw}}/N_{\mathrm{dat}}$) before (after) reweighting 
the prior samples with the RHIC inclusive jet data.
Results are presented for both separate and total data sets and for each
prior discussed in Sect.~\ref{sec:prior}.}
\label{tab:est1-ALL}
\end{table}

We have included the jet data of Tab.~\ref{tab:jet-data} 
by reweighting the prior sets discussed in Sect.~\ref{sec:prior}. The 
$\chi^2$ per data point before and after reweighting 
are listed in Tab.~\ref{tab:est1-ALL},  for each data set and a
combined set including all these data, for each of the four prior sets.
Various measures of the effectiveness of the reweighting process are
listed in Tab.~\ref{tab:est2-ALL}: the effective number of replicas
$N_{\rm eff}$, 
and the modal value of the $\mathcal{P}(\alpha)$ distribution,
$\la\alpha\ra$, defined by Eq.~(12) of Ref.~\cite{Ball:2010gb}.
The parameter $\alpha$ measures the consistency of the data which are
used for reweighting with those included in the prior set: $\alpha$ is
the
factor by which the uncertainty on the new data must 
be rescaled in order 
for both sets to be consistent with each other. A value of $\alpha$ 
close to one  means that the uncertainties have been correctly estimated.

Inspection of Tabs.~\ref{tab:est1-ALL}-\ref{tab:est2-ALL} allows us to
draw the following conclusions:
\begin{itemize}

\item Results are essentially independent of the choice of
prior. This was to be expected, given the very mild sensitivity of this
observable to the polarized flavor-antiflavor decomposition.

\item The effective number of replicas after reweighting
is always significantly lower than the  
size of the initial sample $N_{\mathrm{rep}}=1000$, suggesting that
the RHIC data do have a significant impact on the fit.
The most constraining data sets, for which
$N_{\mathrm{eff}}$ is smallest, are STAR 1j-09A 
and STAR 1j-09B: this is to be expected
since these are the measurements with  smallest statistical
and systematic uncertainties.

\item The effective number of replicas after reweighting is
  nevertheless  always rather larger than a hundred, which is a
  typical size needed for the final replica sample to provide accurate
  results. This means that the size of the prior sample is large enough
  for final results to be reliable.

\item 
The modal value
of the $\mathcal{P}(\alpha)$ distribution for all the STAR data as
well as for the global data set is always close to one, 
suggesting correct uncertainty estimation
(even for the earlier data sets for which no information on correlated
systematics is available). However,
the modal value of $\alpha$ is significantly below one for the PHENIX
data,
suggesting that for this data set uncertainties are overestimated,
possibly due to the missing information on correlated systematics.

\item The $\chi^2$ after reweighting is of order one for the data sets
  for which information on correlated systematics is available, and
  it shows a significant improvement, with an especially remarkable
  agreement for the STAR 1j-09B which, as mentioned, has the smallest
  uncertainties. This suggests that these data are bringing in
  significant new information.
\end{itemize}

\begin{table}[!t]
\small 
\centering
\begin{tabular}{ll|c|cccc|cccc}
\toprule
Experiment & Set & $N_{\mathrm{dat}}$ 
& \multicolumn{4}{c|}{$N_{\mathrm{eff}}$}
& \multicolumn{4}{c}{$\langle\alpha\rangle$}\\
\midrule
& & & $1\sigma$ & $2\sigma$ & $3\sigma$ & $4\sigma$
    & $1\sigma$ & $2\sigma$ & $3\sigma$ & $4\sigma$\\
STAR &            & $41$ & $256$ & $256$ & $256$ & $260$
                         & $1.01$ & $1.10$ & $1.02$ & $1.01$\\
     & STAR 1j-05 & $10$ & $931$ & $931$ & $931$ & $931$
                         & $1.03$ & $1.02$ & $1.05$ & $1.06$ \\
     & STAR 1j-06 &  $9$ & $621$ & $623$ & $622$ & $621$
                         & $0.97$ & $0.98$ & $0.95$ & $0.96$ \\
     & STAR 1j-09A& $11$ & $244$ & $244$ & $244$ & $244$ 
                         & $1.20$ & $1.20$ & $1.15$ & $1.17$\\
     & STAR 1j-09B& $11$ & $299$ & $299$ & $299$ & $300$ 
                         & $1.10$ & $1.05$ & $1.05$ & $1.10$\\                         
\midrule
PHENIX & & & & & & & & & & \\
     &  PHENIX 1j &  $6$ & $740$ & $740$ & $740$ & $741$
                         & $0.55$ & $0.50$ & $0.50$ & $0.55$ \\
\midrule
     &            & $47$ & $334$ & $338$ & $334$ & $333$
                         & $1.00$ & $1.02$ & $0.99$ & $0.99$ \\
\bottomrule
\end{tabular}
\caption{\small The effective number of replicas after reweighting, 
$N_{\mathrm{eff}}$ (the starting sample has $N_{\mathrm{rep}}=1000$), and 
the modal value
of the $\mathcal{P}(\alpha)$ distribution, $\la\alpha\ra$ (note that
here and henceforth  $\la\alpha\ra$ denotes the mode, not the mean of
the $P(\alpha)$ distribution).
Results are quoted for separate and combined data sets and for each of the
different
prior PDF ensembles.}
\label{tab:est2-ALL}
\end{table}

\begin{figure}[p]
\centering
\epsfig{width=0.4\textwidth,figure=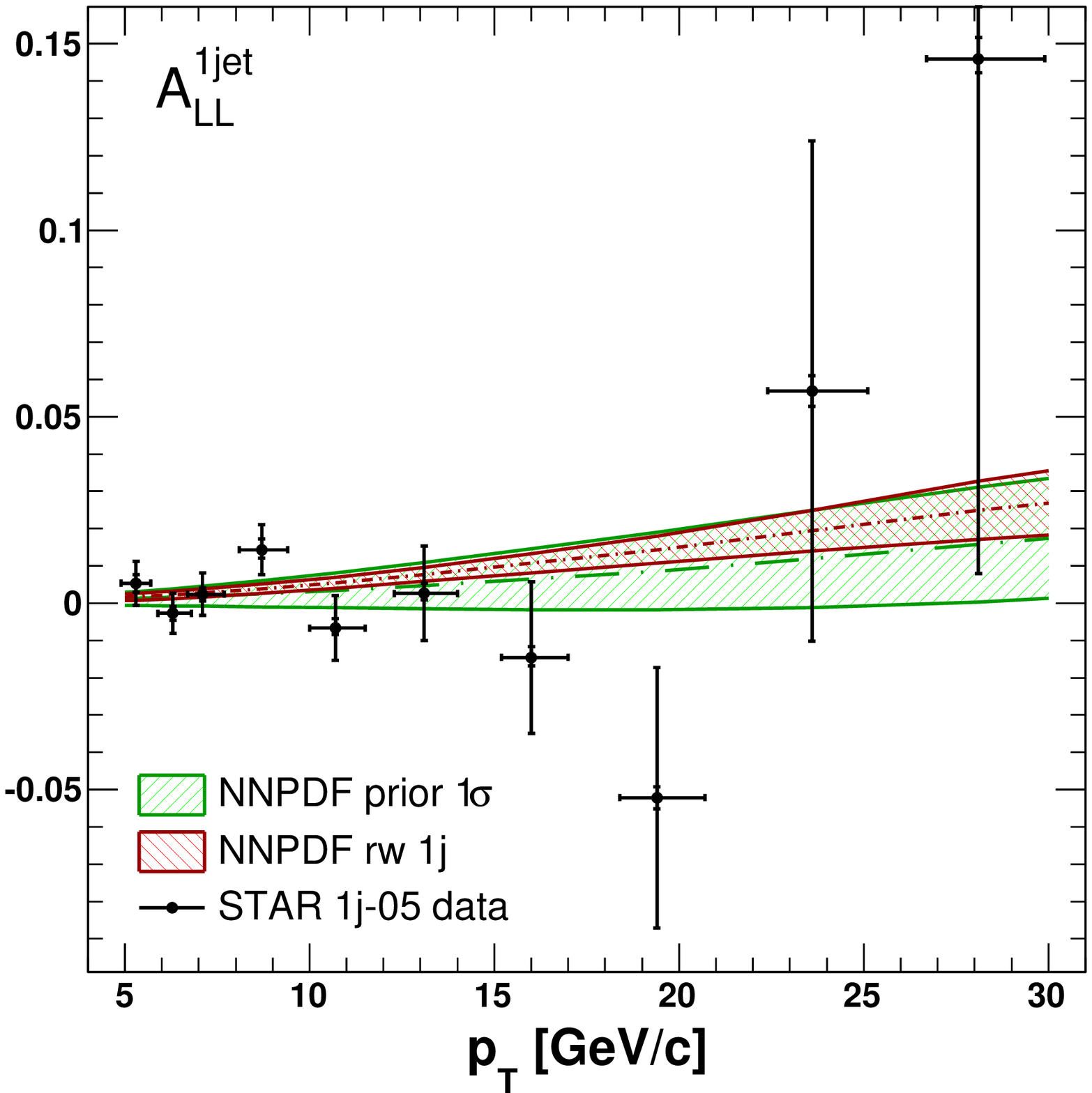}
\epsfig{width=0.4\textwidth,figure=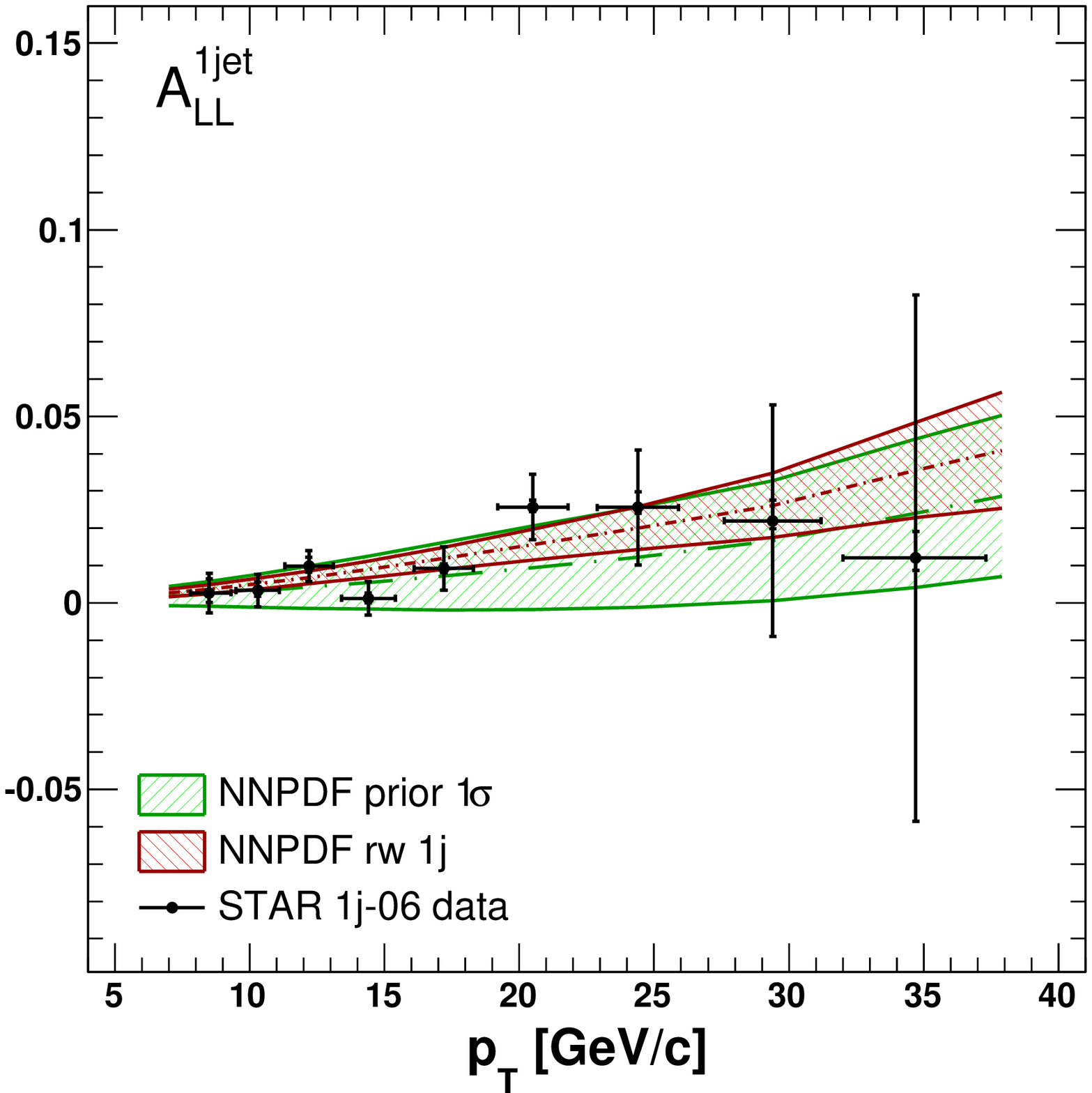}\\
\epsfig{width=0.4\textwidth,figure=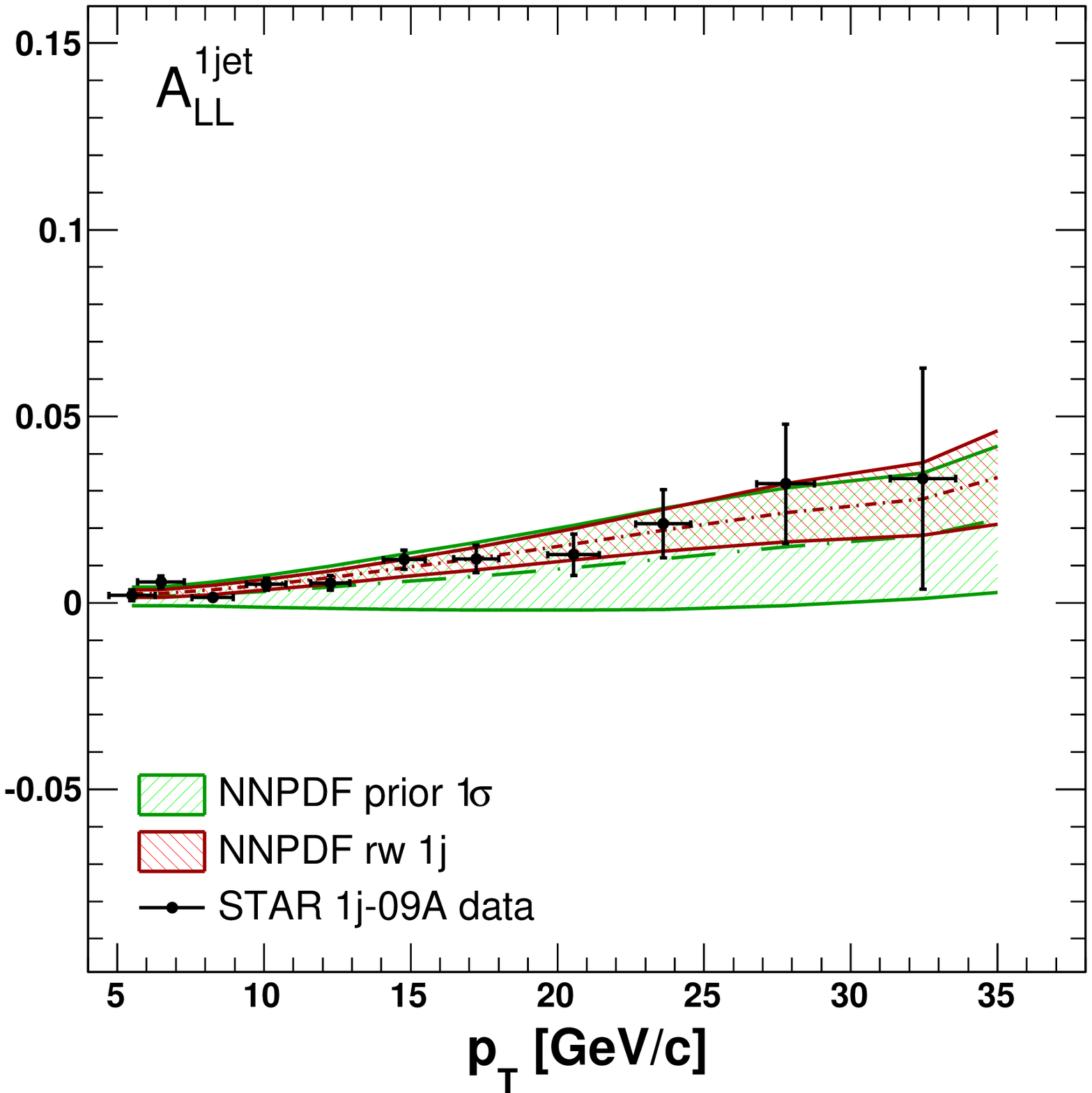}
\epsfig{width=0.4\textwidth,figure=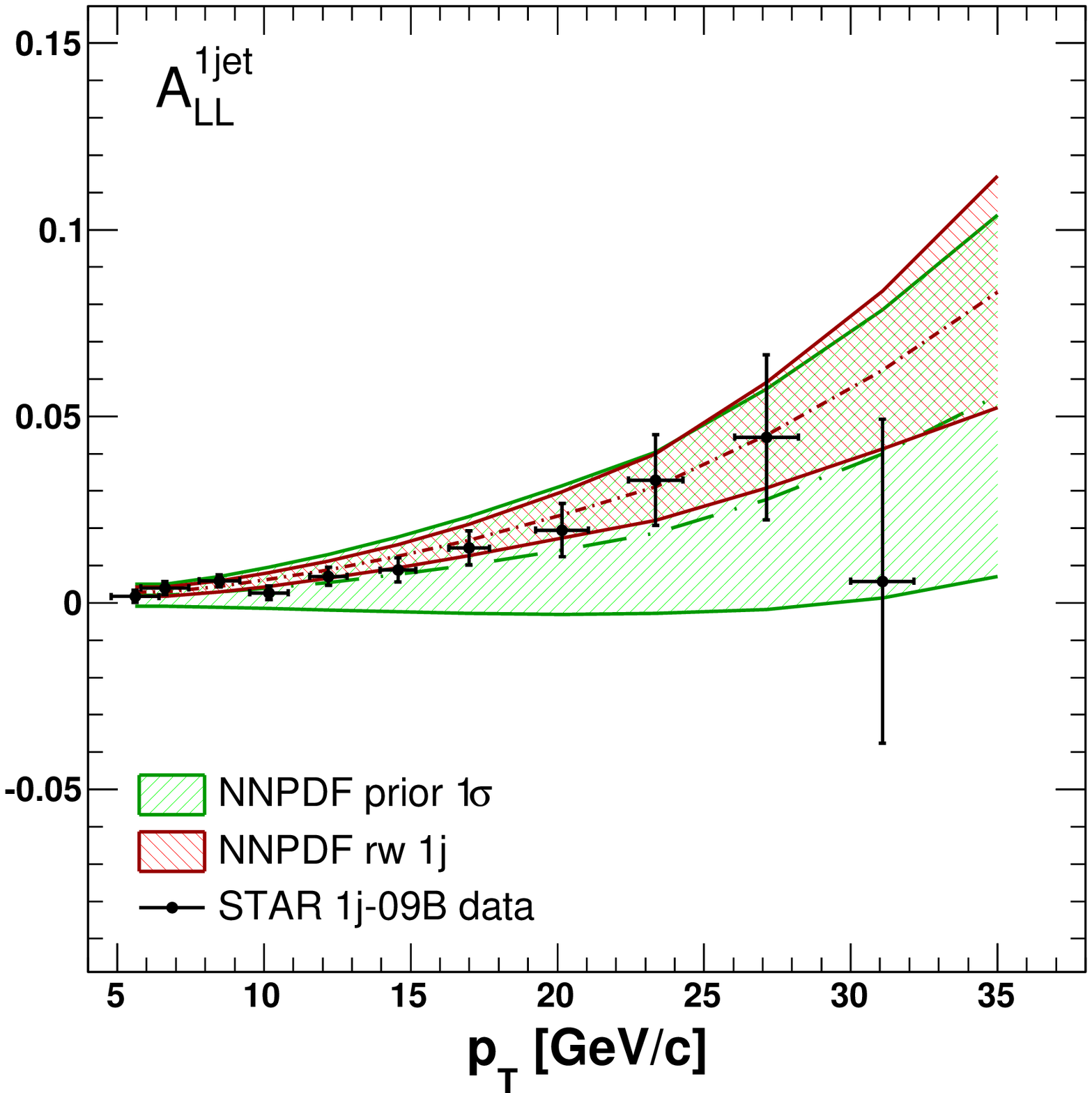}\\
\epsfig{width=0.4\textwidth,figure=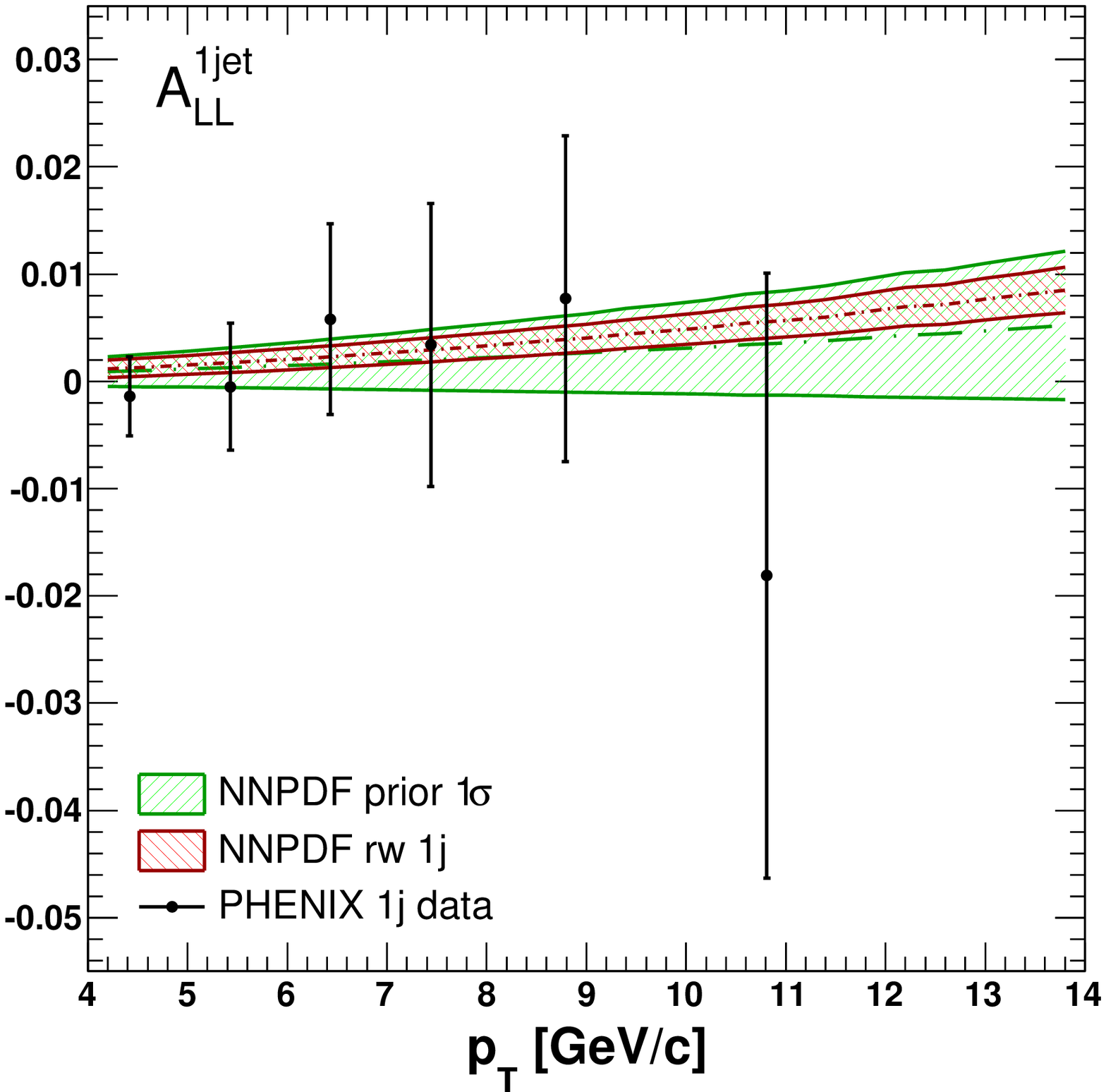}
\caption{\small Predictions for the double longitudinal spin asymmetry
for single-inclusive jet production $A_{LL}^{\rm 1-jet}$,
Eq.~(\ref{eq:ALL-jet}), as a function of the
jet $p_T$, before and after reweighting, compared to the 
corresponding RHIC data.
Note the different scale of vertical axis for the PHENIX
data as compared to the STAR data.}
\label{fig:ALL-plot}
\end{figure}

Predictions for the asymmetry $A_{LL}^{\rm 1-jet}$
Eq.~(\ref{eq:ALL-jet})
obtained using the \texttt{NNPDFpol1.0} PDF set before and after
reweighting are compared to the RHIC data
in Fig.~\ref{fig:ALL-plot}. The curves shown correspond to  the
$1\sigma$ prior PDF ensemble, reweighted with the complete data set of Tab.~\ref{tab:jet-data}. 
The improvement in experimental uncertainties in STAR 1j-09A
and STAR 1j-09B as compared to all
other data sets is clearly visible, and it leads to an equally visible
reduction of the uncertainty on the theoretical prediction, as well as
a significant change of its central value.

\begin{figure}[!t]
\centering
\epsfig{width=0.4\textwidth,figure=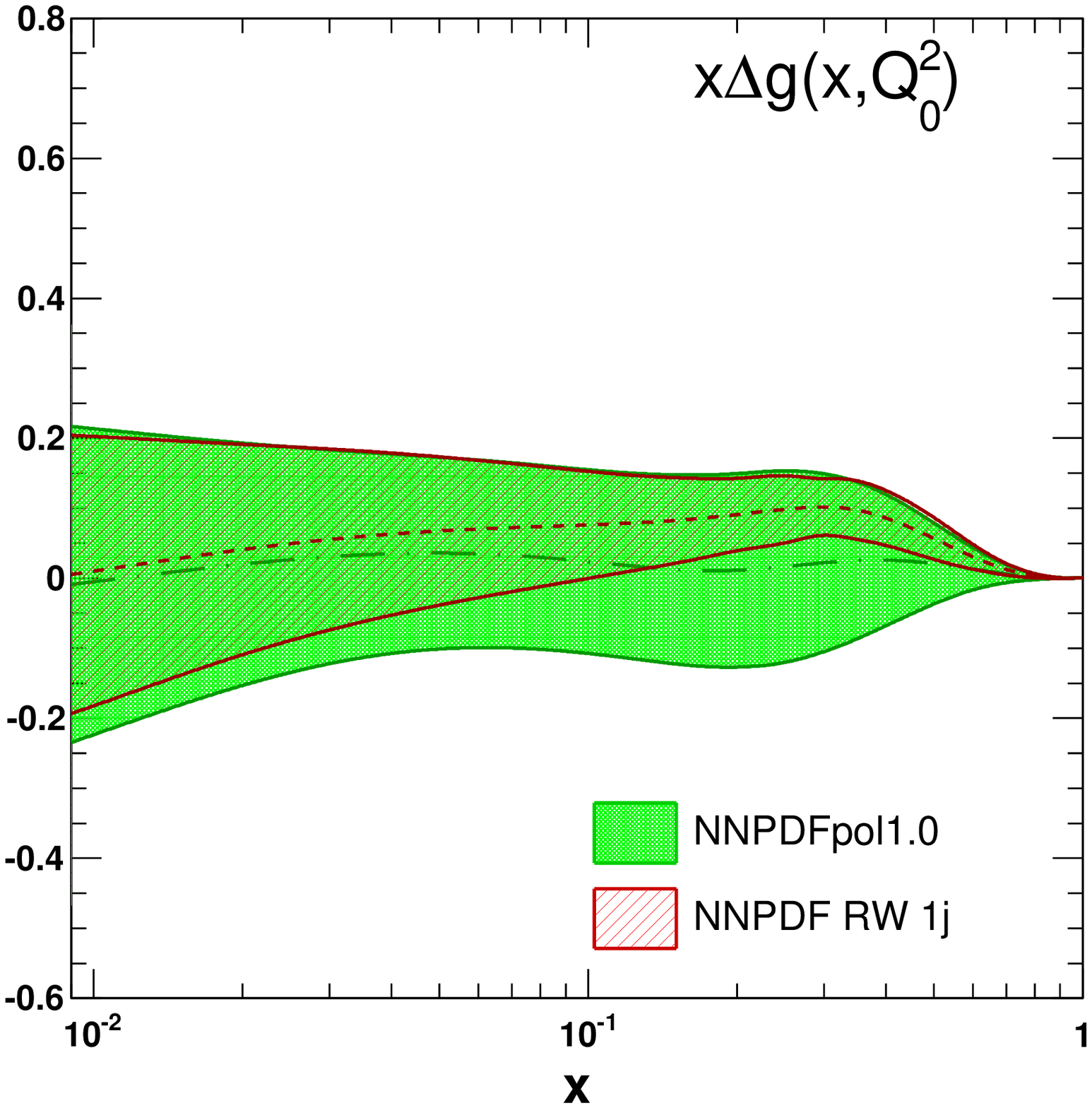}
\epsfig{width=0.4\textwidth,figure=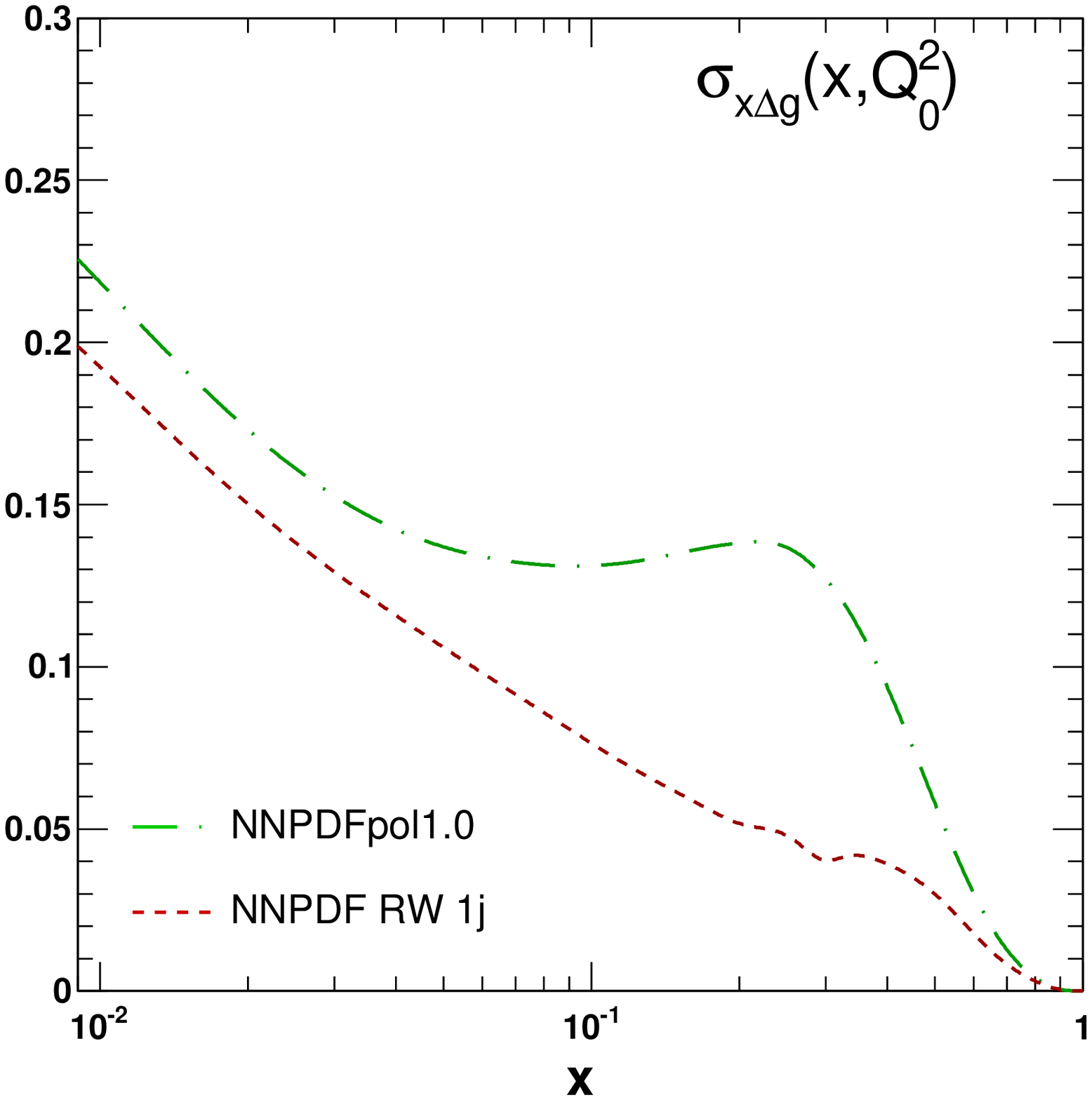}
\caption{\small  The \texttt{NNPDFpol1.0} polarized gluon distribution $x\Delta g(x,Q_0^2)$
at $Q_0^2=1$ GeV$^2$ before and after reweighting with the RHIC jet
data, labeled NNPDF RW 1j (left); the absolute uncertainty on the
gluon is also shown (right).
}
\label{fig:gluonrw-jets}
\end{figure}

This is mostly due to a corresponding reduction in uncertainty and
change of shape of the polarized gluon upon reweighting: this can be 
seen 
in Fig.~\ref{fig:gluonrw-jets}, where the polarized gluon 
distributions before and after reweighting are compared. Here
too the result corresponds to  the
$1\sigma$ prior PDF ensemble, reweighted with the complete data set of Tab.~\ref{tab:jet-data}. 
We have explicitly checked that the reweighted gluon is independent of
the choice of prior.
In the kinematic range probed at RHIC,
$x \in \lc 0.04,0.2\rc$ (see Fig.~\ref{fig:NNPDFpol11-kin}),
the polarized gluon
PDF tends to become positive and its uncertainty is reduced by
more than a factor two for $x\sim 0.3$.
This feature is qualitatively consistent with that reported by the DSSV group
in Ref.~\cite{Aschenauer:2013woa,deFlorian:2014yva},
based on the analysis of the same data. 
The other PDFs are essentially unaffected by the inclusion of the RHIC
jet data.

\section{Polarized quark-antiquark separation: the $W$ asymmetry}
\label{sec:flavour}
 
We now turn to polarized hadron collider data that constrain 
the flavor separation
of polarized quarks and antiquarks.
We consider in particular $W$ production. Because of the chiral nature
of the weak interactions, the polarized parton content may be accessed
both through (parity violating) single-spin and (parity conserving) 
double-spin asymmetries.
The single-spin asymmetry is defined as
\begin{equation}
 A_L
\equiv 
\frac{\sigma^+ - \sigma^-}{\sigma^++\sigma^-}
=
\frac{\Delta\sigma}{\sigma}\mbox{, }
\label{eq:Wasy}
\end{equation}
where $\sigma^{+(-)}$ denotes the cross-section for $W$
production when  colliding   positive (negative)
longitudinally polarized protons off unpolarized protons, and the
double-spin asymmetry
\begin{equation}
A_{LL}\equiv\frac{\sigma^{++}-\sigma^{+-}}{\sigma^{++}+\sigma^{+-}}
\,\mbox{,}
 \label{eq:ALLdef}
\end{equation}
where $\sigma^{++}$ ($\sigma^{+-}$) is the cross-section for $W$
production with equal (opposite) proton beam polarizations.

At leading order, neglecting Cabibbo-suppressed channels,
the former is given by
\begin{equation}
A_L^{W^+}\approx\frac{\Delta u(x_1)\bar{d}(x_2) - \Delta\bar{d}(x_1) u(x_2)}
{u(x_1)\bar{d}(x_2) + \bar{d}(x_1)u(x_2)}\, .
\label{eq:Wasy+pm}
\end{equation}
and the latter is
\begin{equation}
A_{LL}^{W^+}
\approx
-\frac{\Delta u(x_1)\Delta\bar{d}(x_2)+\Delta\bar{d}(x_1)\Delta u(x_2)}{u(x_1)\bar{d}(x_2)+\bar{d}(x_1)u(x_2)}
\,\mbox{,}
 \label{eq:ALLLO}
\end{equation}
where for fixed $W$ rapidity $y_W$,
the momentum fractions $x_{1,2}$ carried by the colliding
partons are given by
\begin{equation}
 x_{1,2}=\frac{M_W}{\sqrt{s}}e^{\pm y_W}\mbox{.}
 \label{eq:x1x2}
\end{equation}
It is thus clear that first, each of these asymmetries is sensitive to
the flavor decomposition of polarized quark and antiquark
distributions and second, that their simultaneous measurement provides
an especially strong constraint, as both linear and quadratic
combinations of polarized PDFs are being measured. This last fact has
the interesting implication that
the single- and double-spin asymmetries
Eqs.~(\ref{eq:Wasy})-(\ref{eq:ALLdef}) must satisfy a nontrivial
positivity bound, derived
in Ref.~\cite{Kang:2011qz}
\begin{equation}
 1\pm A_{LL}(y_W) > \left| A_L(y_W)\pm A_L(-y_W) \right| \, ,
\label{eq:posbound}
\end{equation}
where $y_W$ is the $W$ boson rapidity (not to be confused with 
$\eta_l$, the pseudo-rapidity of the lepton from the $W$ decay, which
is used for the experimental measurements).

Both the STAR and PHENIX collaborations have presented measurements
of the parity-violating spin asymmetry $A_L^{W^\pm}$, Eq.~(\ref{eq:Wasy}), 
based on the RHIC 2009 run at 
$\sqrt{s}=500$ GeV.~\cite{Aggarwal:2010vc,Adare:2010xa}.
These measurements are based on
data sets with low integrated luminosities:
$\mathcal{L}=12$ pb$^{-1}$ and $\mathcal{L}=8.6$ pb$^{-1}$
respectively for 
STAR and PHENIX. Each of these two experiments provides a 
determination of the asymmetry for a single value of the rapidity and
for outgoing $W^\pm$, but as discussed in Ref.~\cite{Nocera:2013spa} they
are affected by very large uncertainties, do not provide
any significant constraint, and we need not discuss them further.

Also, the STAR collaboration has recently presented~\cite{Adamczyk:2014xyw}
results for both the $A_L^{W^\pm}$ and $A_{LL}^{W^\pm}$
asymmetries, based on $\mathcal{L}=9$
pb$^{-1}$  of  $\sqrt{s}=500$ GeV data from the 2010 and 
 $\mathcal{L}=77$ pb$^{-1}$ of data at
 $\sqrt{s}=510$~GeV from the 2011 run. These data have been combined
into a single data set at the nominal energy of  $\sqrt{s}=510$~GeV, and
have greatly reduced uncertainties as compared to previous measurements.

The STAR data are provided for both $W^+$ and $W^-$ final states,
which we will refer to as STAR-$A_L^{W^\pm}$ and STAR-$A_{LL}^{W^\pm}$,
presented in bins of  $\eta_l$, the  rapidity of the lepton from 
the $W$ decay. In particular  the single-spin
asymmetry data are given in six rapidity bins, and the double-spin asymmetry in
three rapidity bins,
integrated over the lepton transverse momentum in the range
$25<p_T^{l}<50$ GeV. Correlated beam polarization systematics
 ($3.4\%$ and $6.5\%$
respectively for single- and double-spin asymmetries) are provided,
while an additional uncorrelated systematics, due to the relative luminosity, 
affects $A_L^{W^\pm}$~\cite{Adamczyk:2014xyw}. 
Using LO kinematics we see that these
STAR $W$ production data constrain light quark
and antiquark PDFs with
$ 0.05\lesssim x \lesssim 0.4$ and  $Q^2 \sim M_W^2$, 
see Fig.~\ref{fig:NNPDFpol11-kin}.

\begin{figure}[!t]
\begin{center}
\epsfig{width=0.4\textwidth,figure=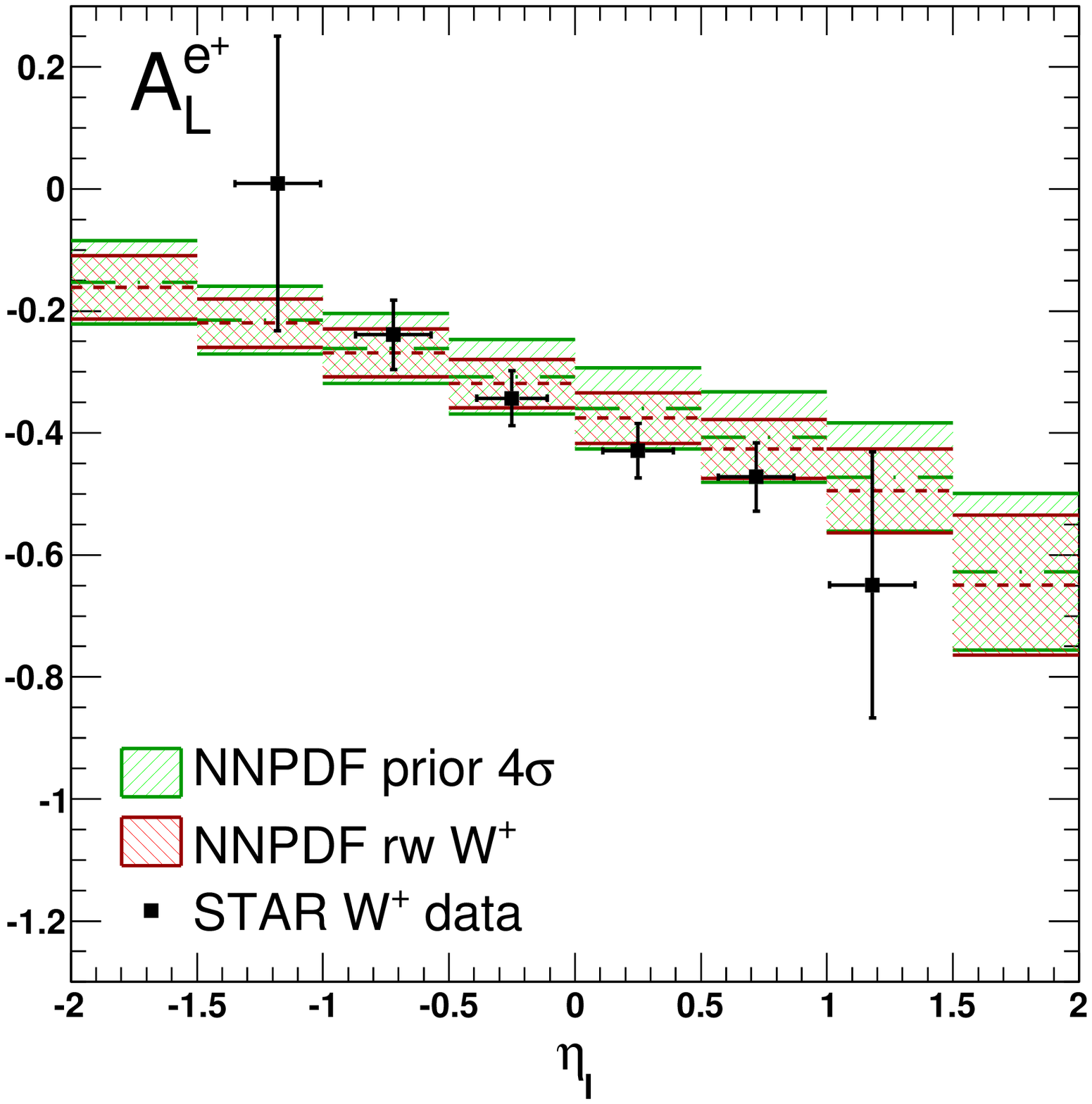}
\epsfig{width=0.4\textwidth,figure=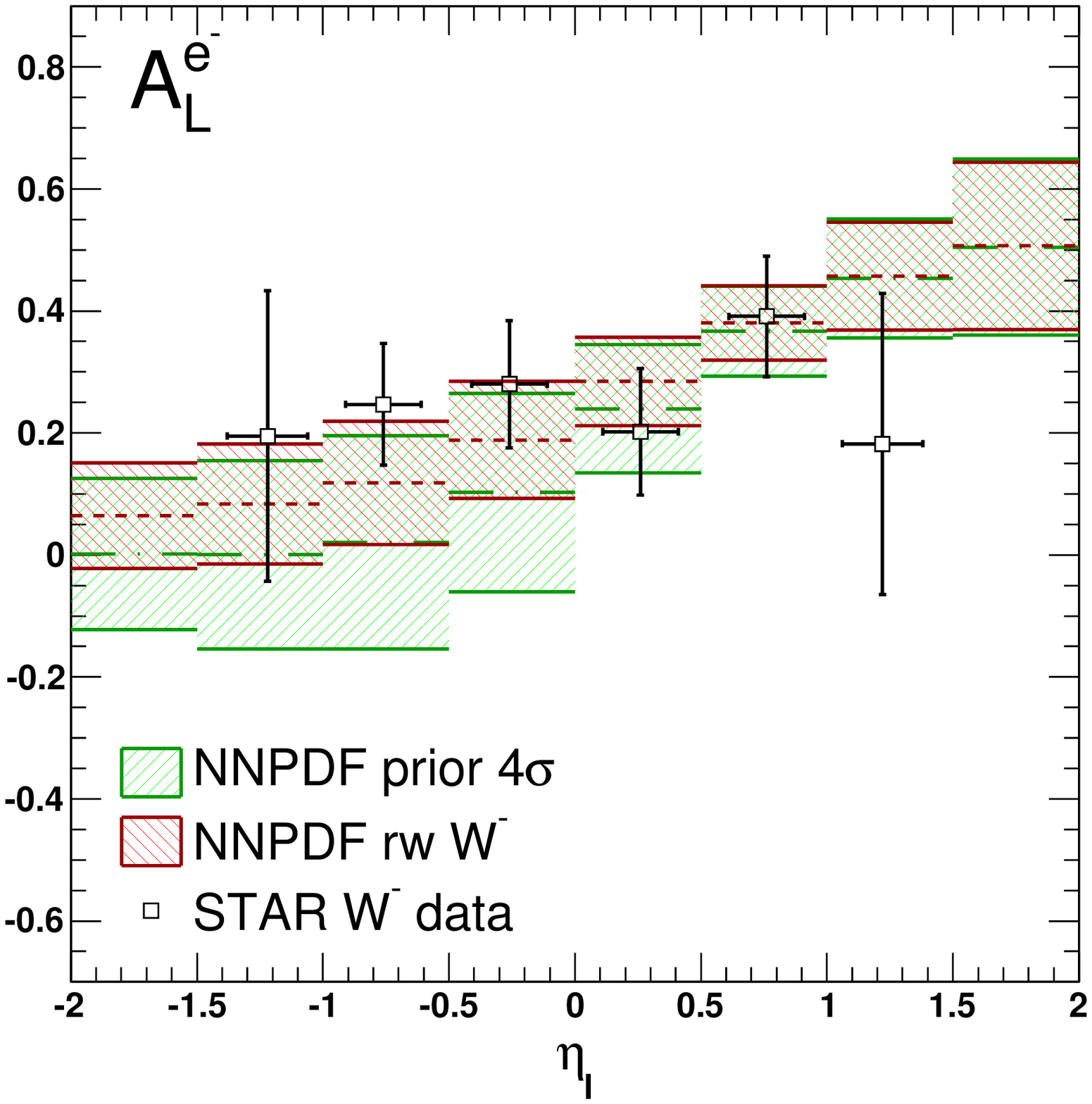}\\
\epsfig{width=0.4\textwidth,figure=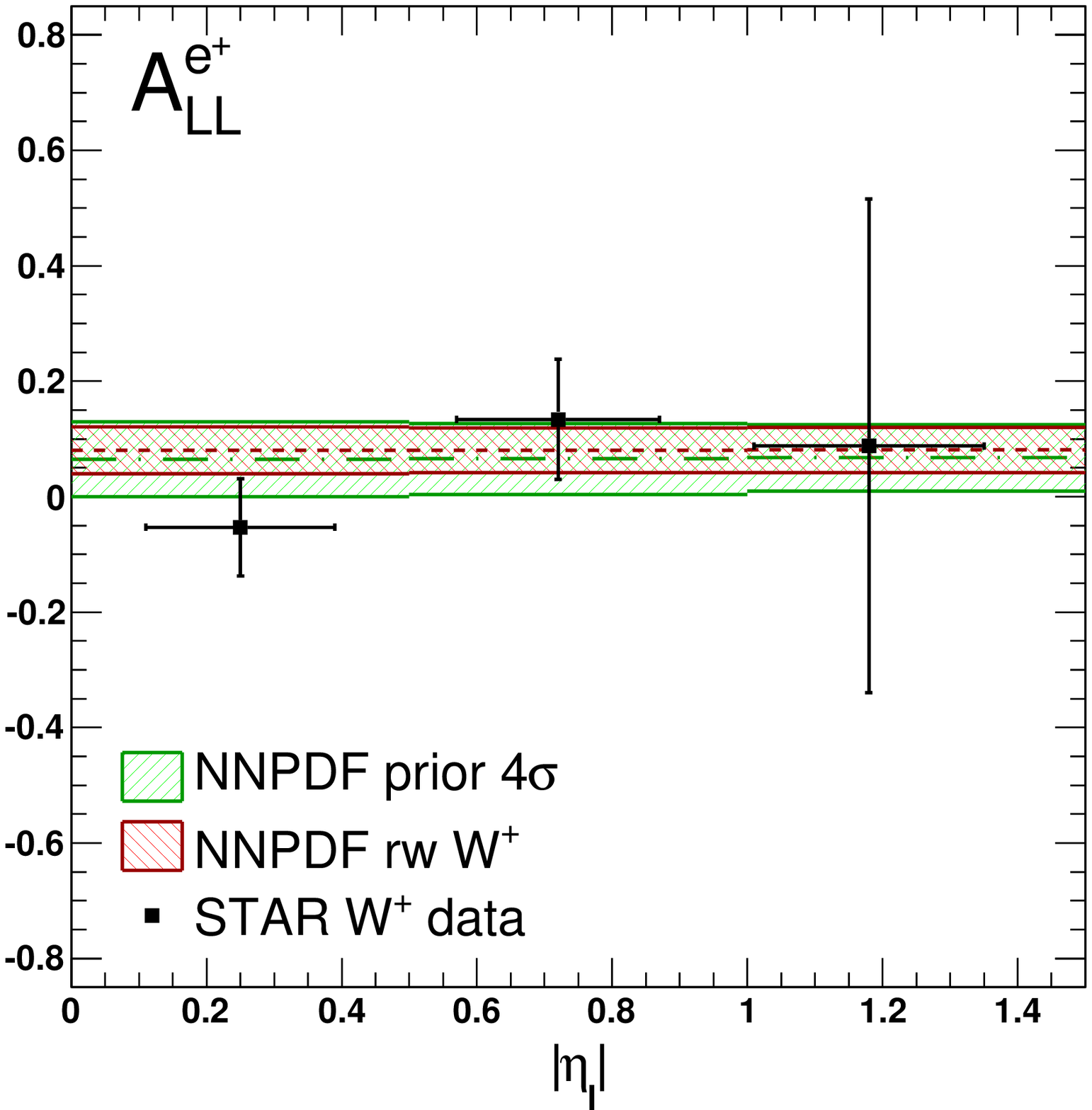}
\epsfig{width=0.4\textwidth,figure=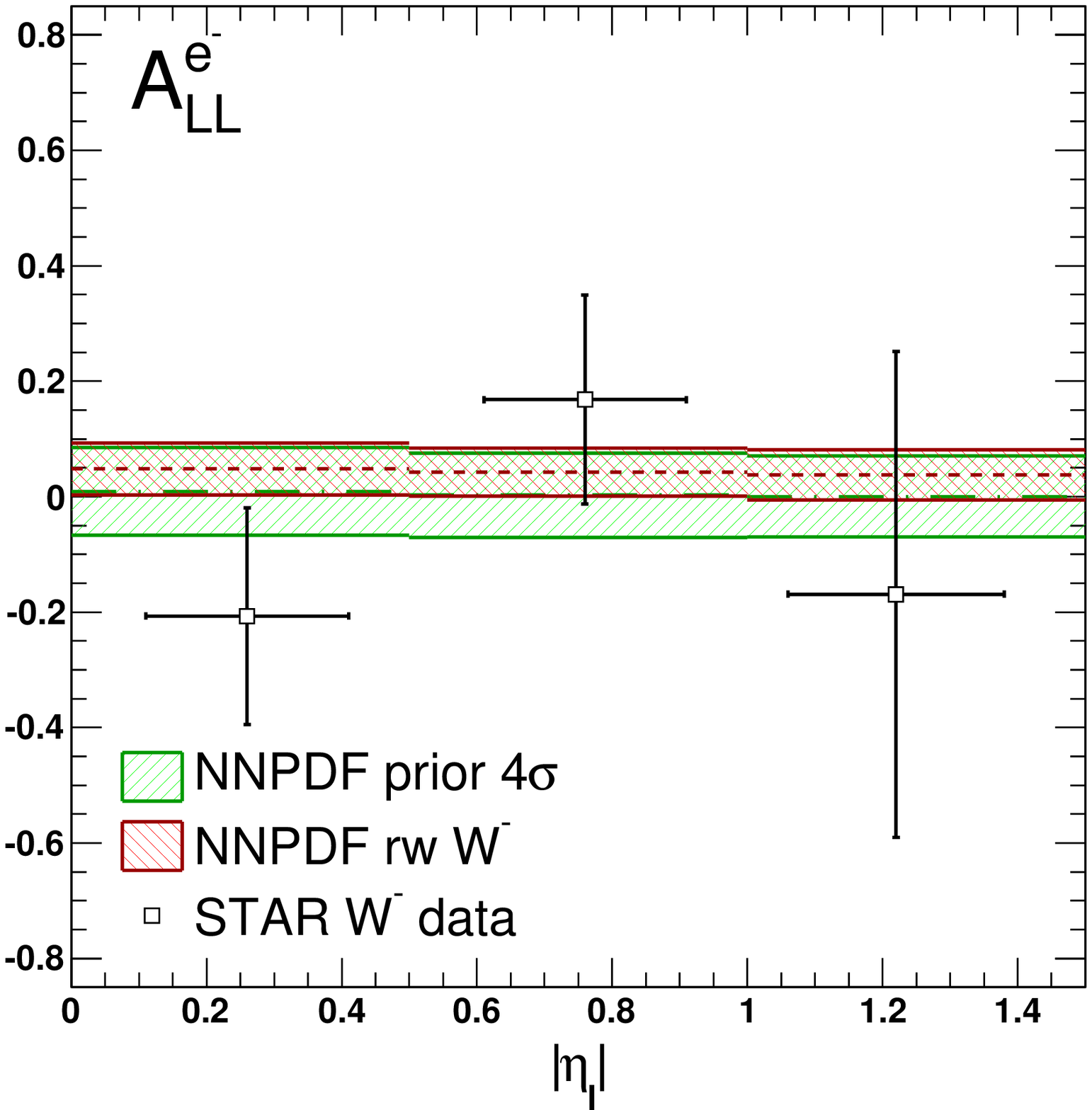}\\
\caption{\small Predictions for the longitudinal positron (left plots) 
and electron 
(right plots) single-spin asymmetries $A_L^{e^+}$ and $A_L^{e^-}$ (top
row) and double-spin asymmetries $A_{LL}^{e^+}$ 
and $A_{LL}^{e^-}$ (bottom row)
before and after reweighting, compared to the STAR data from the 2012 run~\cite{Adamczyk:2014xyw}.
Results obtained with the  $4\sigma$ prior 
are shown. The uncertainties shown are statistical only.}
\label{fig:STAR_afterrw}
\end{center}
\end{figure}

We have included the STAR single- and double-asymmetry data by reweighting.
The asymmetries have been computed using
the {\tt CHE} code~\cite{deFlorian:2010aa},
 suitably modified to handle NNPDF parton sets via the
{\tt LHAPDF} interface.
The values of $\chi^2$ per data-point before and after reweighting
(and the number of data points) are collected in
Tab.~\ref{tab:chi2STAR_after}, while the measures of the reweighting
process (defined as for Tab~\ref{tab:est2-ALL} of Sect.~\ref{sec:gluon}) 
are given in Tab.~\ref{tab:rwSTAR}. In each case, results are given
both for separate and combined STAR data sets, and for all four
priors constructed in Sect.~\ref{sec:prior}.

Inspection of Tabs.~\ref{tab:chi2STAR_after}-\ref{tab:rwSTAR} allows us
to draw the following conclusions:
\begin{itemize}
\item Independence of the choice of prior is achieved between the
  $3\sigma$  and $4\sigma$  priors, both in terms of fit quality
  (Tab.~\ref{tab:chi2STAR_after}) and of reweighting parameters
  (Tab.~\ref{tab:rwSTAR}). We will thus discuss henceforth results
  obtained using the
  $4\sigma$ prior.
\item The effective number of replicas after reweighting is always
  significantly lower than the size of the prior sample for the
  single-spin asymmetry, but quite close to it for the double-spin
  asymmetry, thus suggesting that the former data have a significant 
impact, while the latter does not. This is consistent with 
the fact that statistical uncertainties are smaller for single-spin 
asymmetries.
\item The effective number of replicas, however, even for the
  double-asymmetry data remains rather larger than that found in
  Sect.~\ref{subsec:STARjet} after reweighting with the STAR jet data:
  hence, as the impact of the reweighting is more moderate, the number
  of replicas after reweighting remains adequate for accurate
  phenomenology.  
\item The modal value of the  $\mathcal{P}(\alpha)$ distribution are
  generally close to one, though a bit higher for $A_{L}^{W^-}$
  ($\langle\alpha\rangle\sim1.2$), suggesting a mild underestimation
  of uncertainties, while the effective number of
  replicas after reweighting with this data is smaller than that found when
  reweighting with the  $A_{L}^{W^+}$ data, which should have a very
  similar impact.
\item After reweighting, 
the $\chi^2$ per data-point is below one (in fact, a little more
  than one $\sigma$ below one), thus showing  perfect agreement of the
  reweighted prediction with the data; interestingly the $\chi^2$
  before reweighting for the $A_{L}^{W^-}$ set was much greater than
  one, thus showing that the prior (based on DSSV, but with
  inflated uncertainties) does not agree well with these data.
\end{itemize}

\begin{table}[!t]
  \centering
  \small
  \begin{tabular}{ll|c|cccc|cccc}
   \toprule
    Experiment & Set & $N_{\mathrm{dat}}$ 
    & \multicolumn{4}{c|}{$\chi^2/N_{\mathrm{dat}}$}
    & \multicolumn{4}{c}{$\chi^2_{\mathrm{rw}}/N_{\mathrm{dat}}$}\\
   \midrule
   & & & $1\sigma$ & $2\sigma$ & $3\sigma$ & $4\sigma$ 
       & $1\sigma$ & $2\sigma$ & $3\sigma$ & $4\sigma$\\
     STAR-$A_{L}$      & & 12 & 1.38 & 1.44 &  1.39 & 1.33 
                              & 1.08 & 0.88 &  0.74 & 0.74\\
   & STAR-$A_L^{W^+}$    &  6 & 0.75 & 0.75 &  0.86 & 0.90
                              & 0.75 & 0.75 &  0.68 & 0.70\\
   & STAR-$A_L^{W^-}$    &  6 & 1.92 & 2.03 &  1.82 & 1.67
                              & 1.32 & 1.08 &  0.83 & 0.82\\
   \midrule
     STAR-$A_{LL}$     & &  6 & 0.82 & 0.81 &  0.78 & 0.78 
                              & 0.82 & 0.80 &  0.76 & 0.76\\
   & STAR-$A_{LL}^{W^+}$ &  3 & 0.92 & 0.88 &  0.81 & 0.80
                              & 0.90 & 0.85 &  0.77 & 0.76\\
   & STAR-$A_{LL}^{W^-}$ &  3 & 0.73 & 0.74 &  0.75 & 0.76
                              & 0.73 & 0.74 &  0.75 & 0.76\\   
   \midrule
                       & & 18 & 1.19 & 1.20 &  1.15 & 1.15 
                              & 1.00 & 0.87 &  0.78 & 0.77\\
   \bottomrule
   \end{tabular}
\caption{\small The value of the 
$\chi^2$ per data point $\chi^2/N_{\mathrm{dat}}$
($\chi^2_{\mathrm{rw}}/N_{\mathrm{dat}}$) before (after) reweighting, 
for the different priors. 
The number of data points
$N_{\mathrm{dat}}$ in each of the two STAR data sets is also given.}
\label{tab:chi2STAR_after}
\end{table}
\begin{table}[!t]
  \centering
  \small
  \begin{tabular}{ll|c|cccc|cccc}
   \toprule
    Experiment & Set & $N_{\mathrm{dat}}$ 
    & \multicolumn{4}{c|}{$N_{\mathrm{eff}}$}
    & \multicolumn{4}{c}{$\langle \alpha \rangle$}\\
   \midrule
   & & & $1\sigma$ & $2\sigma$ & $3\sigma$ & $4\sigma$ 
       & $1\sigma$ & $2\sigma$ & $3\sigma$ & $4\sigma$\\
     STAR-$A_{L}$          & & 12 & 697 & 475 &  441 & 497
                                  & 1.27 & 1.25 &  1.10 & 1.10 \\
     & STAR-$A_L^{W^+}$      &  6 & 905 & 852 &  723 & 709
                                  & 0.92 & 0.95 &  0.98 & 0.98\\
     & STAR-$A_L^{W^-}$      &  6 & 529 & 444 &  492 & 516
                                  & 1.45 & 1.43 &  1.20 & 1.20\\
   \midrule
      STAR-$A_{LL}$        & &  6 & 993 & 983 &  968 & 963
                                  & 0.95 & 0.95 &  0.95 & 0.95 \\
     & STAR-$A_{LL}^{W^+}$   &  3 & 972 & 953 &  933 & 924
                                  & 0.98 & 0.97 &  0.95 & 0.97\\
     & STAR-$A_{LL}^{W^-}$   &  3 & 999 & 995 &  994 & 990
                                  & 0.85 & 0.93 &  0.95 & 0.90\\  
   \midrule
                           & & 18 & 717 & 569 &  508 & 501
                                  & 1.11 & 1.10 &  1.15 & 1.11\\
   \bottomrule
   \end{tabular}
\caption{\small The  effective number of replicas
after reweighting, $N_{\mathrm{eff}}$, 
and the modal value of the
$\mathcal{P}(\alpha)$ distribution, for the different priors.}
\label{tab:rwSTAR}
\end{table}

Predictions for 
the longitudinal positron (electron) single- and double-spin asymmetries
$A_L^{e^+}$, $A_{LL}^{e^+}$ ($A_L^{e^-}$, $A_{LL}^{e^-}$) before and
after reweighting  are shown
in Fig.~\ref{fig:STAR_afterrw} as a function  of the lepton rapidity $\eta_l$, and compared
to the STAR data~\cite{Adamczyk:2014xyw}. 
As usual,  the numerator in Eqs.~(\ref{eq:Wasy})-(\ref{eq:ALLdef}) is computed for 
each replica in the PDF ensemble ($N_{\mathrm{rep}}=1000$), while the
denominator is evaluated only for the central unpolarized
replica. The reduction in uncertainty after reweighting is clearly
visible in all plots, and so is the good agreement between the data and the
prediction after reweighting, the improvement being especially clear in
the single-spin asymmetry plots.

\begin{figure}[!t]
\centering
\epsfig{width=0.4\textwidth,figure=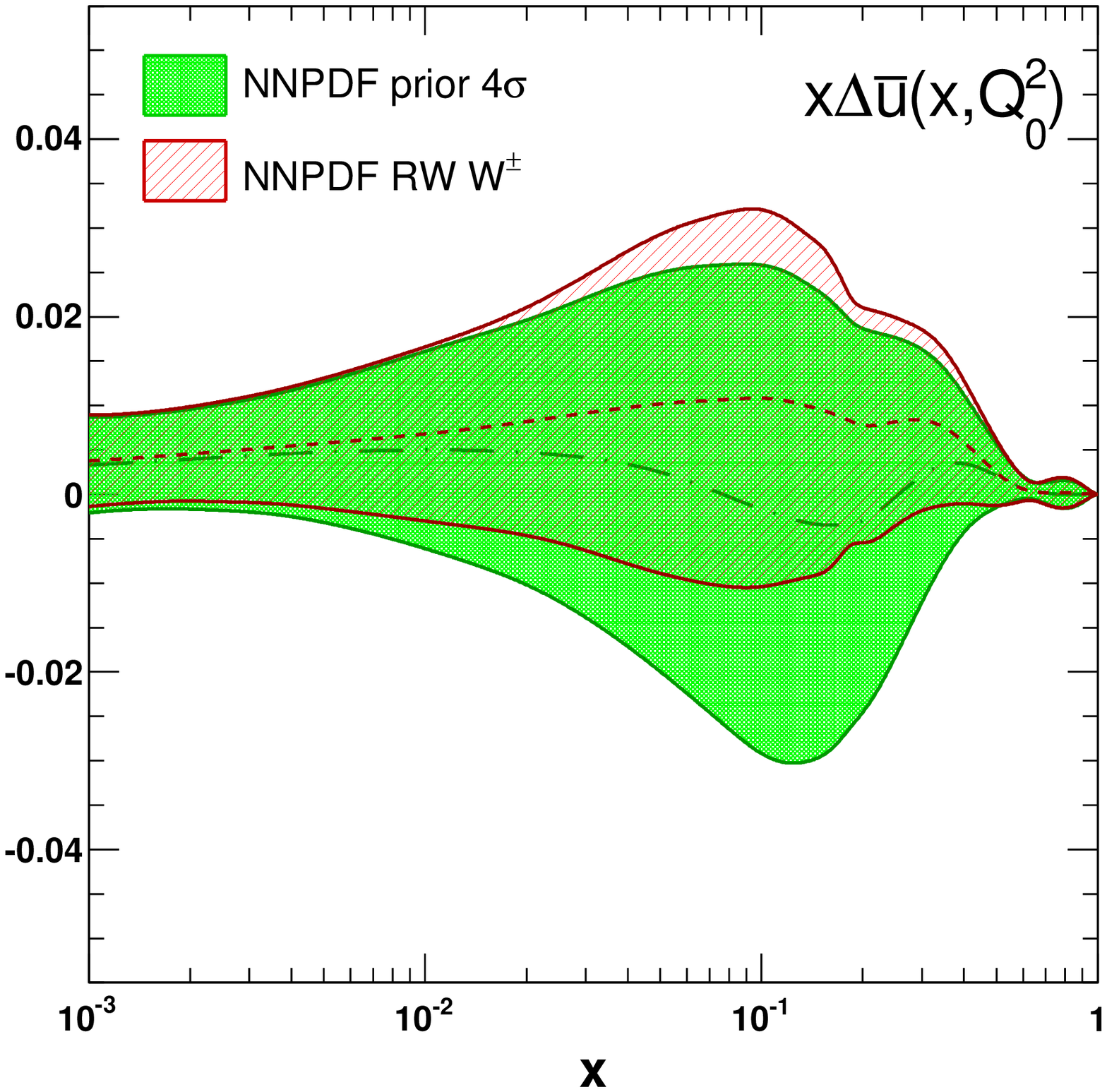}
\epsfig{width=0.4\textwidth,figure=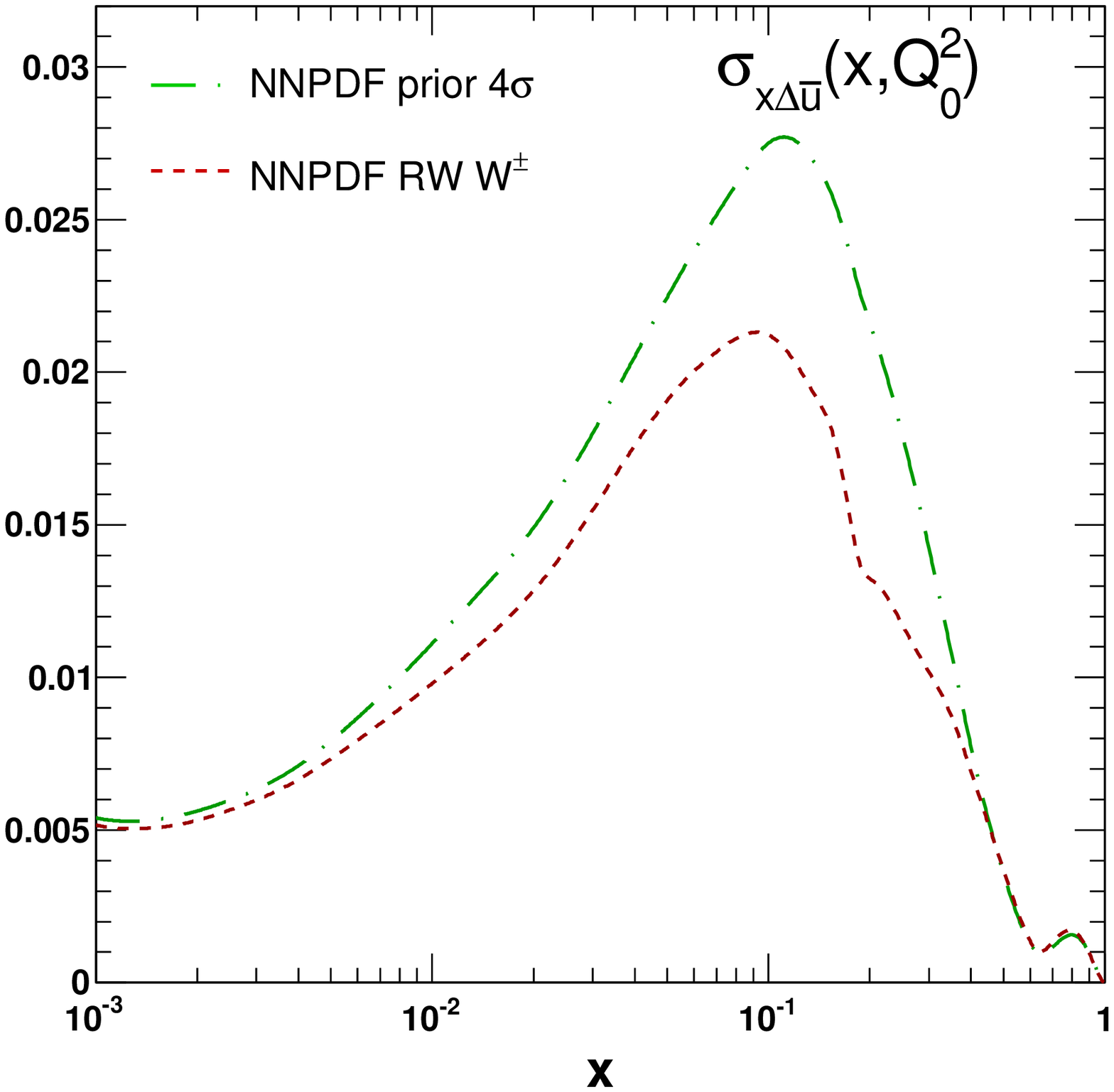}\\
\epsfig{width=0.4\textwidth,figure=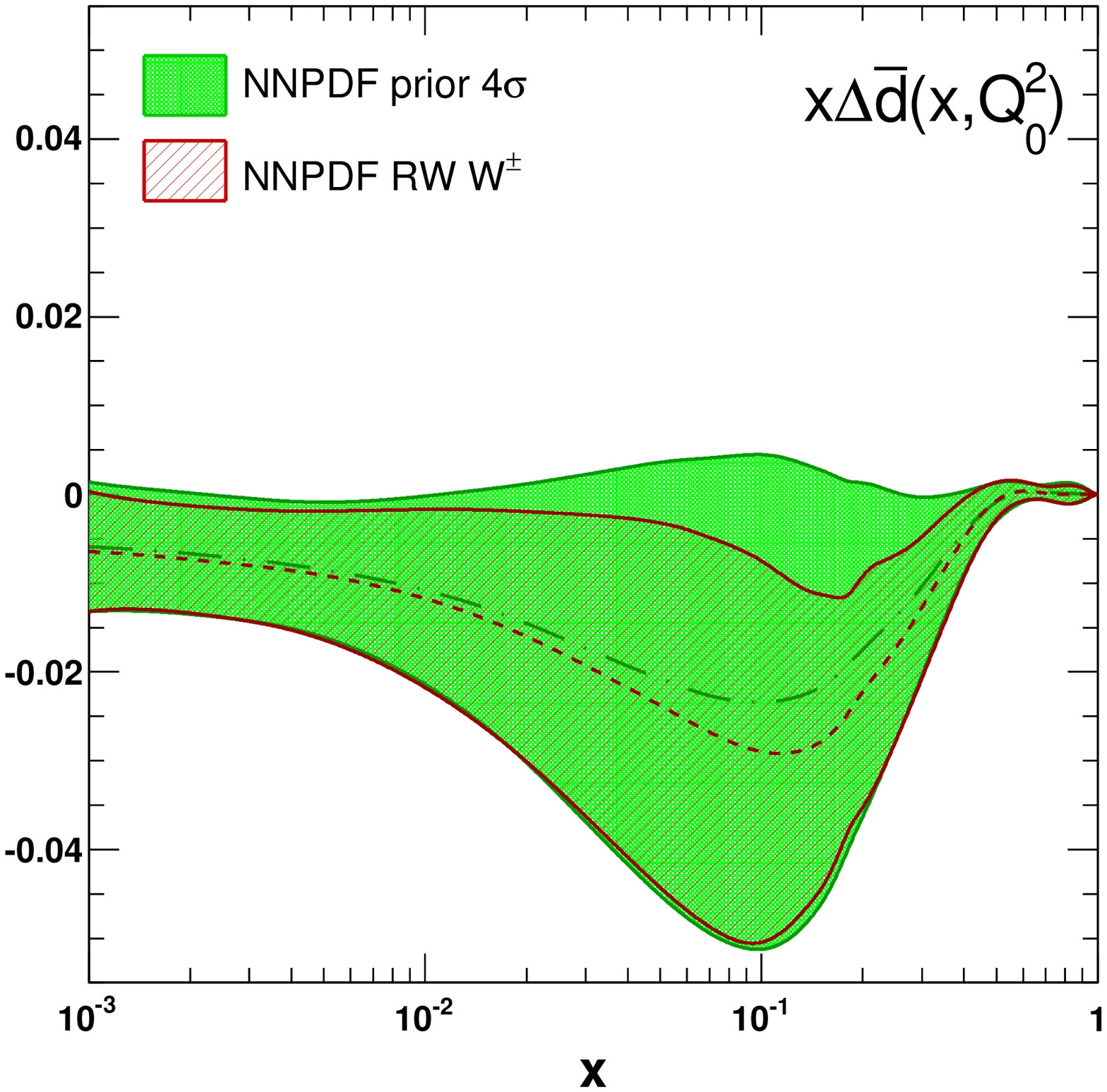}
\epsfig{width=0.4\textwidth,figure=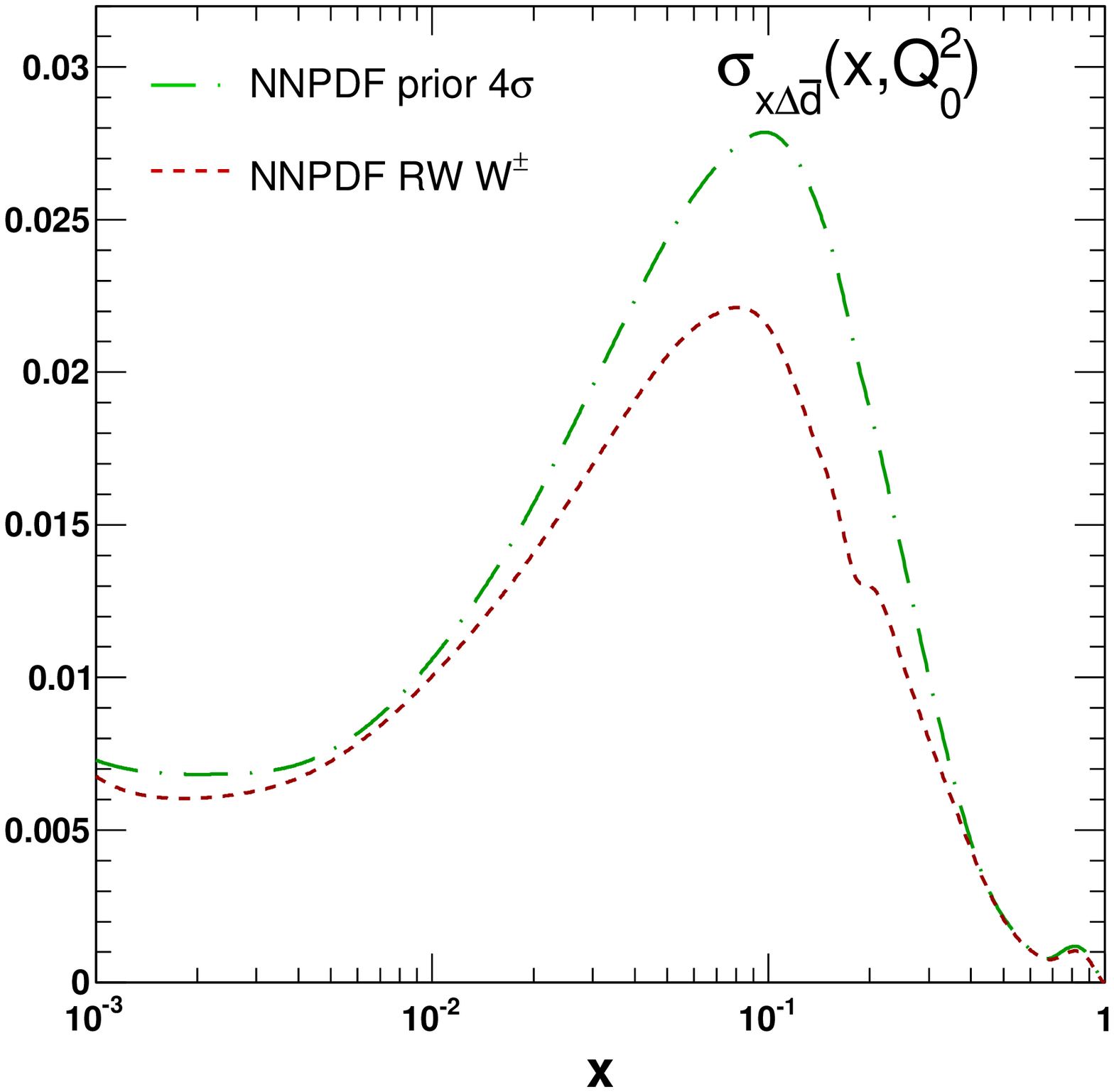}\\
\caption{\small Comparison between the polarized antiquark
sea densities $x\Delta\bar{u}$ (upper plots) and $x\Delta\bar{d}$ (lower plots)
before and after reweighting the $4\sigma$ 
ensemble with complete STAR $W^{\pm}$ data set, at $Q_0^2=1$ GeV$^2$.
The absolute PDF uncertainty is also shown (right plots).}
\label{fig:34sigma-bench}
\end{figure}

The impact of the STAR data on the PDFs is seen in
Fig.~\ref{fig:34sigma-bench}, where we compare the up and down
antiquark polarized PDFs before and after reweighting; the
corresponding absolute uncertainties are explicitly shown in the right
plots. The change in shape in comparison to the prior (whose shape was
determined by that of the DSSV best-fit PDFs) is especially remarkable for the 
$\Delta\bar u$
distribution, and indicates that the
STAR $W$ data pull in a different direction for
$\Delta \bar{u}$ than the SIDIS data used in {\tt DSSV08}.
The reduction in uncertainty is very visible in the
region of the peak, where it amounts to almost 20\%.

\section{The \texttt{NNPDFpol1.1} polarized parton set}
\label{sec:pheno}

In this section, we combine the different pieces of information
obtained in Sects.~\ref{sec:gluon} and~\ref{sec:flavour} and produce a
global polarized PDF set based on the NNPDF methodology,
{\tt NNPDFpol1.1}, which is the main result of this paper.
This set is constructed by simultaneous reweighting of the prior
PDF samples with all the new data from the COMPASS, STAR and PHENIX 
experiments listed in Tab.~\ref{tab:processes} (upper part),
displayed in Fig.~\ref{fig:NNPDFpol11-kin}, and discussed in the
previous sections.

In summary, the {\tt NNPDFpol1.1} set represents the state-of-the-art in
our understanding of the proton's spin content from polarized 
observables which do not entail parton-to-hadron fragmentation.
Further constraints will be provided by a variety of   
semi-inclusive measurements,
which in turn will require the development of a set of parton fragmentation
functions using the NNPDF methodology.
In the long term, the final word on the spin content of the proton will
require
brand new facilities such as an Electron-Ion Collider, which could bring
polarized PDF determinations to a similar level of accuracy as 
their unpolarized counterparts.

Based on our results, we will also reassess the status of
polarized quark and gluon first moments, and, as an example of
application, we will compute
the longitudinal double-spin asymmetry
for single-inclusive particle production in proton-proton collisions,
and  compare results to recent RHIC data.

\subsection{Simultaneous reweighting of RHIC and COMPASS data}
\label{sec:simult}

The {\tt NNPDFpol1.1} parton set is obtained by
performing a global reweighting of the prior polarized 
PDF ensembles (constructed as described in Sect.~\ref{sec:prior})
 with all the relevant data
from the COMPASS, STAR and PHENIX experiments.
The reweighting parameters and the
$\chi^2$ per data point before and after reweighting are
collected in
Tab.~\ref{tab:global1}, to be compared to 
Tabs.~\ref{tab:est1-ALL}-\ref{tab:est2-ALL} of Sect.~\ref{sec:gluon} and
Tabs.~\ref{tab:chi2STAR_after}-\ref{tab:rwSTAR} of
Sect.~\ref{sec:flavour}, where the results of reweighting with
individual data sets were shown.

\begin{table}[!t]
 \centering
 \small
 \begin{tabular}{l|cccc}
   \toprule
 & $1\sigma$ & $2\sigma$ & $3\sigma$ &
   $4\sigma$ \\
\midrule
   $\chi^2/N_{\mathrm{dat}}$ & $1.30$ & $1.31$ & $1.30$ & $1.30$\\ 
   $\chi_{\mathrm{rw}}^2/N_{\mathrm{dat}}$&   $1.09$ & $1.07$ & $1.05$ & $1.05$ \\
\midrule
   $N_{\mathrm{eff}}$ & $181$ & $164$ & $142$ & $135$ \\ 
  $\langle\alpha\rangle$                           & $1.22$ & $1.21$ & $1.23$& $1.22$\\

   \bottomrule
 \end{tabular}
\caption{\small The value of the $\chi^2$ per data point
$\chi^2/N_{\mathrm{dat}}$ ($\chi^2_{\mathrm{rw}}/N_{\mathrm{dat}}$) 
before (after)  reweighting, the effective number of replicas  
after reweighting $N_{\mathrm{eff}}$,  and the modal value
of the $\mathcal{P}(\alpha)$ distribution. All results refer to
reweighting with all the $N_{\mathrm{dat}}=110$ data points corresponding
to the data sets shown in Tab.~\ref{tab:processes} (upper
part),     with different choices for the prior.}
\label{tab:global1}
\end{table}
Inspection of Tab.~\ref{tab:global1} allows us to draw the following
conclusions:
\begin{itemize}
\item Independence of the prior is achieved already with the $3\sigma$
  prior, possibly even between the $2\sigma$ and the $3\sigma$
  prior. The fact that independence of the prior is achieved somewhat
  earlier when performing the global reweighting than when reweighting
  with the most constraining data set (see
  Tabs.~\ref{tab:chi2STAR_after}-\ref{tab:rwSTAR}) is to be expected,
  as a consequence of the fact that 
  the global data set carries more information. 
\item The effective number of replicas after reweighting is by almost
  one order of magnitude smaller than the size of the starting
  data set, thereby showing that the data have a significant impact on
  the fit.
\item The effective number of replicas after reweighting remains
  nevertheless larger than the conventional threshold $N_{\rm
    rep}=100$ which is necessary to ensure reliability of the results
  obtained using the reweighted set.
\item The modal value of the $\mathcal{P}(\alpha)$ distribution is
  just above one, suggesting good compatibility of the new global
  data set with the previous fit, even though, as discussed in
  Sect.~\ref{sec:flavour}, Tab.~\ref{tab:rwSTAR}, there is some
  evidence for moderate uncertainty underestimation  
 for the $A_{L}^{W^-}$ STAR data, which may explain the value slightly
 above one.
\item The  $\chi^2$ per data point improves substantially upon
  reweighting, and is of order one after reweighting. Note however
  that these values should be treated with care because information on
  correlated systematics is not available for all experiments, and
  thus for all experiments for which statistical and systematic
  uncertainties must be added in quadrature the $\chi^2$ per data
  point might be expected to be less than one.
\end{itemize}

In the sequel, we will choose the PDF set obtained by reweighting the
$4\sigma$ prior as our default {\tt NNPDFpol1.1} PDF set:  this guarantees
maximal independence of the choice of prior, without significant loss
of accuracy, given that the values of
$N_{\rm eff}$ for the $3\sigma$ and $4\sigma$ prior are roughly equal.
As a further check of independence of the
prior, in Fig.~\ref{fig:distances-34} we display 
the distance $d(x,Q^2)$, as defined in Appendix~A of 
Ref.~\cite{Ball:2010de}, between PDFs obtained with the $3\sigma$ 
and $4\sigma$ priors, at $Q^2=10$ GeV$^2$. 
This statistical estimator is expected to take the value $d\sim 1$ 
when two samples of $N_{\mathrm{rep}}$ 
replicas are extracted from the same underlying probability distribution, 
while it is $d=\sqrt{N_{\mathrm{rep}}}$ when the two samples are extracted
from two distributions which differ on average by one standard deviation. 
As we see in  Fig.~\ref{fig:distances-34},
the two ensembles are indeed statistically 
equivalent.
\begin{figure}[!t]
\begin{center}
\epsfig{width=0.80\textwidth,figure=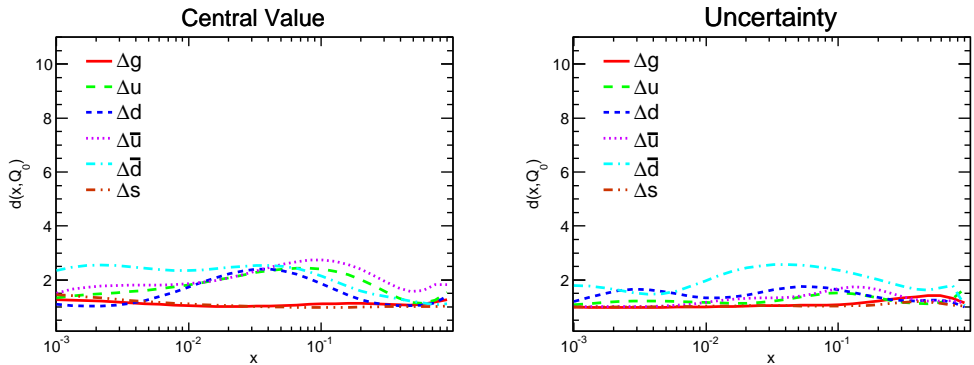}
\end{center}
\caption{\small Distances between parton sets obtained via global 
reweighting of $3\sigma$ and $4\sigma$ prior PDF ensembles at $Q^2=10$ GeV$^2$.}
\label{fig:distances-34}
\end{figure}

In order to construct the \texttt{NNPDFpol1.1} set, made of
$N_{\mathrm{rep}}=100$ replicas, the results
from the global reweighting discussed above are unweighted
following the procedure in Ref.~\cite{Ball:2011gg}.
We show in  Tab.~\ref{tab:chi2oldnew}
the $\chi^2$ of each of the 
experiments included in the \texttt{NNPDFpol1.1} analysis  computed
with the \texttt{NNPDFpol1.1} PDF set; for experiments which were
already included in  \texttt{NNPDFpol1.0} the value obtained using
that set is also shown.
It is clear that the description of inclusive DIS data 
in \texttt{NNPDFpol1.1} is as good as in the original set, and that
the description of each individual set used for reweighting is as good
in the combined reweighted set as it was when reweighting with each
individual set (as shown in Tab.~\ref{tab:est1-ALL} and~\ref{tab:chi2STAR_after} of Sects.~\ref{sec:gluon}
and~\ref{sec:flavour} respectively). The comparison of predictions
obtained using {\tt
  NNPDFpol1.1} 
to the data is  accordingly very similar to that shown in
Figs.~\ref{fig:COMPASS_afterrw},~\ref{fig:ALL-plot},~\ref{fig:STAR_afterrw},
and is therefore not shown.

\begin{table}[!t]
\centering
\small
\begin{tabular}{lcc}
\toprule
Experiment & \texttt{NNPDFpol1.0} & \texttt{NNPDFpol1.1}\\
\midrule
EMC       & 0.44 & 0.43\\
SMC       & 0.93 & 0.90\\
SMClowx   & 0.97 & 0.97\\
E143      & 0.64 & 0.67\\
E154      & 0.40 & 0.45\\
E155      & 0.89 & 0.85\\
COMPASS-D & 0.65 & 0.70\\
COMPASS-P & 1.31 & 1.38\\
HERMES97  & 0.34 & 0.34\\
HERMES    & 0.79 & 0.82\\
\midrule
COMPASS (OC) & --- & 1.22\\
STAR (jets)  & --- & 1.05\\
PHENIX (jets)& --- & 0.24\\
STAR-$A_L$   & --- & 0.72\\
STAR-$A_{LL}$& --- & 0.75\\
\bottomrule
\end{tabular}
\caption{\small The $\chi^2$ per data point of all the experiments included in the 
\texttt{NNPDFpol1.1} analysis computed with the corresponding PDF set;
for experiments which were already included in \texttt{NNPDFpol1.0} 
the value obtained using that set is also shown.}
\label{tab:chi2oldnew}
\end{table}

\begin{figure}[!t]
\begin{center}
\epsfig{width=0.4\textwidth,figure=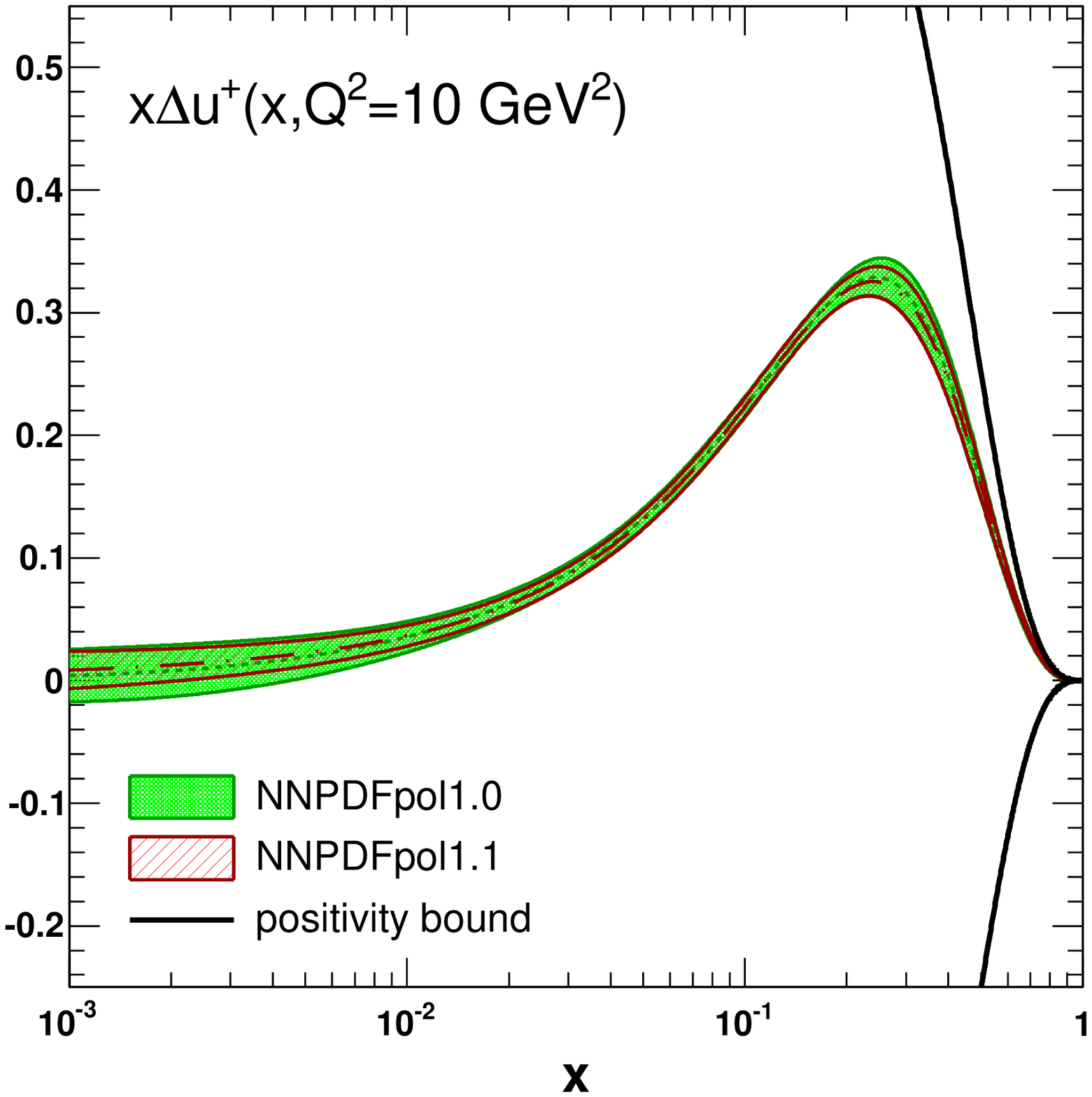}
\epsfig{width=0.4\textwidth,figure=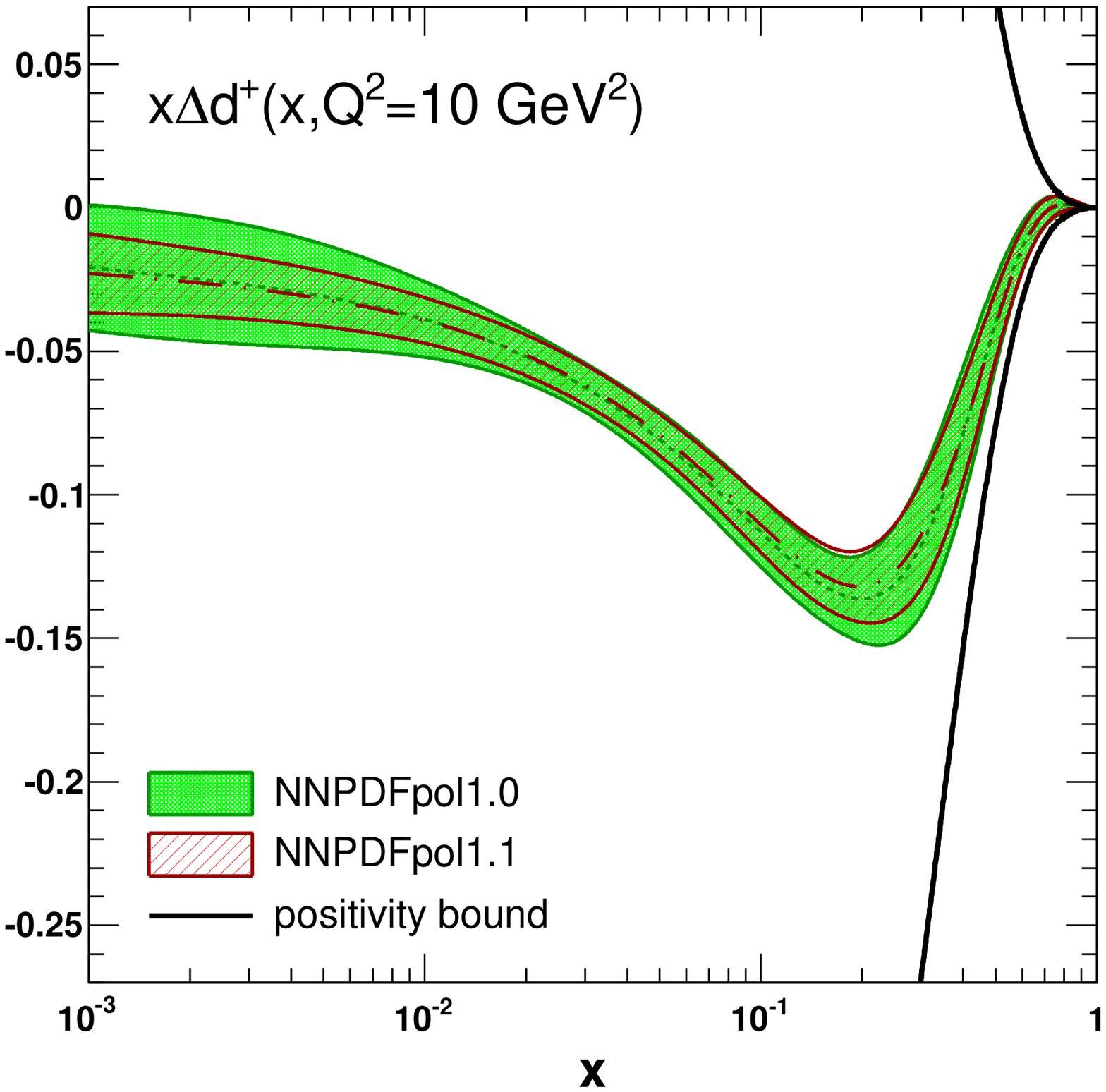}\\
\epsfig{width=0.4\textwidth,figure=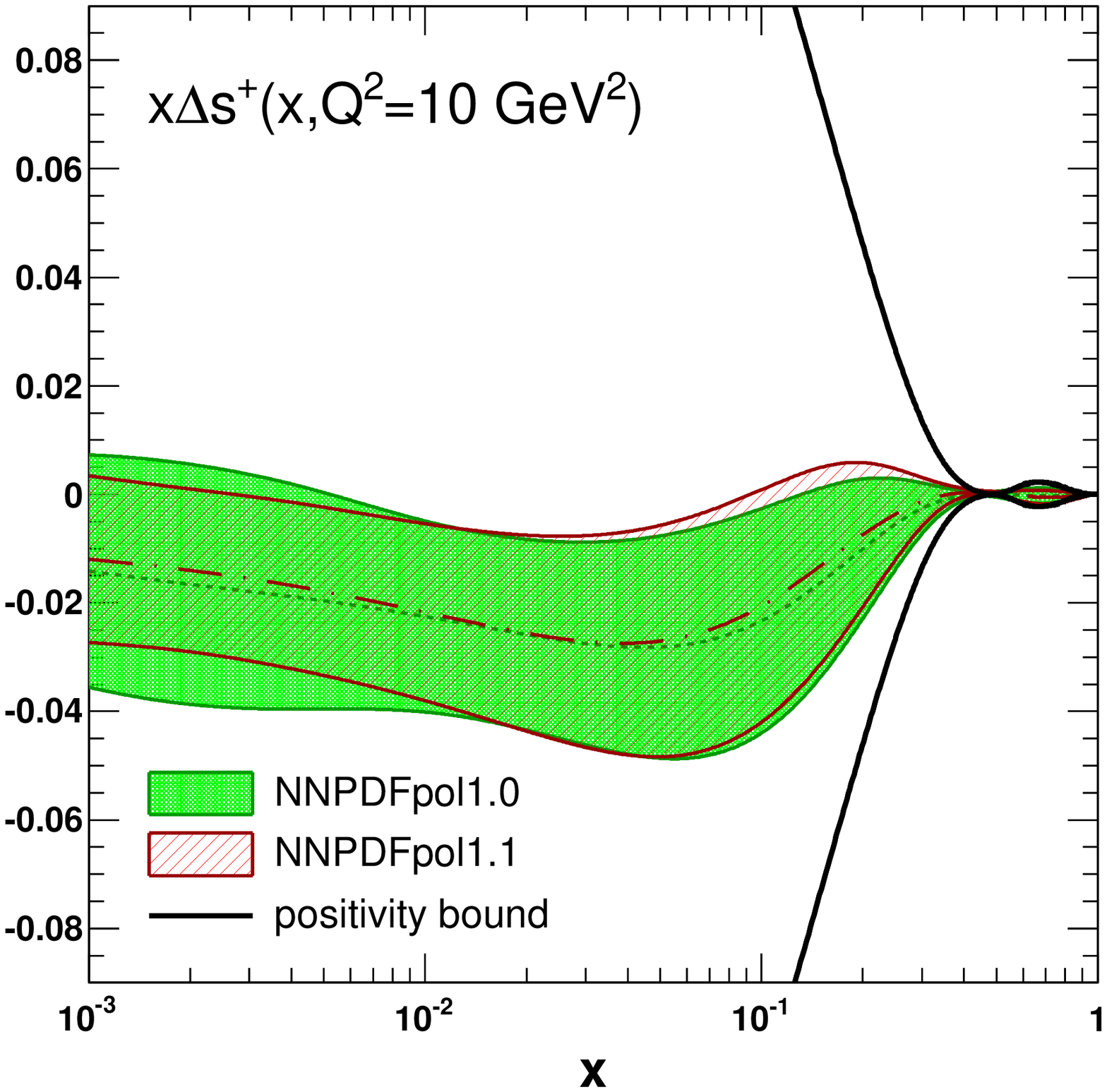}
\epsfig{width=0.4\textwidth,figure=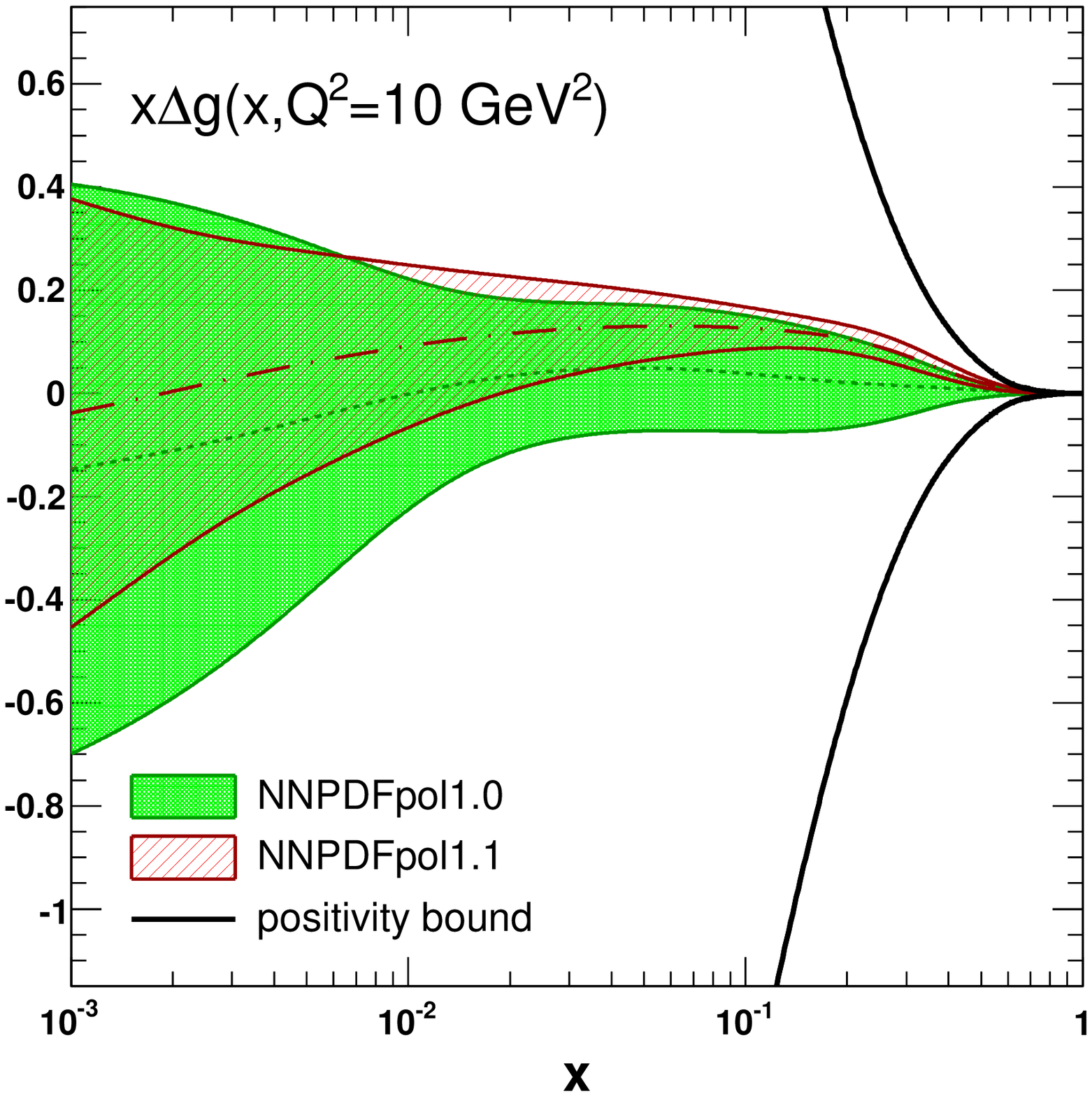}\\
\end{center}
\caption{\small Comparison between \texttt{NNPDFpol1.0} and \texttt{NNPDFpol1.1}
parton sets at $Q^2=10$ GeV$^2$ for the $x \Delta u^+$, 
$x \Delta d^+$, $x \Delta s^+$ and $x \Delta g$ polarized PDFs.
The positivity bound from the unpolarized {\tt NNPDF2.3} set is also shown.
}
\label{fig:pdfs11}
\end{figure}

\begin{figure}[!t]
\begin{center}
\epsfig{width=0.80\textwidth,figure=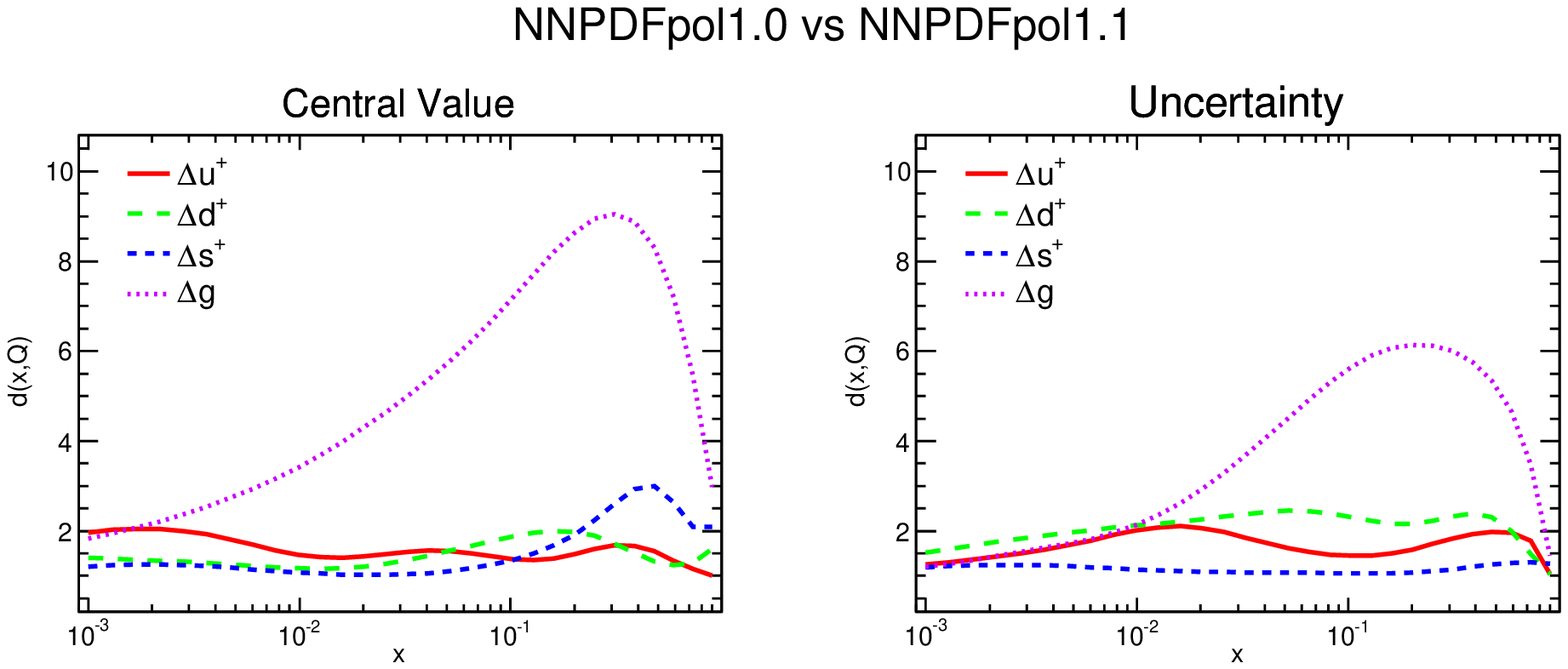}
\end{center}
\caption{\small Distances $d(x,Q^2)$ 
between \texttt{NNPDFpol1.0} and \texttt{NNPDFpol1.1} at $Q^2=10$ GeV$^2$.}
\label{fig:distances-oldnew}
\end{figure}

We now turn to the  \texttt{NNPDFpol1.1} PDFs which we first compare
in Fig.~\ref{fig:pdfs11}
with those of the previous set, \texttt{NNPDFpol1.0}. Because in
\texttt{NNPDFpol1.0} only the PDF combinations $\Delta u^+$, $\Delta d^+$,
$\Delta s^+$, and the gluon PDF, $\Delta g$, are determined, only these
can be compared directly. The comparison is shown at $Q^2=10$ GeV$^2$,
where the   positivity bound coming from the unpolarized 
{\tt NNPDF2.3} set is also displayed.
In order to quantitatively assess the impact of the new data for
these PDF combinations, in Fig.~\ref{fig:distances-oldnew}
we also plot the distance $d(x,Q)$ between \texttt{NNPDFpol1.0} and
\texttt{NNPDFpol1.1}  at $Q^2=10$ GeV$^2$.

The main differences are found for $\Delta g$
in the region 
$0.02\lesssim x\lesssim 0.5$ covered by the 
STAR inclusive jet production data.
In this region, the gluon from 
\texttt{NNPDFpol1.1} parton set is positive 
at the one-sigma level, and it has
a much reduced uncertainty in comparison to its \texttt{NNPDFpol1.0}
counterpart. At small $x$, outside the kinematical coverage of
the STAR jet data,
the \texttt{NNPDFpol1.0} and \texttt{NNPDFpol1.1} are again 
statistically equivalent.
On the other hand, the total quark-antiquark combinations 
$\Delta u^+$, $\Delta d^+$ and $\Delta s^+$
are only moderately affected by the new data, 
and they are still mostly constrained
by the polarized inclusive DIS data, with the RHIC data only leading
to a minor reduction in uncertainty mostly for the  $\Delta d^+$ 
distribution and in the small $x$ region. 

\begin{figure}[!p]
\begin{center}
\epsfig{width=0.4\textwidth,figure=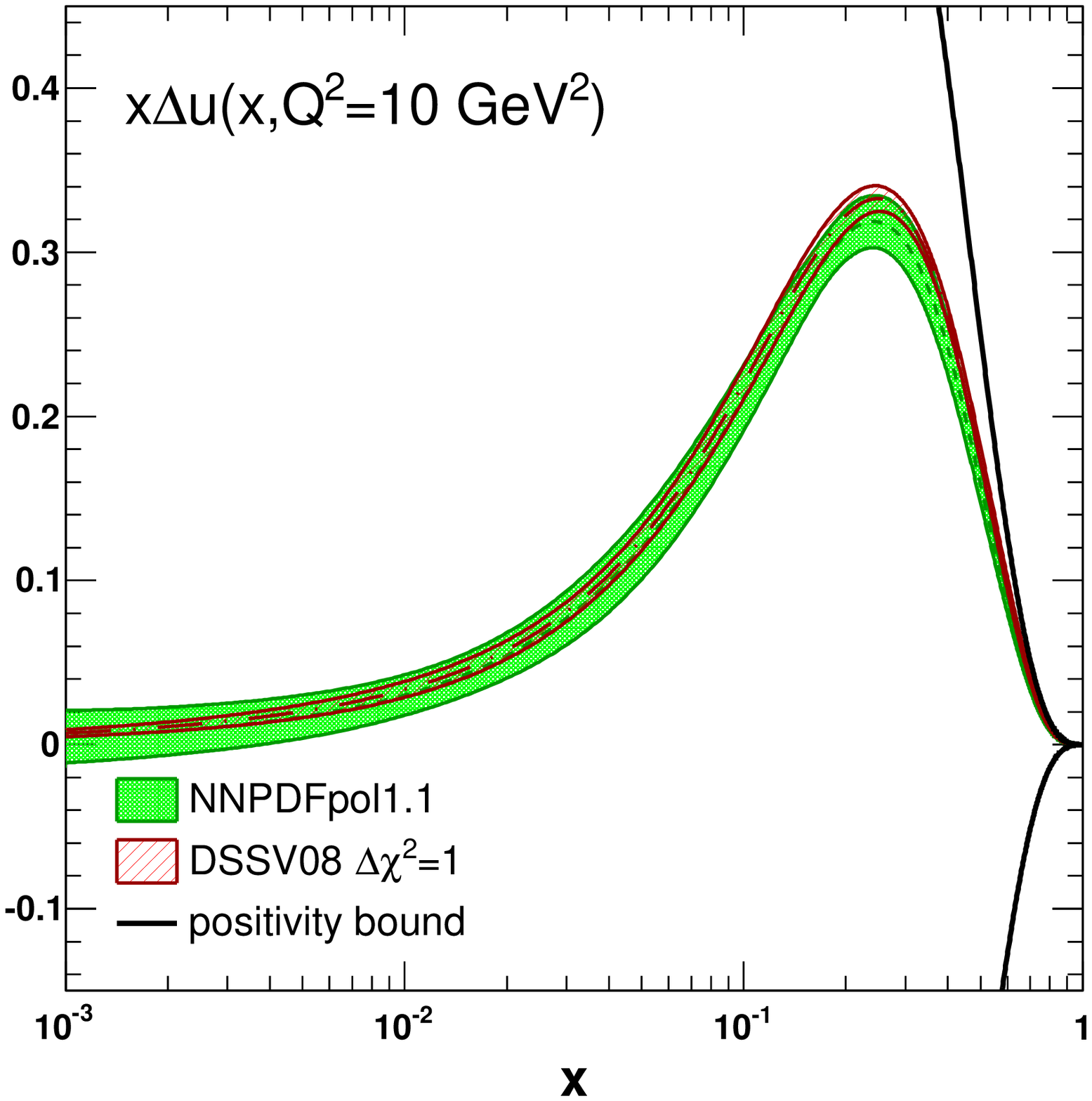}
\epsfig{width=0.4\textwidth,figure=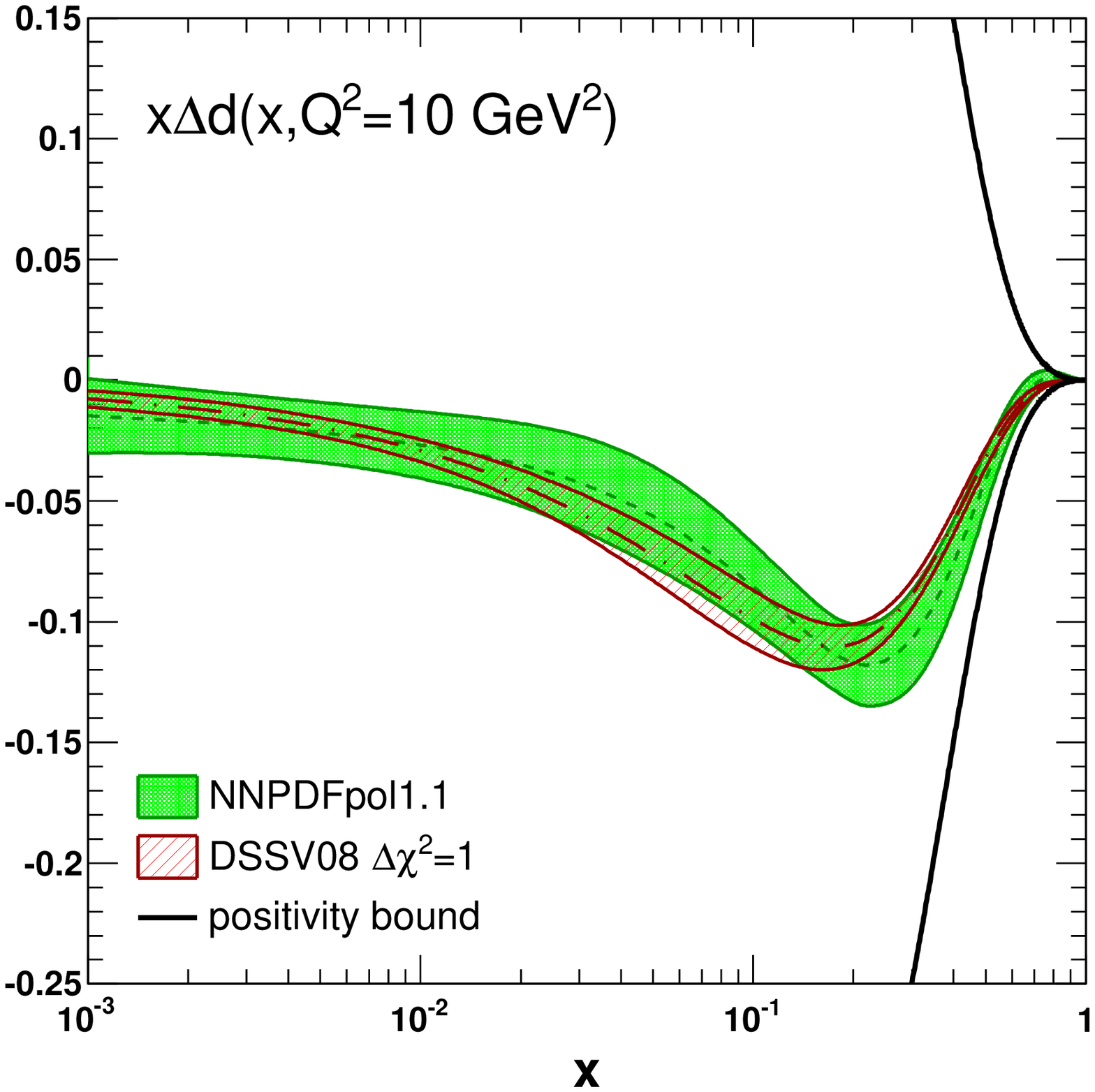}\\
\epsfig{width=0.4\textwidth,figure=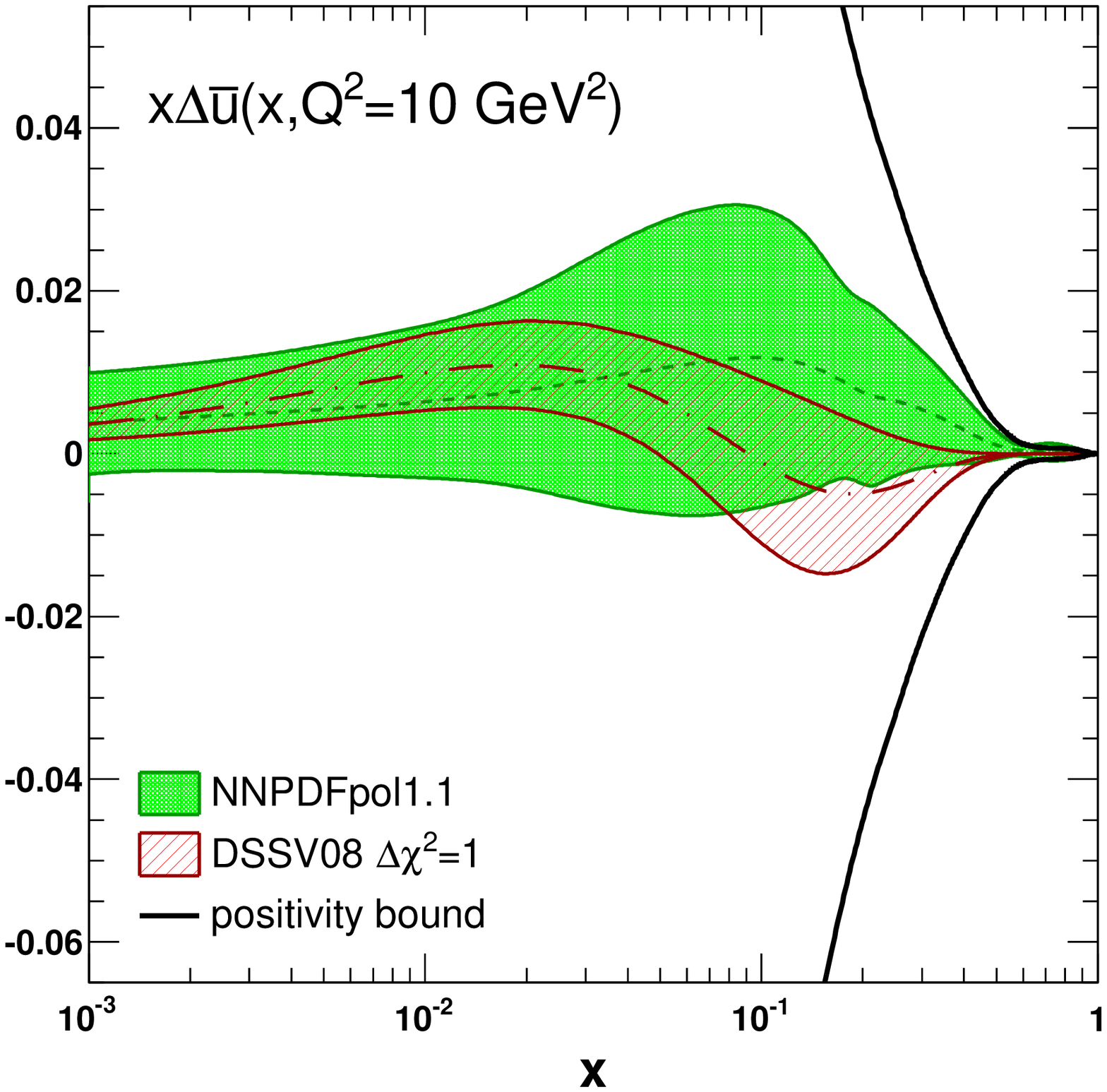}
\epsfig{width=0.4\textwidth,figure=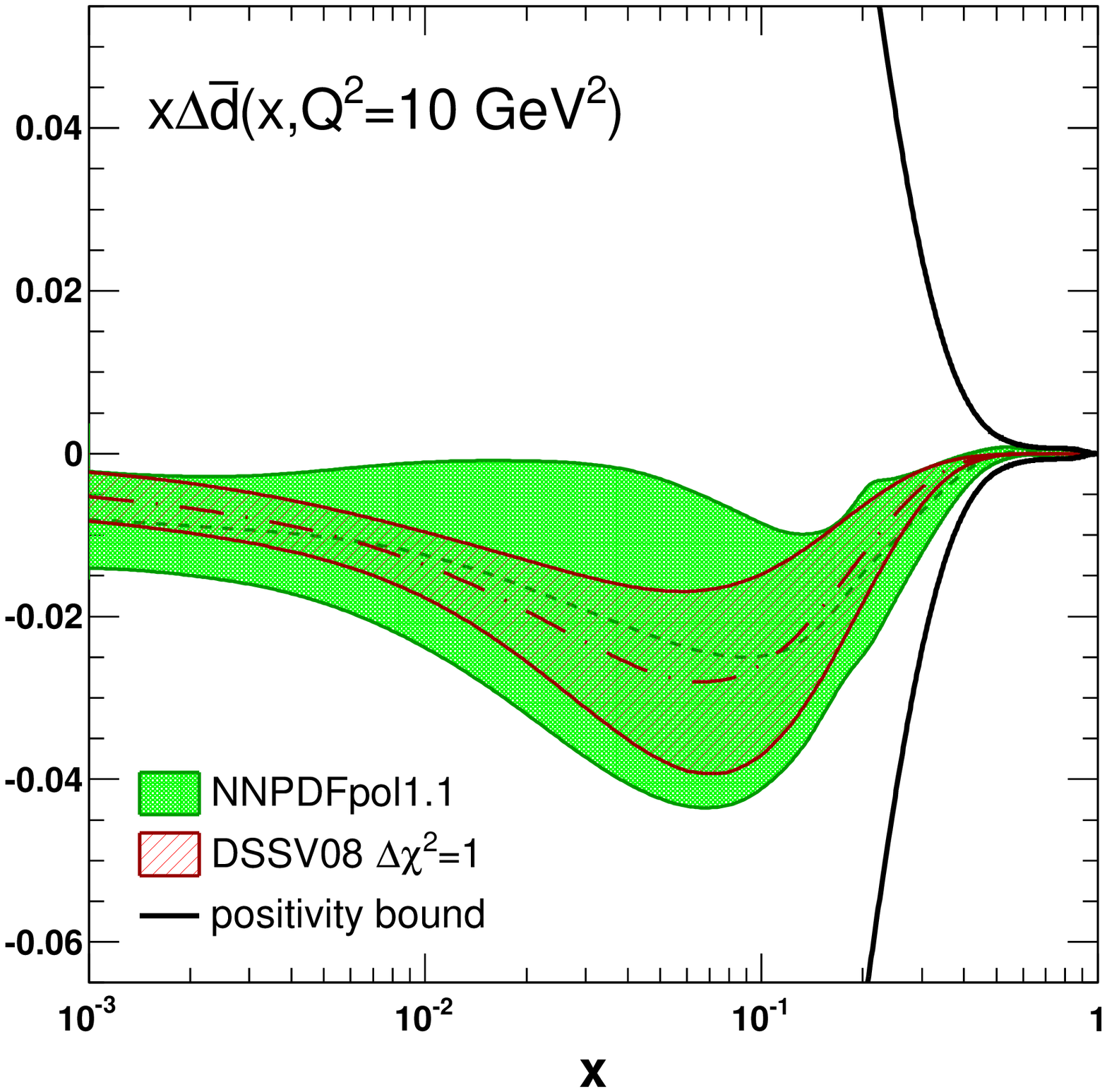}\\
\epsfig{width=0.4\textwidth,figure=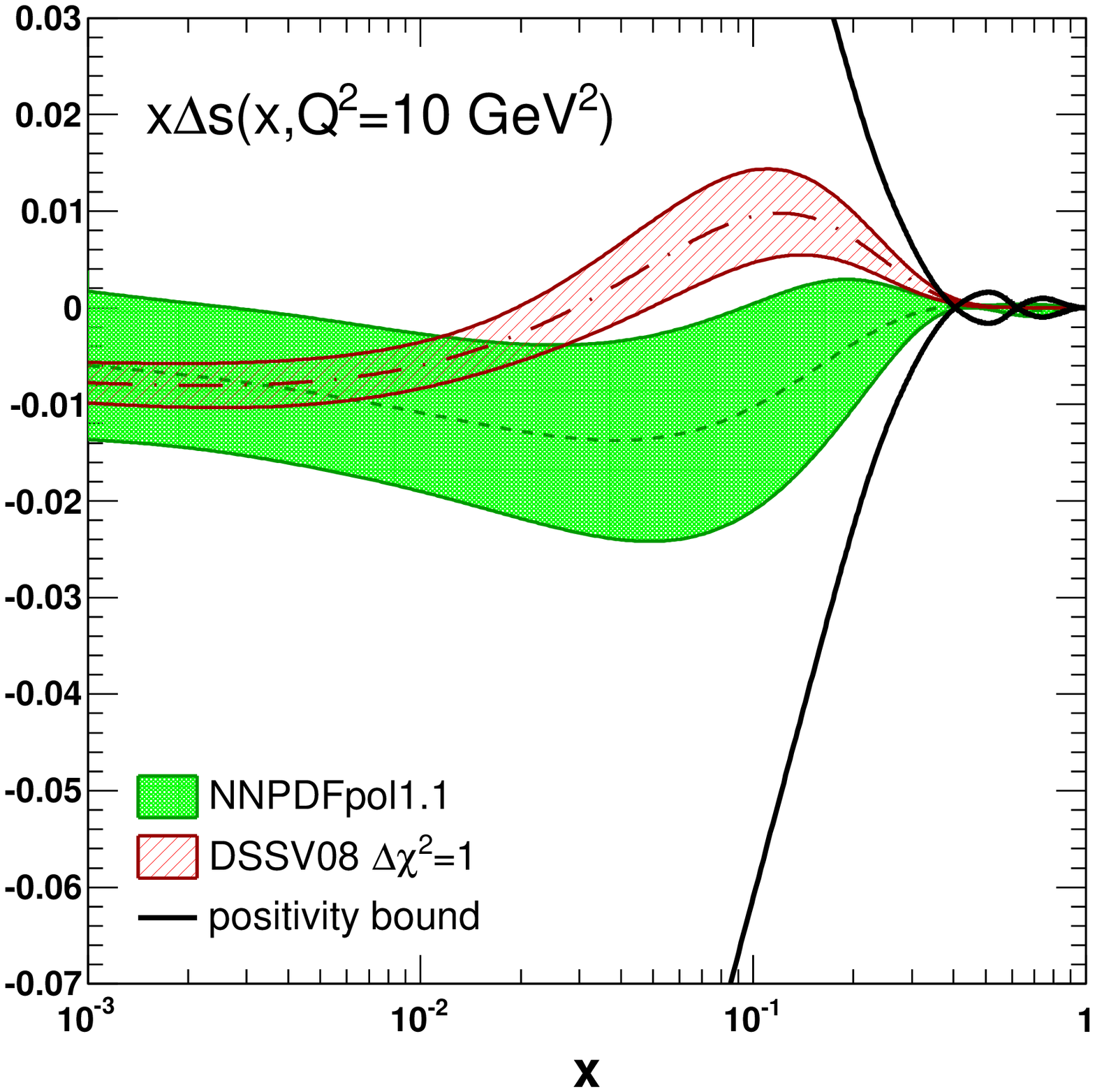}
\epsfig{width=0.4\textwidth,figure=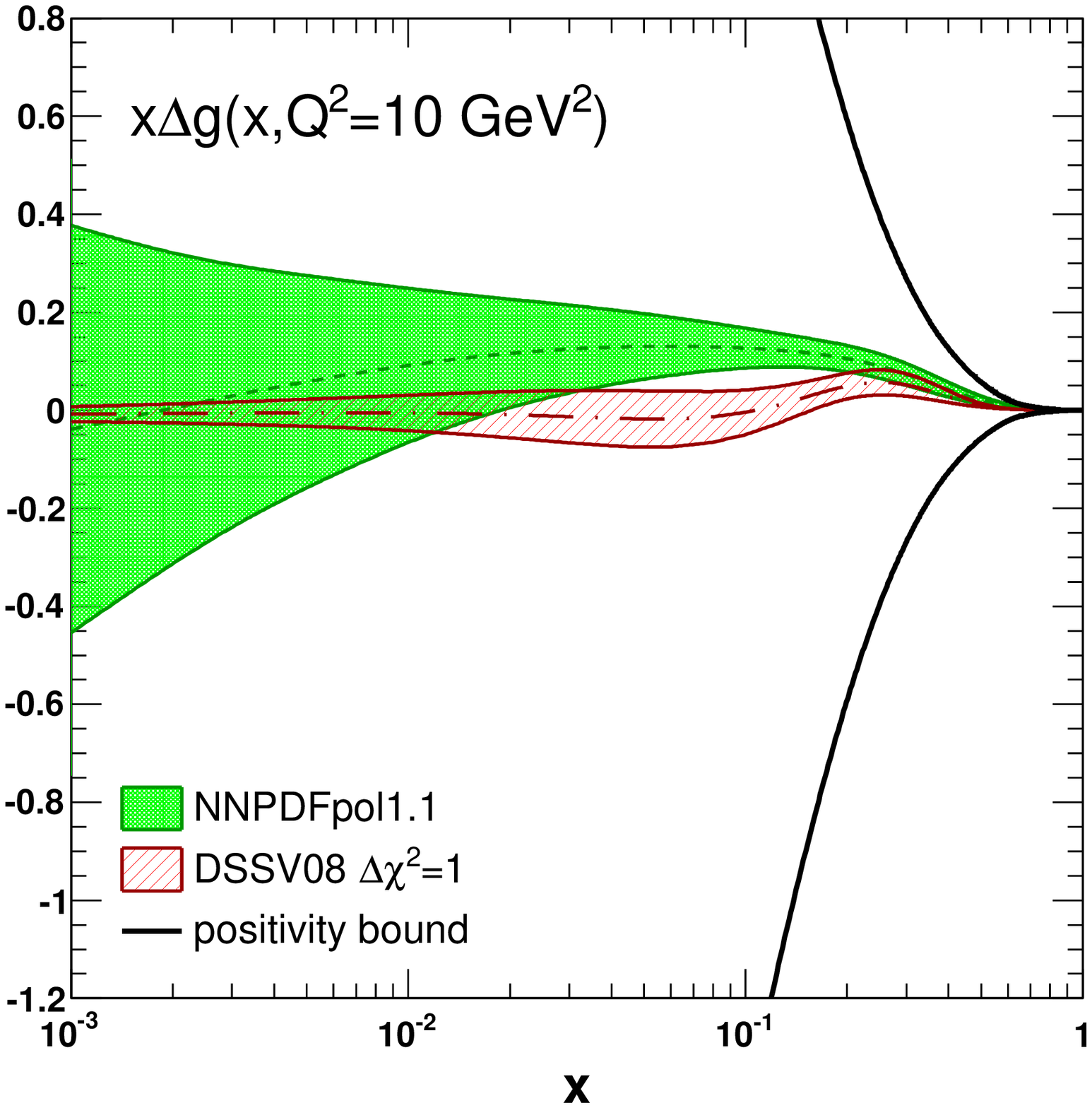}
\caption{\small The \texttt{NNPDFpol1.1} parton set
compared to \texttt{DSSV08}~\cite{deFlorian:2009vb} at $Q^2=10$ GeV$^2$.
}
\label{fig:xpdfs}
\end{center}
\end{figure}

We now turn to the individual \texttt{NNPDFpol1.1} PDFs, which are
compared in Fig.~\ref{fig:xpdfs} to those 
from the global {\tt DSSV08} fit~\cite{deFlorian:2009vb}
at $Q^2=10$ GeV$^2$;
PDF uncertainties are nominal one-sigma error bands for \texttt{NNPDFpol1.1}, 
and Hessian uncertainties ($\Delta\chi^2=1$) for {\tt  DSSV08}.
The main conclusions from the comparison in Fig.~\ref{fig:xpdfs} 
are the following:
\begin{itemize}

\item The
$\Delta u$ and $\Delta d$ PDFs of the \texttt{NNPDFpol1.1} and {\tt
  DSSV08} are qualitatively similar, though for {\tt NNPDFpol1.1} 
uncertainties are typically larger. Note however that the default
$\Delta\chi^2=1$ adopted by \texttt{DSSV08} may lead to
uncertainty  underestimation: it is well known that in Hessian global
unpolarized fits a tolerance $\Delta\chi^2=T$ with $T>1$ is needed for
faithful uncertainty estimation, as also recognized in
Ref.~\cite{deFlorian:2009vb} (see also the discussion of first moments
in Sect.~\ref{sec:pspin} below).

 \item The  {\tt NNPDFpol1.1} polarized gluon PDF is consistent
at the one-sigma level with its \texttt{DSSV08} 
counterpart in the large-$x$ region $x\gtrsim 0.2$, where they have
similar uncertainties. However, for $x<0.2$, $\Delta g$ has a node in the
\texttt{DSSV08} determination, while it is clearly positive 
in \texttt{NNPDFpol1.1}.
This result is mostly driven by the recent precise 
inclusive jet production data from STAR
(STAR 1j-09A and STAR 1j-09B of Sect.~\ref{sec:gluon}), 
which were not available at the time of the original \texttt{DSSV08} 
analysis shown in  Fig.~\ref{fig:xpdfs}  (only STAR 1j-05 and STAR 1j-06
were included).
Recent updates of the \texttt{DSSV08} fit including also STAR 1j-09A
and STAR 1j-09B data sets~\cite{Aschenauer:2013woa,deFlorian:2014yva} suggest
a positive $\Delta g$ consistent with the {\tt NNPDFpol1.1} result.

 \item As already noticed in Sect.~\ref{sec:flavour}, inclusion of
   the
$W^\pm$ data (not included in the \texttt{DSSV08} fit) visibly affects
   the shape of the
$\Delta\bar{u}$ distribution, especially above $x\sim 3\cdot10^{-2}$,
   which thus differs from that in the  \texttt{DSSV08} set. This
   might be a signal of tension between $W^\pm$ and semi-inclusive DIS
   data, possibly due to limited knowledge of fragmentation functions.  
A similar change in the shape of $\Delta\bar{u}$, making its peak less
negative,  was
also found in a preliminary global fit including STAR data in the DSSV
framework~\cite{Stratmann:2013DIS}. 

 \item Since $W$ boson production data in the kinematic regime probed by STAR
are not sensitive to strangeness, the discrepancy 
between NNPDF and DSSV determinations of $\Delta s$, already found in 
Ref.~\cite{Ball:2013lla}, is still present. As discussed in our
previous work~\cite{Ball:2013lla}, in the NNPDF analysis 
the polarized strange PDF is obtained from inclusive DIS data through its 
$Q^2$ evolution and assumptions about flavor symmetry of the proton sea
enforced by experimentally measured baryon octet decay constants.
On the other hand, the DSSV determination of polarized PDFs also 
includes semi-inclusive data with identified kaons in final states, 
which are directly sensitive to strangeness, but subject  to 
the uncertainty in the kaon fragmentation functions 
(which is difficult to quantify).
\end{itemize}

\begin{figure}[!t]
\begin{center}
\epsfig{width=0.4\textwidth,figure=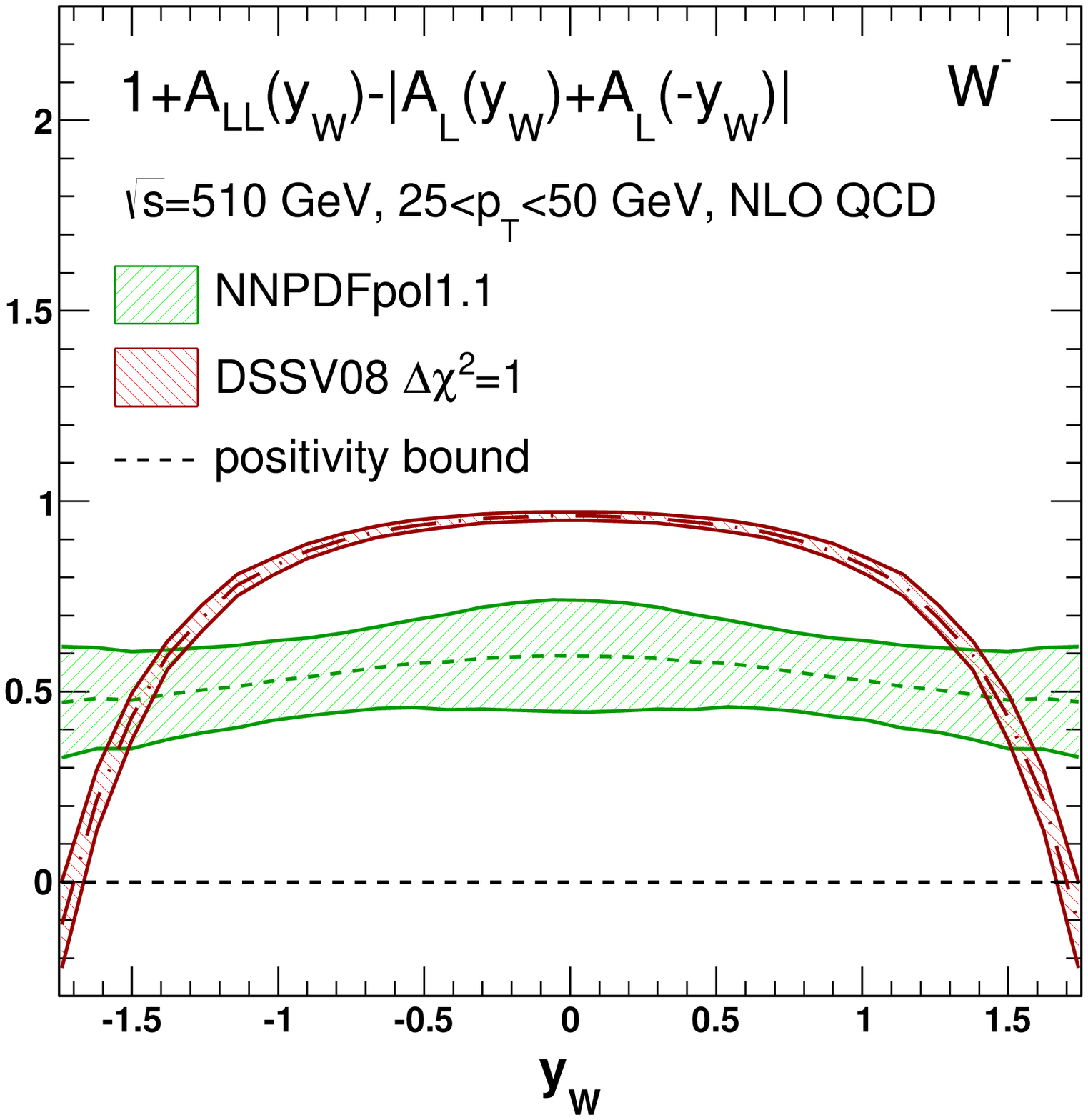}
\epsfig{width=0.4\textwidth,figure=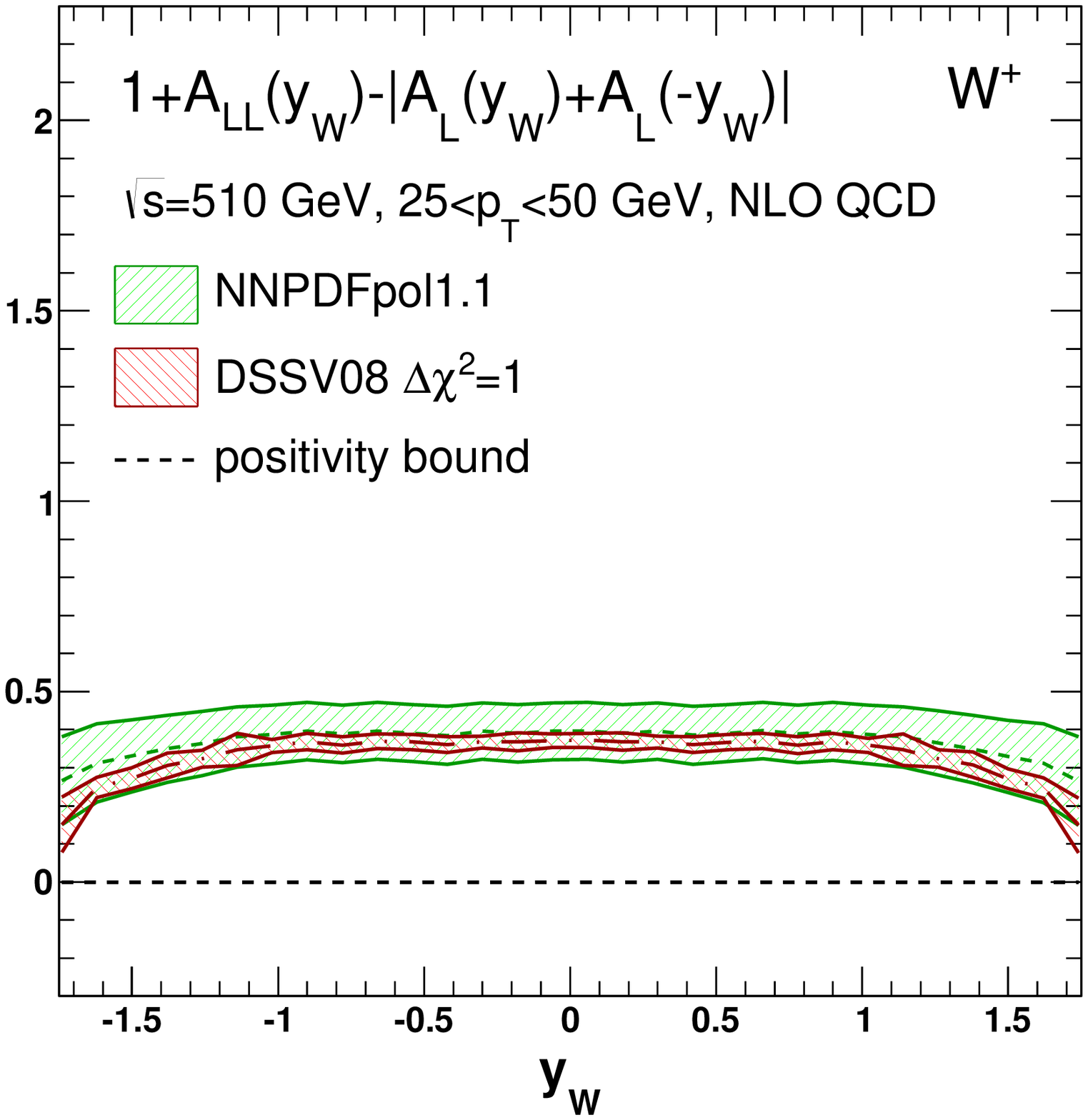}\\
\epsfig{width=0.4\textwidth,figure=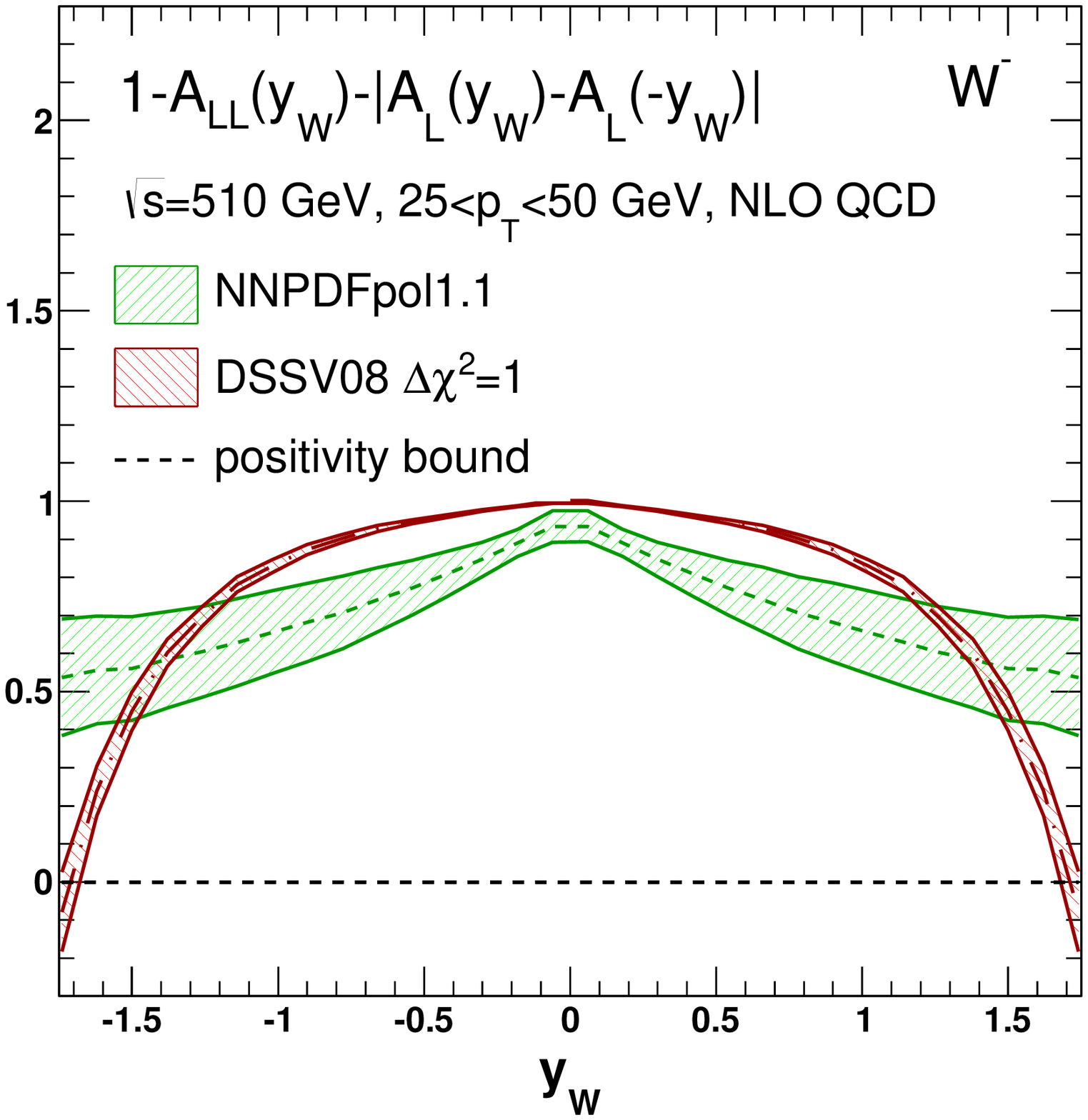}
\epsfig{width=0.4\textwidth,figure=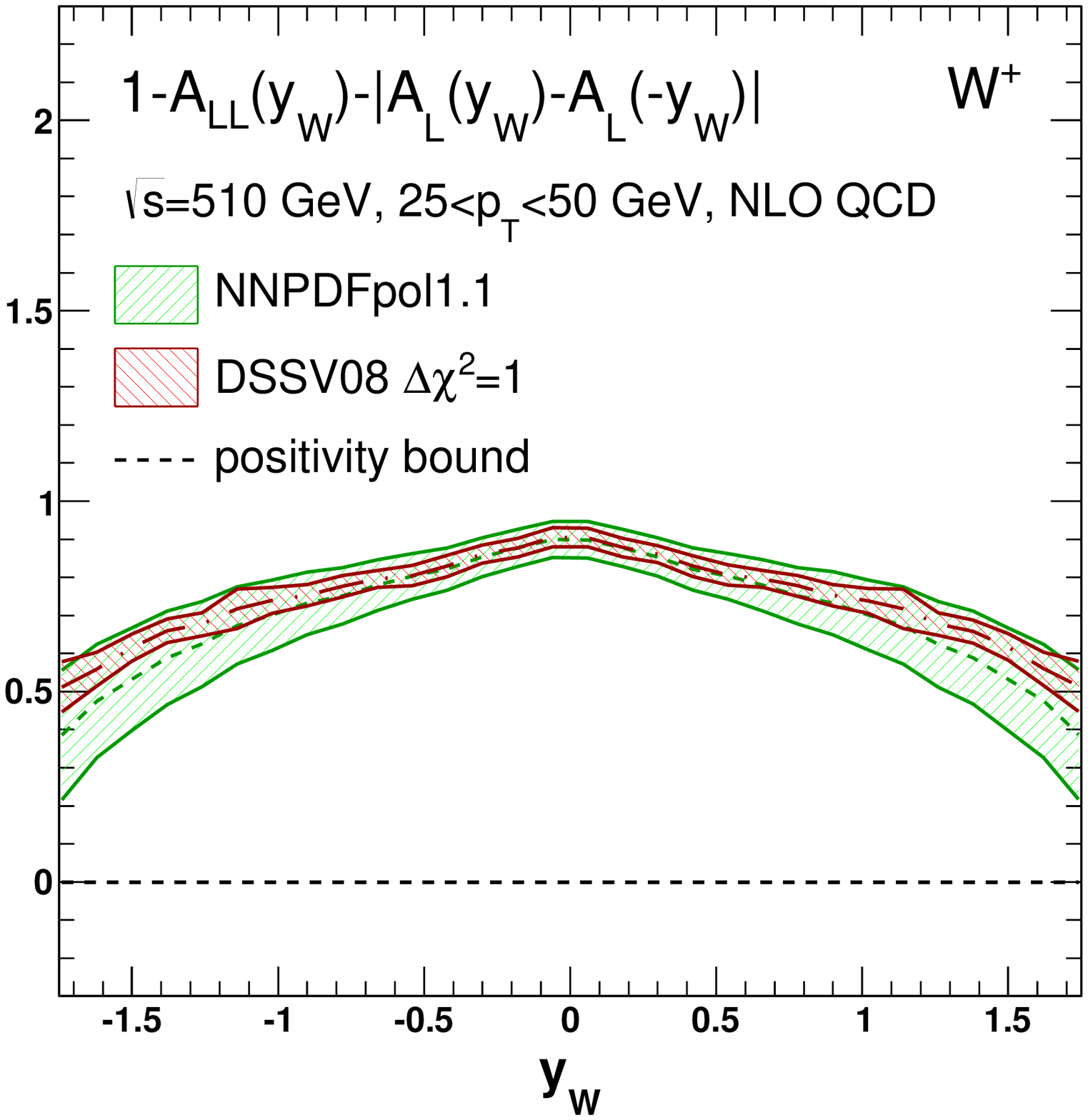}\\
\caption{\small The combinations 
$1\pm A_{LL}(y_W)-\left|A_L(y_W)\pm A_L(-y_W)\right|$
of longitudinal single- and double- spin asymmetries, $A_{LL}$ and $A_{L}$,
for $W$ production in polarized proton-proton collisions
plotted as a function of the $W$ boson rapidity $y_W$. Results are shown 
for both the \texttt{NNPDFpol1.1} and \texttt{DSSV08} parton sets. 
The positivity bounds (15) are satisfied whenever the curves are positive.}
\label{fig:asybound}
\end{center}
\end{figure}

In view of the comparison with other PDF sets, it is interesting to
examine positivity bounds which, as discussed in
Sect.~\ref{sec:prior}, our PDFs satisfy by construction for each
individual flavor, as is also clear from
Fig.~\ref{fig:xpdfs}.  Note that positivity 
also implies that single- and double-spin asymmetries
must satisfy the bounds
Eq.~(\ref{eq:posbound})~\cite{Kang:2011qz}.
Results for the combinations of asymmetries which are bounded to be
non-negative are shown in Fig.~\ref{fig:asybound},
determined using \texttt{NNPDFpol1.1} and \texttt{DSSV08} PDFs (with,
as in Sect.~\ref{sec:flavour} the  \texttt{NNPDF2.3}~\cite{Ball:2012cx} and 
\texttt{MRST02}~\cite{Martin:2002aw} respectively as corresponding
unpolarized sets), for the  RHIC center-of-mass energy
$\sqrt{s}=510$ GeV, integrated over the lepton transverse momentum
in the range $25<p_T<50$, for  $W^-$ and $W^+$ production.

Interestingly, while \texttt{NNPDFpol1.1} satisfies the bounds (as it
must to by construction) \texttt{DSSV08} PDFs appear to violate 
both the bounds for $W^-$ production when  $|y_W|\gtrsim 1.6$,
which roughly corresponds to momentum fractions $x\gtrsim 0.8$. This
ties in neatly with the observation that
while \texttt{NNPDFpol1.1} and \texttt{DSSV08} results are in good 
agreement for  $W^+$ production, they differ for $W^-$, reflecting  
the different shape of the 
$\Delta\bar{u}$ distribution, also seen in Fig.~\ref{fig:xpdfs}.

\subsection{The spin content of the proton revisited}
\label{sec:pspin}
The first moments of the polarized PDFs can be directly related to 
the fraction of the proton spin carried by individual partons.
In Ref.~\cite{Ball:2013lla} we presented a detailed analysis
of the first moments of various PDF combinations from the fit
to polarized inclusive DIS data only.
In this section we revisit this analysis with the
\texttt{NNPDFpol1.1} parton set, and quantify the impact of the RHIC
data 
on the proton
spin content.
We define the (truncated) first moments of the polarized PDFs 
$\Delta f(Q^2,x)$ in the region $[x_{\mathrm{min}},x_{\mathrm{max}}]$ 
\begin{equation}
 \langle \Delta f(Q^2)\rangle^{[x_{\mathrm{min}},x_{\mathrm{max}}]}
 \equiv
 \int_{x_{\mathrm{min}}}^{x_{\mathrm{max}}}dx\, \Delta f(x,Q^2) \ .
 \label{eq:moments}
\end{equation}
We will provide results for both full
moments, $\langle\Delta f(Q^2)\rangle^{[0,1]}$, 
and truncated moments, restricted to 
the $x$ region which roughly corresponds to the kinematic  coverage of
experimental data, 
\textit{i.e.} $\langle\Delta f(Q^2)\rangle^{[10^{-3},1]}$, see
Fig.~\ref{fig:NNPDFpol11-kin}.

We consider first polarized quark and
antiquark PDFs. In contrast to the previous \texttt{NNPDFpol1.0}
determination, we can now 
determine the first moment of individual light flavor and
antiflavors. In Tab.~\ref{tab:qqmomenta} we show
results for the $C$-even combinations
$\Delta u^+$ and $\Delta d^+$, for the light antiquarks
 $\Delta\bar{u}$ and $\Delta\bar{d}$,  
for the polarized strangeness $\Delta s$ (recall that we assume
$\Delta s=\Delta \bar s$),  and for
the singlet PDF combination 
$\Delta\Sigma=\sum_{q=u,d,s}\Delta q^+$.
The corresponding central values and one-sigma
PDF uncertainties obtained from the
$N_{\mathrm{rep}}=100$ replicas of the \texttt{NNPDFpol1.1} parton set 
at $Q^2=10$ GeV$^2$ are collected in Tab.~\ref{tab:qqmomenta}.
We compare our results to both \texttt{NNPDFpol1.0} and
\texttt{DSSV08}.
In the latter case, we quote the PDF uncertainty obtained from the
Lagrange multiplier method with $\Delta\chi^2/\chi^2=2\%$, which is
recommended in Ref.~\cite{deFlorian:2009vb} as more reliable. This
corresponds to a value of the tolerance parameter, defined above in
Sect.~\ref{sec:simult}, of
order $T\sim8$, hence to uncertainties rather larger than those shown for
DSSV PDFs in Fig.~\ref{fig:xpdfs}.  
For  {\tt DSSV08}  we also show in parenthesis the contribution  that
must be added to the truncated moment given in the table in order to
obtain the first moment: this contribution comes from extrapolation in
the unmeasured region, and it should be assigned 100\%
uncertainty~\cite{deFlorian:2009vb}.  
\begin{table}[!t]
 \centering
 \small
 \begin{tabular}{c|cc|ccc}
 \toprule
 & \multicolumn{2}{c|}{$\langle\Delta f(Q^2)\rangle^{[0,1]}$}
 & \multicolumn{3}{c}{$\langle\Delta f(Q^2)\rangle^{[10^{-3},1]}$}\\
 $\Delta f$ 
 & \texttt{NNDPFpol1.0}
 & \texttt{NNPDFpol1.1}
 & \texttt{NNDPFpol1.0}
 & \texttt{NNPDFpol1.1}
 & \texttt{DSSV08} \\
 \midrule
 $\Delta u^+$
 & $+0.77\pm 0.10$
 & $+0.79\pm 0.07$
 & $+0.76\pm 0.06$
 & $+0.76\pm 0.04$
 & $+0.793^{+0.028}_{-0.034}\,(+0.020)$\\
 $\Delta d^+$
 & $-0.46\pm 0.10$
 & $-0.47\pm 0.07$
 & $-0.41\pm 0.06$
 & $-0.41\pm 0.04$
 & $-0.416^{+0.035}_{-0.025}\,(- 0.042)$\\
 $\Delta\bar{u}$
 & ---
 & $+0.06\pm 0.06$ 
 & ---
 & $+0.04\pm 0.05$
 & $+0.028^{+0.059}_{-0.059}\,(+ 0.008)$\\
 $\Delta\bar{d}$
 & ---
 & $-0.11\pm 0.06$
 & ---
 & $-0.09\pm 0.05$
 & $-0.089^{+0.090}_{-0.080}\,(- 0.026)$\\
 $\Delta s$
 & $-0.07\pm 0.06$
 & $-0.07\pm 0.05$
 & $-0.06\pm 0.04$
 & $-0.05\pm 0.04$
 & $-0.006^{+0.028}_{-0.031}\,(- 0.051)$\\
 $\Delta\Sigma$
 & $+0.16\pm 0.30$
 & $+0.18\pm 0.21$
 & $+0.23\pm 0.15$
 & $+0.25\pm 0.10$
 & $+0.366^{+0.042}_{-0.062}\,(+ 0.124)$\\
 \bottomrule
 \end{tabular}
 \caption{\small Full and truncated first moments of the polarized quark 
distributions, Eq.~(\ref{eq:moments}), at $Q^2=10$ GeV$^2$,
for \texttt{NNPDFpol1.1}, 
 \texttt{NNPDFpol1.0} (when available) and \texttt{DSSV08}.
The uncertainties shown are one-sigma for  NNPDF and  Lagrange
multiplier with
 $\Delta\chi^2/\chi^2=2\%$
for DSSV. The  number in parenthesis for \texttt{DSSV08} is the
contribution  that should be added to the truncated moment in order to
obtain the full moment.}
 \label{tab:qqmomenta}
\end{table}

The first moments obtained
from \texttt{NNPDFpol1.1} and \texttt{NNPDFpol1.0} are
perfectly consistent
with each other, as expected based on the  agreement of the 
 PDFs seen in Fig.~\ref{fig:pdfs11}.
The new constraints on polarized quark PDFs from the
RHIC $W$ data lead to a substantial reduction, by almost a factor two,
of the
PDF uncertainties on the first moments of $\Delta u^+$ and
$\Delta d^+$.
The contribution to the uncertainty coming from the data and
extrapolation
regions are of comparable size in
  \texttt{NNPDFpol1.1}, just as in
\texttt{NNPDFpol1.0}. This means that the  uncertainty due to the
extrapolation  has  also decreased  by almost a factor two
in \texttt{NNPDFpol1.1} in comparison to \texttt{NNPDFpol1.0}, despite
the fact that the kinematic coverage of the data at small $x$ is not
significantly  extended by the hadronic data (recall
Fig.~\ref{fig:NNPDFpol11-kin}): this is because the lower uncertainty
in the data region also limits the spread of acceptable small-$x$
extrapolations.
Interestingly, all quark truncated first moments are also compatible
between \texttt{NNPDFpol1.1} and 
\texttt{DSSV08}, despite the differences in shape in the
$\Delta\bar{u}$ PDF.

We now turn to the gluon.  
Results for full and truncated moments at $Q^2=10$ GeV$^2$
are presented in Tab.~\ref{tab:gmomenta}, where, on top of
\texttt{NNPDFpol1.1}, \texttt{NNPDFpol1.0}, and {\tt DSSV08} results,
we also show the result found with a
recent update of the DSSV family, 
\texttt{DSSV++}~\cite{Aschenauer:2013woa}. The interest of this
comparison lies in the fact that \texttt{DSSV++} also includes
jet production data from RHIC, so the 
data which
constrain the gluon are essentially the same as in \texttt{NNPDFpol1.1}  
(the exception being pion production data from RHIC, included
in \texttt{DSSV++} but not in \texttt{NNPDFpol1.1}).
We show full and truncated moments in the
measured region $[10^{-3},1]$, and also truncated moments in 
the region $x \in [0.05,0.2]$, which corresponds to the
range covered by the RHIC
inclusive jet data (see
Fig.~\ref{fig:NNPDFpol11-kin}).
\begin{table}[!t]
 \centering
 \small
 \begin{tabular}{l|ccc}
 \toprule
 & $\langle \Delta g(Q^2)\rangle^{[0,1]}$
 & $\langle \Delta g(Q^2)\rangle^{[10^{-3},1]}$
 & $\langle \Delta g(Q^2)\rangle^{[0.05,0.2]}$ \\
 \midrule
 \texttt{NNPDFpol1.0}
 & $-0.95\pm 3.87$
 & $-0.06\pm 1.12$
 & $+0.05\pm 0.15$\\
 \texttt{NNPDFpol1.1}
 & $+0.03\pm 3.24$
 & $+0.49\pm 0.75$
 & $+0.17\pm 0.06$\\ 
 \texttt{DSSV08}
 & ---
 & $0.013^{+0.702}_{-0.314}\,(+ 0.097)$
 & $0.005^{+0.129}_{-0.164}$\\
 \texttt{DSSV++}
 & ---
 & ---
 & $0.10^{+0.06}_{-0.07}$\\ 
 \bottomrule
 \end{tabular}
 \caption{\small Same as Tab.~\ref{tab:qqmomenta}
but for  the gluon. Results are also shown for the truncated moment in
the range of the RHIC jet data and for \texttt{DSSV++}.
 \label{tab:gmomenta}}
\end{table}

The significant improvement in the uncertainties on the first moment
of $\Delta g$ in the data region when going from \texttt{NNPDFpol1.0} to
\texttt{NNPDFpol1.1} is apparent; clearly, the improvement is
concentrated in the region of the RHIC jet data. The uncertainty from
the extrapolation, however, does not improve significantly and it
dominates the result for the full first moment, which remains
essentially undetermined. This strongly suggests that only with 
a wider kinematic coverage, such as 
would be obtained at a polarized Electron-Ion Collider, could a
significantly more accurate determination of the polarized gluon first
moment be achieved~\cite{Aschenauer:2012ve,Ball:2013tyh}. A
moderate improvement might also be possible from RHIC jet data 
at higher center-of-mass energy, up to $\sqrt{s}=500$
GeV. Interestingly, the
contribution to the first moment from the region of the RHIC jet data
is clearly positive. The result found in this region is in good
agreement (both in terms of central value and uncertainty) with that of
\texttt{DSSV++}~\cite{Aschenauer:2013woa}. This provides some evidence
for a positive polarized gluon, though unfortunately the large
uncertainty due to the extrapolation does not allow one to draw firm
conclusions about the full first moment.

\subsection{Single-particle inclusive production asymmetries at RHIC}
\label{sec:pionpheno}

As an illustration of the predictive power of the
{\tt NNPDFpol1.1} set, in this section we compute the
longitudinal double-spin asymmetry for single-hadron
production in polarized proton-proton collisions, which has also
been measured at RHIC recently.
As for inclusive jets, Eq.~(\ref{eq:ALL-jet}), 
this asymmetry is defined as
\begin{equation}
A_{LL}^{h}=\frac{\sigma^{++}-\sigma^{+-}}{\sigma^{++}+\sigma^{+-}}
\equiv \frac{d\Delta\sigma}{d\sigma}
\mbox{ ,}
\label{eq:RHICasy}
\end{equation}
where $\sigma^{++}$ ($\sigma^{+-}$) is the cross-section for the process 
with equal (opposite) proton beam polarizations.
Theoretical predictions for the polarized 
and unpolarized differential cross-sections $d\Delta\sigma$ and
$d\sigma$ may be obtained using factorized expression with the general
structure
\begin{eqnarray}
d\sigma&=&\sum_{a,b,c=q,\bar{q},g} f_a \otimes  f_b 
\otimes D_c^H \otimes d\hat{\sigma}^{c}_{ab}
\mbox{ ,}
\label{eq:numRHIC}
\\
d\Delta\sigma&=&\sum_{a,b,c=q,\bar{q},g}\Delta f_a \otimes \Delta f_b 
\otimes D_c^H \otimes d\Delta\hat{\sigma}^{c}_{ab}
\mbox{ ,} \nonumber
\label{eq:denRHIC}
\end{eqnarray}
where the sum runs over all
initial- and final-state partonic channels, 
$d\Delta\hat{\sigma}^{c}_{ab}$, $d\hat{\sigma}^{c}_{ab}$ are
respectively the polarized and unpolarized cross-sections for the
partonic subprocess 
\be
a + b \to \lp c \to H \rp  + X, \, 
\ee
and $D_c^H$ is the fragmentation function for parton $c$ into hadron
$H$. 

\begin{figure}[!t]
\begin{center}
\epsfig{width=0.4\textwidth,figure=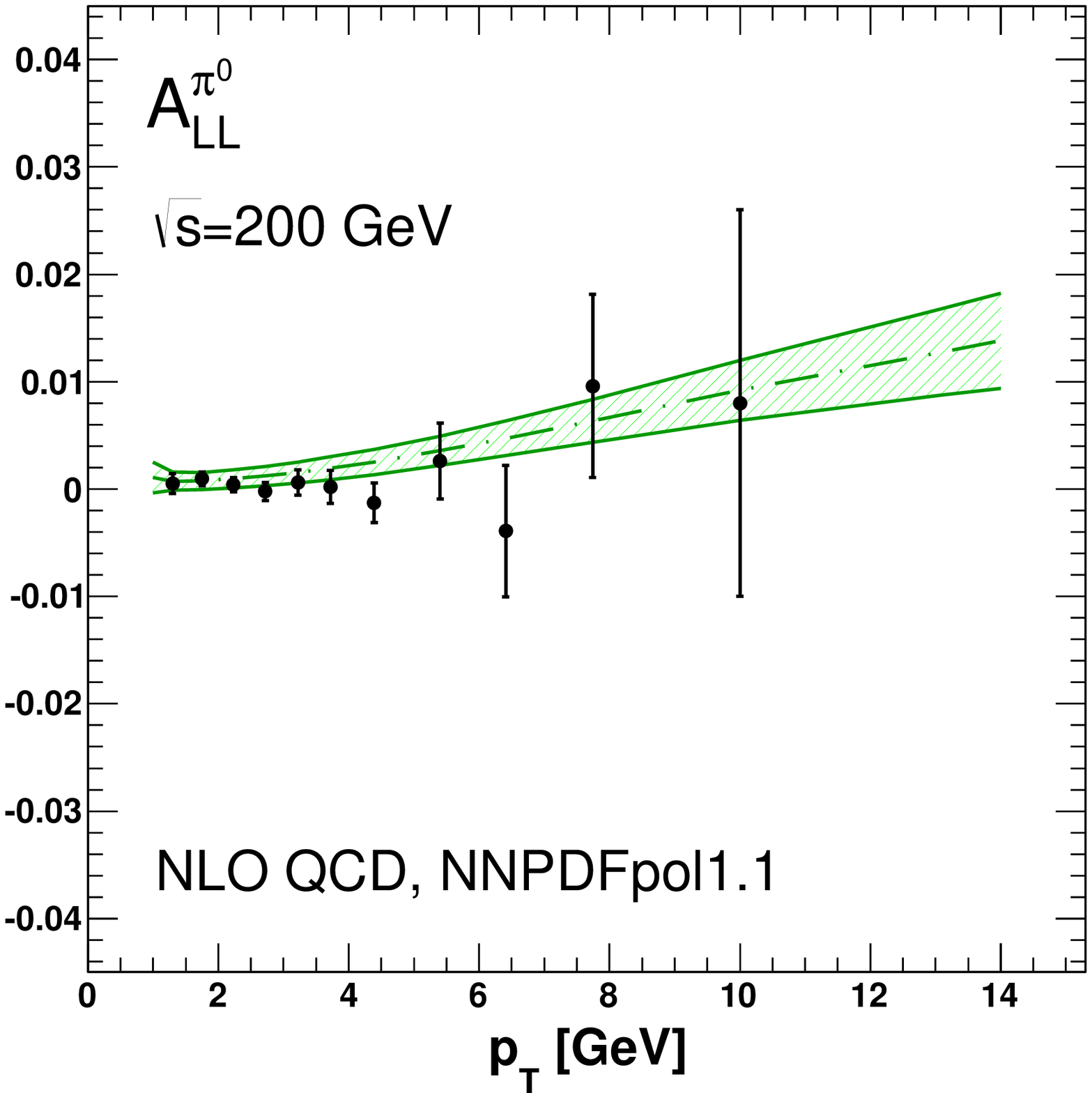}
\epsfig{width=0.4\textwidth,figure=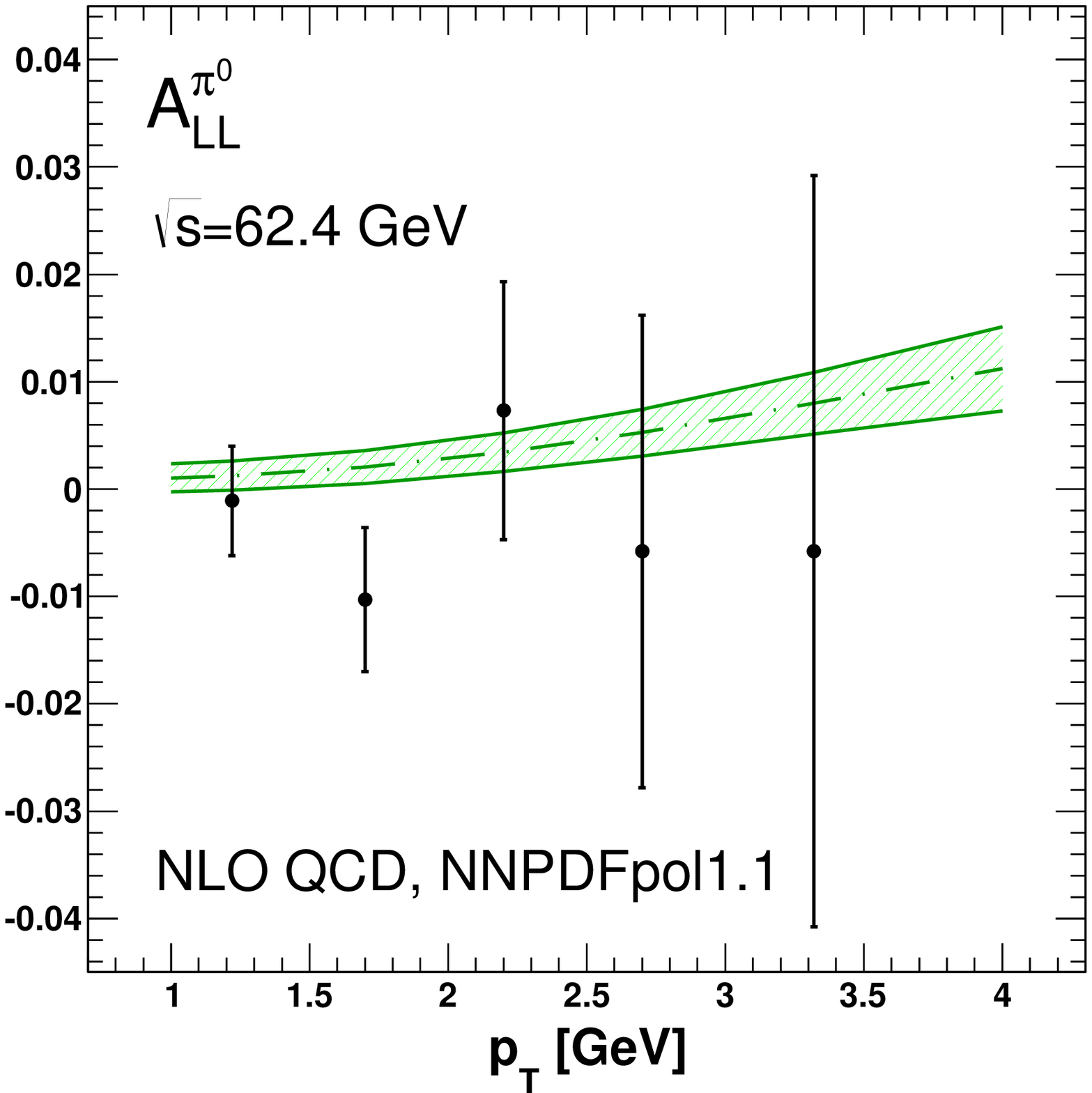}\\
\epsfig{width=0.4\textwidth,figure=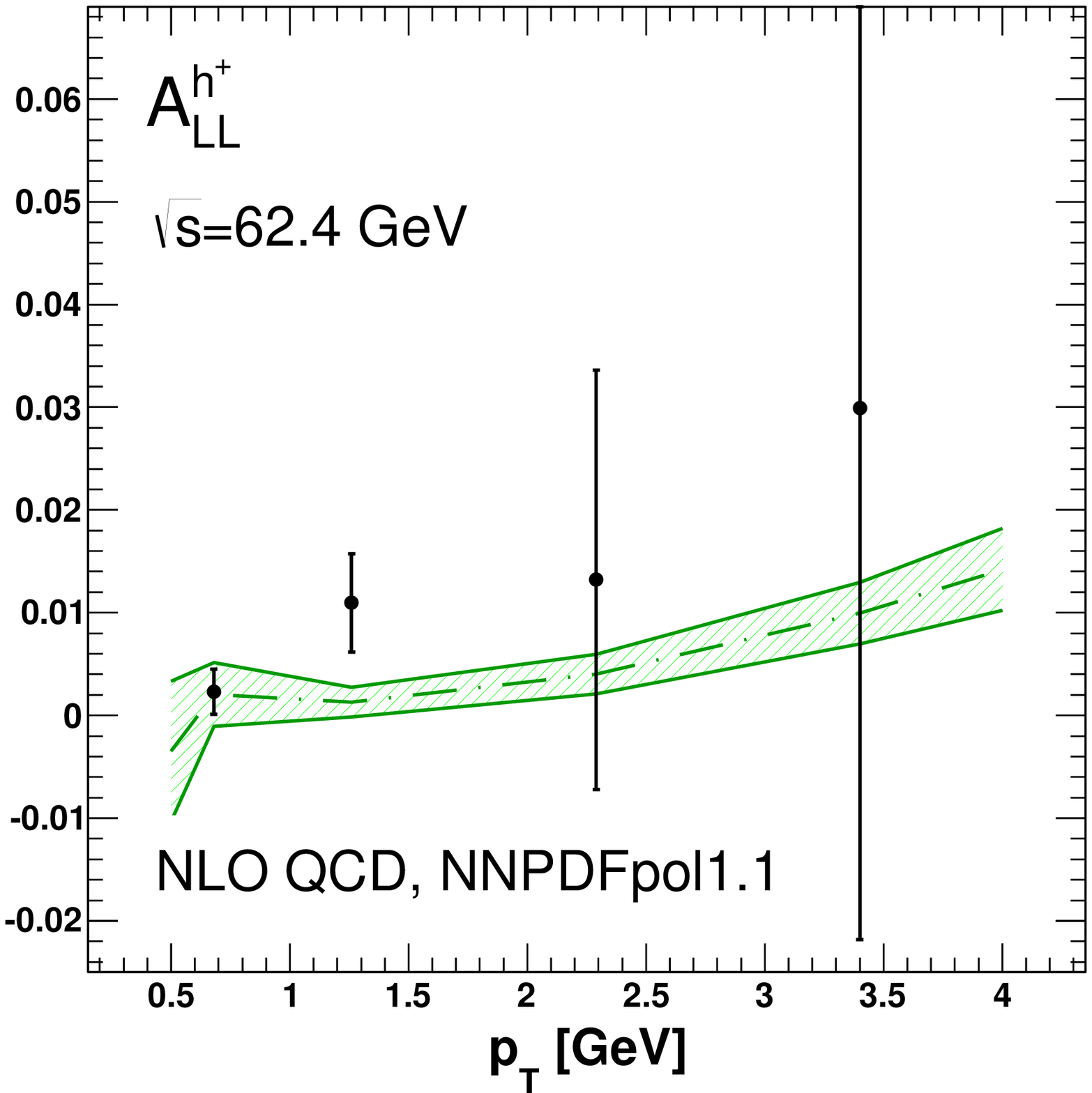}
\epsfig{width=0.4\textwidth,figure=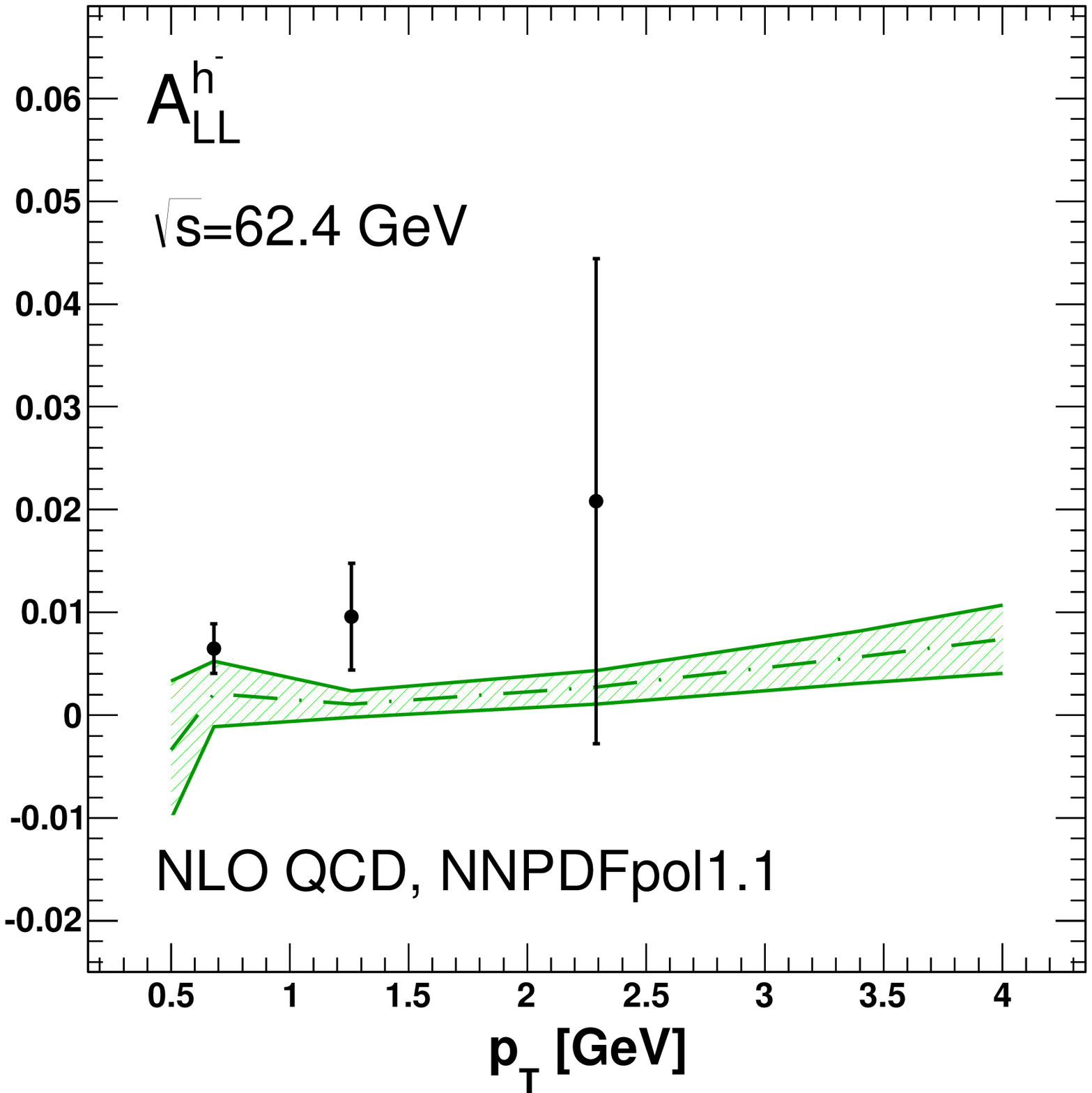}
\caption{\small NLO predictions for neutral-pion (upper plots) 
and positively and negatively charged hadrons (lower plots) 
spin asymmetries using {\tt NNPDFpol1.1} with the
corresponding PDF uncertainty band, compared to available data from 
the PHENIX experiment presented in
Refs.~\cite{Adare:2008aa,Adare:2008qb,Adare:2012nq,Adare:2014hsq}.}
\label{fig:phenixasy1}
\end{center}
\end{figure}

Predictions for these processes are subject to a significant uncertainty due
to lack of knowledge of the relevant fragmentation functions:
therefore, we did
not use them for PDF determination in order not to introduce a
theoretical uncertainty over which we have poor control. 
On the other hand, the relevant partonic cross-sections are known up
to NLO, and may be obtained using the code of
Ref.~\cite{Jager:2002xm}.

\begin{figure}[!t]
\begin{center}
\epsfig{width=0.4\textwidth,figure=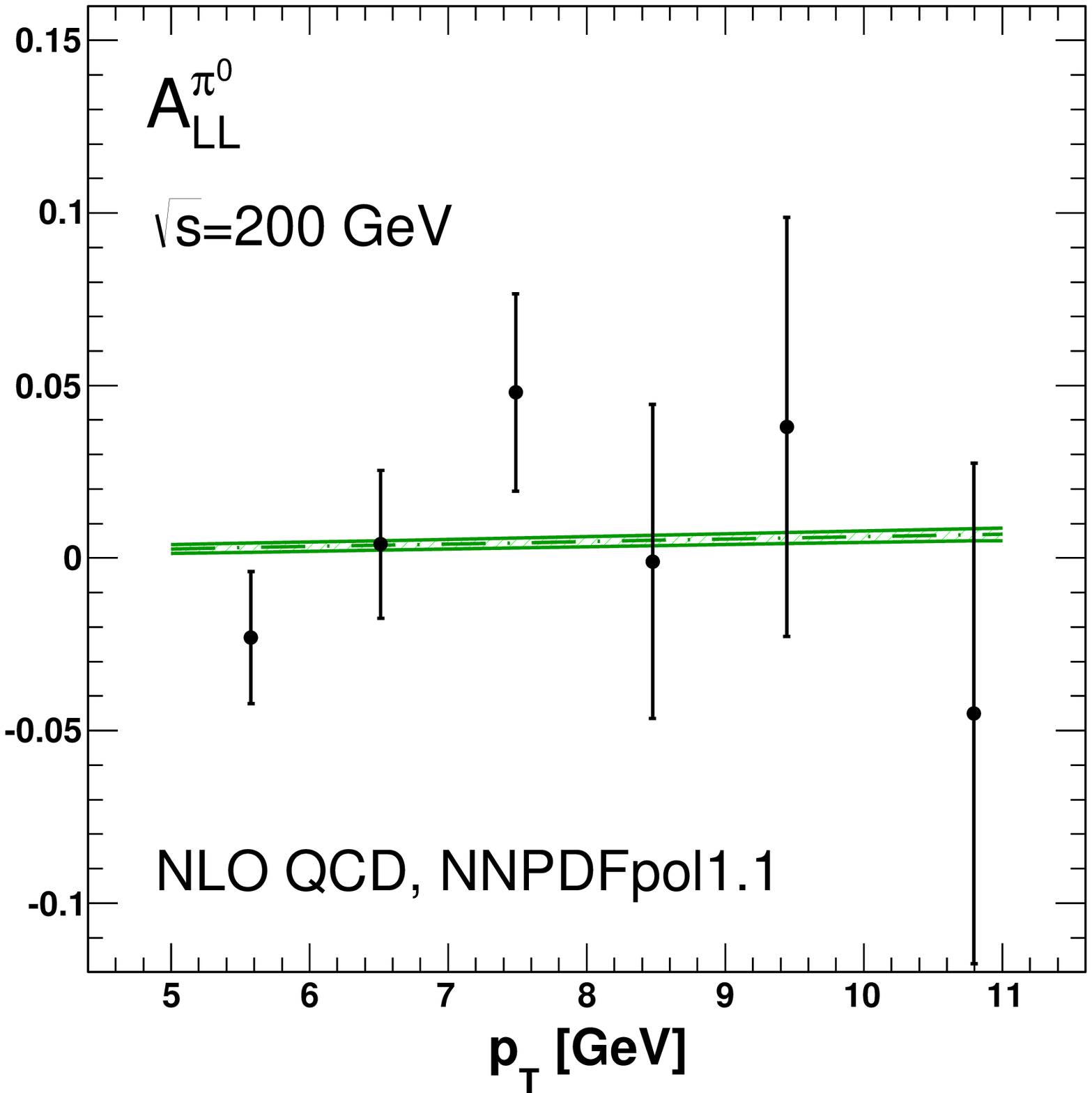}
\epsfig{width=0.4\textwidth,figure=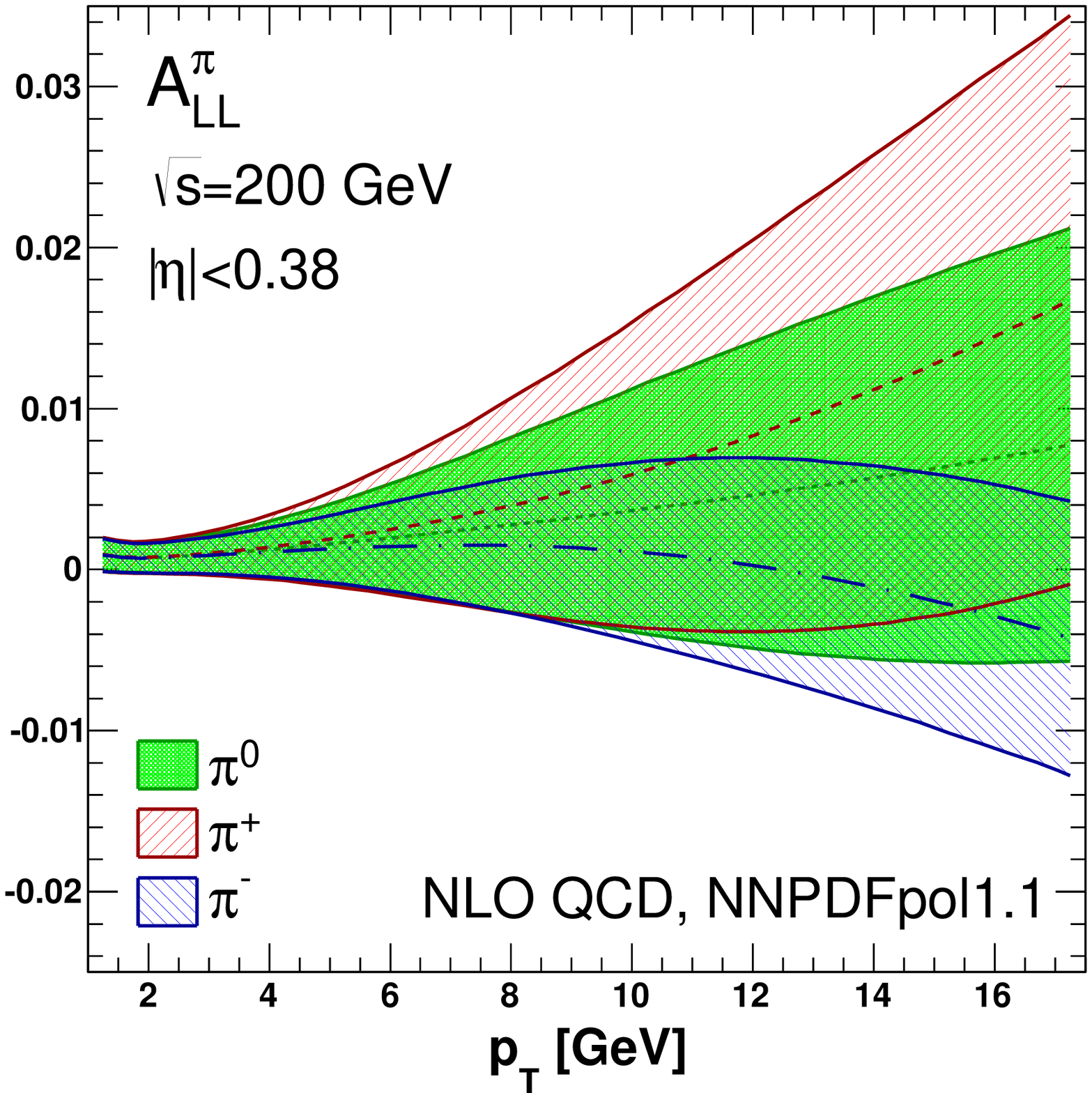}\\
\caption{\small (Left panel) Predictions for the neutral-pion spin asymmetry 
compared to data measured by STAR~\cite{Adamczyk:2013yvv}.
(Right panel) Prediction for the neutral- and charged-pion spin
asymmetries in the kinematic range accessed by upcoming 
PHENIX measurements.}
\label{fig:phenixasy2}
\end{center}
\end{figure}

In  Figs.~\ref{fig:phenixasy1}-\ref{fig:phenixasy2} we show
predictions for several of
these 
 double-spin asymmetries, compared to RHIC data, obtained using
the \texttt{NNPDFpol1.1} parton set. We use the
fragmentation functions from the {\tt DSS07}
set~\cite{deFlorian:2007aj}.
All uncertainties shown are PDF uncertainties  from the
$N_{\mathrm{rep}}=100$ polarized replica set (with, as usual, the
unpolarized denominator computed using the central PDF, as its
uncertainty is negligible). Of course, an extra unknown uncertainty
from the fragmentation function should also be included, but it cannot
be reliably estimated at present.

We specifically compare to PHENIX data for 
neutral pion production at
$\sqrt{s}=200$ GeV~\cite{Adare:2008aa,Adare:2014hsq} and 
$\sqrt{s}=62.4$ GeV~\cite{Adare:2008qb}, and 
mid-rapidity ($|\eta|<0.35$) charged hadron production at $\sqrt{s}=62.4$ 
GeV~\cite{Adare:2012nq}, and to  STAR data for 
neutral-pion production with
forward rapidity ($0.8<\eta<2.0$) 
at $\sqrt{s}=200$ GeV~\cite{Adamczyk:2013yvv}. 
Earlier PHENIX data for
neutral pion production~\cite{Adler:2004ps,Adler:2006bd,Adare:2007dg}, with 
significantly larger uncertainties, are not considered.

Our predictions are always
 in good agreement with the data within 
experimental uncertainties; they  suggest that double-spin asymmetries
for single-hadron production remain quite small in all the
available $p_T$ range, typically below the 1\% level.
Our predictions for negatively charged pion asymmetry
is also  small for all transverse momenta, see
Fig.~\ref{fig:phenixasy2}.
In contrast, $A_{LL}^{\pi^+}$ is larger than $A_{LL}^{\pi^0}$.
High-$p_T$ data (both polarized and unpolarized) 
are potentially sensitive to the gluon distribution, hence these data
might eventually provide a further handle on the polarized gluon, if
sufficiently accurate fragmentation functions become available.

\section{Conclusions and outlook}
\label{sec:conclusions}

We have presented a first global polarized PDF determination based on
NNPDF methodology, which includes, on top of the deep-inelastic
scattering data already used in our previous {\tt NNPDFpol1.0}
polarized PDF set, COMPASS charm production data and
all relevant inclusive hadronic data from polarized collisions at RHIC, 
i.e. essentially all available
data which do not require knowledge of light-quark fragmentation functions.
We have thus achieved a significant improvement in accuracy in the
determination of the gluon distribution in the medium and small-$x$
region (from jet data), with evidence for a positive gluon polarization in this
region, and a determination of individual light quark and antiquark PDFs
(from $W^\pm$ productions data). 
Together with the available NNPDF unpolarized PDF sets (currently
{\tt NNPDF2.3}~\cite{Ball:2012cx}) this provides a first global set of
polarized and unpolarized PDFs determined with a consistent
methodology, including mutually consistent constraints from 
cross-section positivity. This provides a reliable
framework for phenomenological applications, also including
possible searches for new physics with polarized beams~\cite{Fuks:2014uka}.

Future inclusive RHIC data (specifically from the PHENIX and STAR 
collaborations)
with improved accuracy and kinematic coverage will lead to even more precise
polarized PDF determinations. In addition, a significant potential for 
improvement
lies in the use of semi-inclusive data, whose availability and
accuracy is
constantly increasing both from
fixed-target~\cite{Ackerstaff:1999ey,Adeva:1997qz,Airapetian:2004zf,Alekseev:2007vi,Alekseev:2009ab}
and RHIC collider~\cite{Adare:2008qb,Adare:2008aa,Adamczyk:2013yvv} data.
However, a consistent inclusion of these data in our 
framework requires a simultaneous determination of
fragmentation functions using NNPDF methodology.

However, 
present and future data from existing facilities are unlikely to
substantially improve our knowledge of polarized first moments, i.e. of
the proton spin structure. Indeed,  the accuracy of present
determinations of polarized first moments is  already limited by the
uncertainties due to extrapolation into the unmeasured
 small-$x$ region. This is especially true for the polarized gluon, for
 which we have now tantalizing evidence of a positive polarization in
 the measured region, which is however completely swamped by an
 uncertainty from the extrapolation region which is larger by one
 order of magnitude. Improved accuracy requires a widening of the
 kinematic coverage: this could be achieved at
a polarized Electron-Ion 
Collider~\cite{Deshpande:2005wd,Accardi:2012hwp,Boer:2011fh}, which
would probe polarized PDFs down
to much smaller values of $x$, as shown 
quantitatively in Refs.~\cite{Aschenauer:2012ve,Ball:2013tyh}, and
would also provide further  constraints on  flavor separation 
from polarized charged-current 
DIS~\cite{Aschenauer:2013iia}.

\bigskip
\bigskip
\begin{center}
\rule{5cm}{.1pt}
\end{center}
\bigskip
\bigskip
The {\tt NNPDFpol1.1} polarized PDFs, with $N_{\rm rep}=100$ replicas,
are available from the NNPDF {\tt HepForge} web site,
\begin{center}
{\bf \url{http://nnpdf.hepforge.org/}~}
\end{center}
in a format compliant with the {\tt LHAPDF} interface~\cite{Whalley:2005nh,web:LHAPDF}. 
In addition, stand-alone Fortran77, C++ and Mathematica driver codes are
 also available there.

\vspace{2cm}
{\bf\noindent  Acknowledgments \\}

We would like to thank M.~Stratmann and D.~de~Florian for useful discussions
and for providing us with code for jet and $W$ boson production
at polarized hadron colliders~\cite{Jager:2004jh,deFlorian:2010aa}, 
and S.~Frixione for providing us with the code of 
Ref.~\cite{deFlorian:1998qp} for jets.
We also would like to thank C.~Aidala, C.~Gagliardi, A.~Vossem and J.~Stevens 
for useful
information about the PHENIX and STAR data, and R.~Fatemi for providing
us with the STAR data used in this publication.
This research was supported by an Italian PRIN2010 grant (SF, ERN,
GR), by a UK STFC grant ST/J000329/1 (RDB), by a European Investment 
Bank EIBURS grant (SF, ERN), and by the European Commission through 
the HiggsTools Initial Training Network PITN-GA-21012-316704.


\end{document}